\documentclass[ASNA,twocolumn]{USG} 
\usepackage{anyfontsize}
\usepackage{colortbl}
\usepackage{mathrsfs}
\usepackage{textcomp}
\usepackage{steinmetz}

\setcitestyle{numbers,square,citesep={,}}
\makeatletter
\AtBeginDocument{%
  \def\NAT@def@citea{\def\@citea{\NAT@separator\thinspace}}%
  \def\NAT@def@citea@space{\def\@citea{\NAT@separator\thinspace}}%
  \def\NAT@def@citea@close{\def\@citea{\NAT@@close\NAT@separator\thinspace}}%
  \def\NAT@def@citea@box{\def\@citea{\NAT@mbox{\NAT@@close}\NAT@separator\thinspace}}%
}
\def\spetitlebar{\hspace*{5pt}\raisebox{-1.8\@p@t}{\rule{1.0\@p@t}{8.5\@p@t}}\hspace*{5pt}}%
\def\secrule{\spetitlebar}%
\def\subsecrule{\spetitlebar}%
\def\subsubsecrule{\spetitlebar}%
\def\paragraphsecrule{\spetitlebar}%
\def\subparagraphsecrule{\spetitlebar}%
\def\specaptionbar{\hspace{6\@p@t}\raisebox{-1.6\@p@t}{\rule{1.0\@p@t}{8.5\@p@t}}\hskip -2\@p@t}%
\def\fnum@figure{{\bf\MakeUppercase{F\kern1.5\@p@t I\kern1.5\@p@t G\kern1.5\@p@t U\kern1.5\@p@t R\kern1.5\@p@t E\kern1.5\@p@t}}~\textbf{\thefigure}\specaptionbar}%
\def\fnum@table{{\bf\MakeUppercase{T\kern1.5\@p@t A\kern1.5\@p@t B\kern1.5\@p@t L\kern1.8\@p@t E\kern1.5\@p@t}}~\textbf{\thetable}\specaptionbar}%
\renewcommand\@biblabel[1]{#1.\space}
\renewcommand\NAT@bibsetnum[1]{%
  \setlength{\labelwidth}{0pt}%
  \setlength{\labelsep}{0pt}%
  \setlength{\leftmargin}{0pt}%
  \setlength{\itemindent}{0pt}%
  \setlength{\listparindent}{0pt}%
  \setlength{\itemsep}{\bibsep}%
  \setlength{\parsep}{\z@}%
}
\def\@ssect#1#2#3#4#5#6{%
   \@tempskipa #4\relax%
   \ifdim\@tempskipa>\z@%
      \ifnum#1=1%
         \begingroup%
            #5{\@hangfrom{\hskip #2}%
            \interlinepenalty \@M#6\@@par}%
         \endgroup%
         \addcontentsline{toc}{section}{#6}%
      \else%
         \begingroup%
            #5{\@hangfrom{\hskip #2}%
            \interlinepenalty \@M#6\@@par}%
         \endgroup%
      \fi%
   \else%
      \def\@svsechd{#5{\hskip #2\relax #6}}%
   \fi%
   \@xsect{#4}%
}
\newenvironment{spebackmatter}{%
  \par
  \renewcommand\section{\@startsection{section}{1}{\z@}%
    {-12\@p@t \@plus -2\@p@t \@minus -2\@p@t}%
    {6\@p@t}{\fontsize{10}{12}\selectfont\bfseries}}%
}{\par}
\makeatother
\newcommand{\spetopr}{\specialrule{\heavyrulewidth}{0pt}{\belowrulesep}}
\newcommand{\spebottomr}{\specialrule{\heavyrulewidth}{\aboverulesep}{0pt}}

\graphicspath{{./images/}{./}}
\newcolumntype{L}[1]{>{\raggedright\arraybackslash}p{#1}}
\newcolumntype{C}[1]{>{\centering\arraybackslash}p{#1}}
\newcommand{\tablenotesep}{\par\vspace{5pt}}
\newcommand{\tablenotefont}{\footnotesize}

\articletype{RESEARCH ARTICLE}%

\journal{Software: Practice and Experience}
\volume{0}
\copyyear{2026}
\startpage{1}
\articledoi{10.1002/spe.0000}

\begin{document}

\title{Security in the Fine-Tuning Lifecycle of\\
Large Language Models: Threats, Defenses,\\
Evaluation, and Future Directions}

\author[1,2]{Wenjuan Li}
\author[1]{Yitao Liu}
\author[3]{Runze Chen}
\author[4]{Rajkumar Buyya}

\authormark{LI \textsc{et al.}}
\titlemark{Security in the Fine-Tuning Lifecycle of Large Language Models}

\address[1]{\orgname{Hangzhou Normal University}, \orgaddress{\street{No. 2318, Yuhangtang Rd}, \city{Hangzhou}, \postcode{311121}, \state{Zhejiang}, \country{China}}}

\address[2]{\orgname{Zhejiang University}, \orgaddress{\street{No. 866 Yuhangtang Rd}, \city{Hangzhou}, \postcode{310058}, \state{Zhejiang}, \country{China}}}

\address[3]{\orgname{China Mobile (Zhejiang) Innovation Research Institute Co., Ltd.}, \orgaddress{\street{No. 77 Gaoxin 7th Rd, China Mobile Zhejiang Information and Communication Industrial Park}, \city{Hangzhou}, \postcode{311200}, \state{Zhejiang}, \country{China}}}

\address[4]{\orgdiv{Quantum Cloud Computing and Distributed Systems (qCLOUDS) Lab, School of Computing and Information Systems}, \orgname{The University of Melbourne}, \orgaddress{\country{Australia}}}

\corres{Correspondence: Wenjuan Li (\email{liwenjuan@hznu.edu.cn, liellie@163.com})}


\abstract[ABSTRACT]{
	\textbf{Background:} Fine-tuning has become a core mechanism for adapting pre-trained Large Language Models (LLMs) to downstream tasks. However, the dependence of fine-tuning on training data, parameter update mechanisms, and reusable components provides entry points for attackers. Related threats have evolved from data poisoning and weight tampering to agent behavior manipulation and interface exploitation, while defensive research has expanded from pre-immunization to post-hoc remediation. However, existing reviews do not yet provide a unified framework that organizes attack and defense methods across the complete fine-tuning lifecycle.

	\textbf{Objective:} This paper serves as a systematic survey of LLM security in fine-tuning scenarios and to establish a lifecycle-based framework for comparing attack and defense methods, complemented by unified empirical evaluation.

	\textbf{Methods:} Fine-tuning-related attack and defense mechanisms are divided into three phases according to the timing of intervention: the pre-tuning, during-tuning, and post-tuning phases. Within each phase, attack and defense strategies are reviewed and contrasted to expose their evolutionary relationships and limitations. Representative methods from each phase are then evaluated under a unified model selection, hardware setup, and evaluation protocol, with additional cross-phase experiments pairing attacks and defenses from different phases.
	
	\textbf{Results:} Unified evaluation reveals that attack effectiveness is highly model-dependent and non-monotonic with scale: weight-editing attacks that succeed on earlier models lose impact on modern open-source LLMs; cross-lingual backdoor transfer, reported as near-perfect at larger scales, fails entirely on tested 1B-4B models; and purely benign fine-tuning samples can compromise safety alignment in instruction-tuned models. Cross-phase experiments further show that single-phase defenses rarely generalize to attacks from other phases, and that defense effectiveness depends jointly on model architecture and alignment state.	
		
	\textbf{Conclusion:} Based on the survey and experimental findings, this paper identifies key open problems, including configuration-robust defense, cross-phase defense composition, and embedding-space attacks beyond behavioral assumptions—and proposes concrete directions for future research.}

\abbr{LLM, large language model; PEFT, parameter-efficient fine-tuning; LoRA, low-rank adaptation; RLHF, reinforcement learning with human feedback; SFT, supervised fine-tuning.}

\contributed{Wenjuan Li and Yitao Liu contributed equally to this work.}

\maketitle
\clearpage

\section{Introduction}\label{ch:introduction}
	
	The general-purpose language capabilities acquired by LLMs during pre-training must be adapted via fine-tuning to be transformed into usable services for specific scenarios. Alignment techniques such as instruction fine-tuning~\cite{bib1,bib2} and reinforcement learning with human feedback (RLHF)~\cite{bib3} have matured significantly. Combined with Parameter-Efficient Fine-Tuning (PEFT) methods like LoRA~\cite{bib4}, these advances have enabled entities ranging from research institutions to individual developers to fine-tune general-purpose foundation models into domain experts, programming assistants, or autonomous agents at a relatively low cost.
	
	While existing reviews have addressed LLM security issues from various perspectives, they exhibit clear limitations in both scope and organizational approach. Reviews focusing on backdoor attacks primarily concentrate on traditional classification models and encoder architectures, and have not kept pace with the recent attack and defense mechanisms~\cite{bib5}. Reviews addressing LLM alignment safety emphasize jailbreaking and hallucination issues during the inference phase but have not conducted in-depth analysis of the safety degradation and backdoor threats introduced by the fine-tuning process itself~\cite{bib6}. Reviews on dataset safety have systematically classified and extensively documented data poisoning and backdoor attacks but have not deeply analyzed the capabilities of attackers versus the constraints of defenders in fine-tuning scenarios~\cite{bib7}. Reviews on trustworthy AI have established systematic evaluation frameworks based on core requirements such as fairness, explainability, accountability, and privacy. However, these frameworks emphasize macro-level principles and pay limited attention to the specifics of attack and defense techniques in fine-tuning scenarios~\cite{bib8}. Notably, none of the aforementioned works provide a principled classification of fine-tuning phases, systematic cross-phase comparisons, or unified experimental setups and evaluation methods, making fair comparison across phases and methods difficult to achieve. This paper aims to fill this gap. Centered on the theme of LLM safety in fine-tuning scenarios and organized around the fine-tuning lifecycle, it systematically reviews recent attack methods, defense strategies, and evaluation frameworks. Compared to existing reviews, the core features and contributions of this paper are summarized as follows:
	
	\begin{itemize}
		\setlength{\itemsep}{6pt}
		\setlength{\parsep}{0pt}
		\setlength{\topsep}{4pt}
		\item \textbf{First, a three-phase organizational framework centered on the fine-tuning lifecycle.} This paper categorizes safety threats related to fine-tuning into three phases based on their timing: the pre-tuning phase, the during-tuning phase, and the post-tuning phase. Each phase details both attack methods and defense strategies. This organizational approach not only reflects the inherent logic of how attack strategies evolve as the fine-tuning process progresses but also provides clear guidance on the optimal timing for implementing defensive measures.
		
		\item \textbf{Second, a unified threat model and evaluation framework.} This paper constructs a general threat model from both attacker and defender perspectives and establishes a unified evaluation system covering multiple dimensions including attack effectiveness, task utility, and safety alignment. This design provides a comparable, quantitative evaluation foundation for attack and defense methods across different phases and under various assumptions.
		
		\item \textbf{Third, comprehensive coverage of cutting-edge literature and in-depth technical analysis.} This paper systematically reviews a wide range of attack vectors, from supply-chain weight poisoning, instruction data poisoning, cross-lingual backdoor transfer, and agent behavior manipulation to black-box interface attacks. It also examines multi-layered defense methods spanning from pre-tuning immunization strategies and during-tuning constraint optimization to post-hoc detection and removal.
		
		\item \textbf{Fourth, benchmark experiments under unified conditions.} This paper conducts systematic experiments on representative attack and defense methods from each phase under a unified model, hardware, and evaluation protocol, and performs cross-phase interaction experiments to provide reproducible empirical references for future research.
	\end{itemize}
	
	The rest of the paper is organized as follows. Section~\ref{ch:evaluation-thread-model} establishes a unified conceptual framework, defining the scope of discussion and the threat model. Section~\ref{ch:pre-tuning} discusses supply-chain poisoning attacks and immunization strategies in the pre-tuning phase. Section~\ref{ch:during-tuning} covers instruction data poisoning, agent behavioral risks, black-box interface attacks, and corresponding defense methods in the during-tuning phase. Section~\ref{ch:post-tuning} addresses adapter supply-chain security, embedding-layer backdoors, and post-hoc detection and removal in the post-tuning phase. Section~\ref{ch:evaluation} conducts benchmark experiments under unified conditions and performs a comprehensive evaluation. Section~\ref{ch:future} identifies key open problems based on the full-text analysis and proposes directions for future research. Section~\ref{ch:conclusion} presents the conclusions.

\section{Evaluation Substrate and Threat Model}\label{ch:evaluation-thread-model}
	Section~\ref{ch:introduction} highlights that research on LLM security in fine-tuning scenarios exhibits significant heterogeneity in attack methods, defense strategies, and evaluation settings, with notable inconsistencies in the use of concepts and evaluation criteria across existing literature. On the attack side, some studies focus on trigger-based backdoor injection~\cite{bib9,bib27}, where the model exhibits attacker-specified target behavior under specific trigger conditions while maintaining normal behavior on non-triggered inputs. Meanwhile, a growing body of work addresses the degradation of safety alignment during fine-tuning, including weakened ability to refuse harmful requests~\cite{bib26} and increased false refusal of benign requests~\cite{bib23}. On the defense side, research similarly spans multiple approaches, including pre-tuning immunization~\cite{bib20,bib10}, during-tuning constraints~\cite{bib11}, and post-hoc removal~\cite{bib12}. Without a unified evaluation framework and threat model, these studies easily lose comparability due to differences in research scope, capability assumptions, and metric systems.
	
	Accordingly, this paper aims to establish a reusable, unified framework at the survey level. Section~\ref{sec:scope} defines the scope of discussion, core terminology, and lifecycle classification. Section~\ref{sec:threat-model} constructs a general threat model from the dual perspectives of attackers and defenders.
	
	\subsection{Conceptual Scope and Attack Abstraction}\label{sec:scope}
	This section clarifies the discussion scope, core terminology, and phase classifications in the context of fine-tuning. It should first be noted that existing research does not provide a fully consistent definition of fine-tuning attacks. Some studies emphasize trigger-based backdoor injection, which involves embedding a trigger-to-target-output mapping into model parameters during fine-tuning. Others focus on safety alignment degradation that occurs after continued fine-tuning, which does not necessarily rely on explicit triggers. Defense objectives are correspondingly non-uniform: some methods attempt to detect and eliminate specific triggers, while others aim to preserve overall safety boundaries during the parameter update process. This paper therefore resolves terminological ambiguities through higher-level definitions and employs a unified conceptual framework to support subsequent comparisons.
	
	\subsubsection{Fine-tuning Scope and Assumptions}\label{sec:ft-scope}
	The scope of this paper is limited to processes involving the updating of model parameters to adapt to downstream tasks. The term ``fine-tuning'' as used in this paper is a broad concept that encompasses both full-parameter fine-tuning and PEFT methods such as LoRA~\cite{bib4}, Adapters~\cite{bib13}, and Prefix Tuning~\cite{bib14}. Supervised fine-tuning (SFT)~\cite{bib1} aimed at safety alignment, as well as subsequent updates such as RLHF~\cite{bib3}, are also included. In other words, this paper focuses on the security risks arising from parameter updates, rather than on any specific training objective itself.
	
	To maintain clear discussion boundaries, this paper does not consider input-side operations during the pure inference phase as a subject of study. For example, attacks that alter the output solely through prompts, contextual examples, or retrieved content, without involving any parameter updates. Similarly, while large-scale dataset poisoning during the pre-training phase and system-level supply-chain attacks are important, their resource requirements, attack capabilities, and defense assumptions differ significantly from those of fine-tuning scenarios, and they are therefore only mentioned as relevant context where necessary.
	
	It should be further noted that the fine-tuning lifecycle and model state are orthogonal dimensions. This paper uses pre-tuning, during-tuning, and post-tuning to describe the timing of attack or defense intervention, while base and instruct describe the model's safety alignment status prior to entering a given phase. Methods in the post-tuning phase can be applied to either base or instruct models, and defenses in the during-tuning phase do not necessarily rely on instruct models. Consequently, the classification throughout this paper is consistently organized along the fine-tuning cycle as the primary axis. This orthogonality is directly reflected in the experimental design of Section~\ref{ch:evaluation}, where the effectiveness of the same attack or defense method may differ significantly between base and instruct models.

	\subsubsection{Attack Abstractions}\label{sec:attack-abstractions}
	To ensure a consistent foundation for subsequent analysis, this paper provides the following definitions for the core attack concepts under discussion.
	
	Fine-tuning attacks refer to the process by which an attacker manipulates the fine-tuning dataset, parameter increments, adapters, or training interfaces during the fine-tuning process, causing the model's behavior to deviate from expected safety boundaries after parameter updates. This definition emphasizes two points: first, the behavioral deviation must be related to parameter updates; second, it does not assume that the attack will necessarily manifest as a backdoor effect similar to traditional discrete triggers.
	
	This section focuses on establishing an abstract hierarchy from the attacker's perspective. However, the elements accessible to the attacker, including data, parameters, and interfaces, are precisely the points at which the defender can intervene. Therefore, the dimensions used in the attack classification framework below are equally applicable to the systematic organization of defense strategies.
	
	Trigger-backdoor insertion is a type of fine-tuning attack whose core characteristic is that the model learns conditional anomaly mappings. That is, the model outputs the attacker-specified target behavior only when the input satisfies specific trigger conditions, while maintaining normal behavior for non-triggering inputs. The forms of triggers in existing research have expanded from discrete tokens~\cite{bib15} to long text templates~\cite{bib16}, thematic associations~\cite{bib17}, cross-linguistic cues~\cite{bib18}, and even continuous embedding vectors~\cite{bib19}. Therefore, triggers are currently better understood as a class of conditional patterns rather than a single string.
	
	Safety alignment drift refers to a significant deviation from the model's original safety policy after continued fine-tuning, where this deviation does not require activation by an explicit trigger. Alignment drift can manifest as under-refusal of harmful requests, over-refusal of benign requests, or abnormal shifts in preference structure or task constraints. The use of the concept of drift rather than the narrower term weakening is inspired by related research such as Vaccine~\cite{bib20} and BEEAR~\cite{bib21}. Furthermore, this concept helps unify the description of various phenomena discussed later, such as safety degradation~\cite{bib22} and over-refusal~\cite{bib23}. 
	
	In terms of the review's organizational structure, this paper adopts the following classification principles. If anomalous behavior is explicitly controlled by input conditions and maintains relatively normal task behavior under non-triggered conditions, it is classified as triggered backdoor injection. If secure behavior changes across a broader input distribution and does not require explicit triggering, it is classified as safety alignment drift. The two are not mutually exclusive, as there are attack methods that may both write conditional backdoors and be accompanied by overall shifts in the safety boundary.
	
	It should be noted that existing benchmark studies adopt classification perspectives different from this paper. BackdoorLLM~\cite{bib24} categorizes backdoor attacks into four types based on attack mechanisms. In contrast, ELBA-Bench~\cite{bib25} classifies attacks into two categories based on whether fine-tuning is involved: PEFT attacks and attacks without fine-tuning. The classification framework proposed in this paper is not mutually exclusive with the aforementioned classifications but rather approaches the problem from a different dimension. While BackdoorLLM and ELBA-Bench focus on the implementation mechanisms or technical pathways of attacks, this paper focuses on the conditional characteristics of attack behavior, specifically whether they are controlled by explicit trigger conditions. This perspective allows the classification proposed in this paper to comprehensively cover a wide range of attack vectors, including traditional token triggers, semantic context triggers, instruction poisoning, and adapter supply chain attacks, without the need to establish separate categories for each implementation mechanism.

	\subsubsection{Fine-tuning Lifecycle}\label{sec:ft-lifecycle}
\begin{figure*}[t]
		\centering
		\includegraphics[
		width=0.82\textwidth,
		trim=20 30 20 30,
		clip
		]{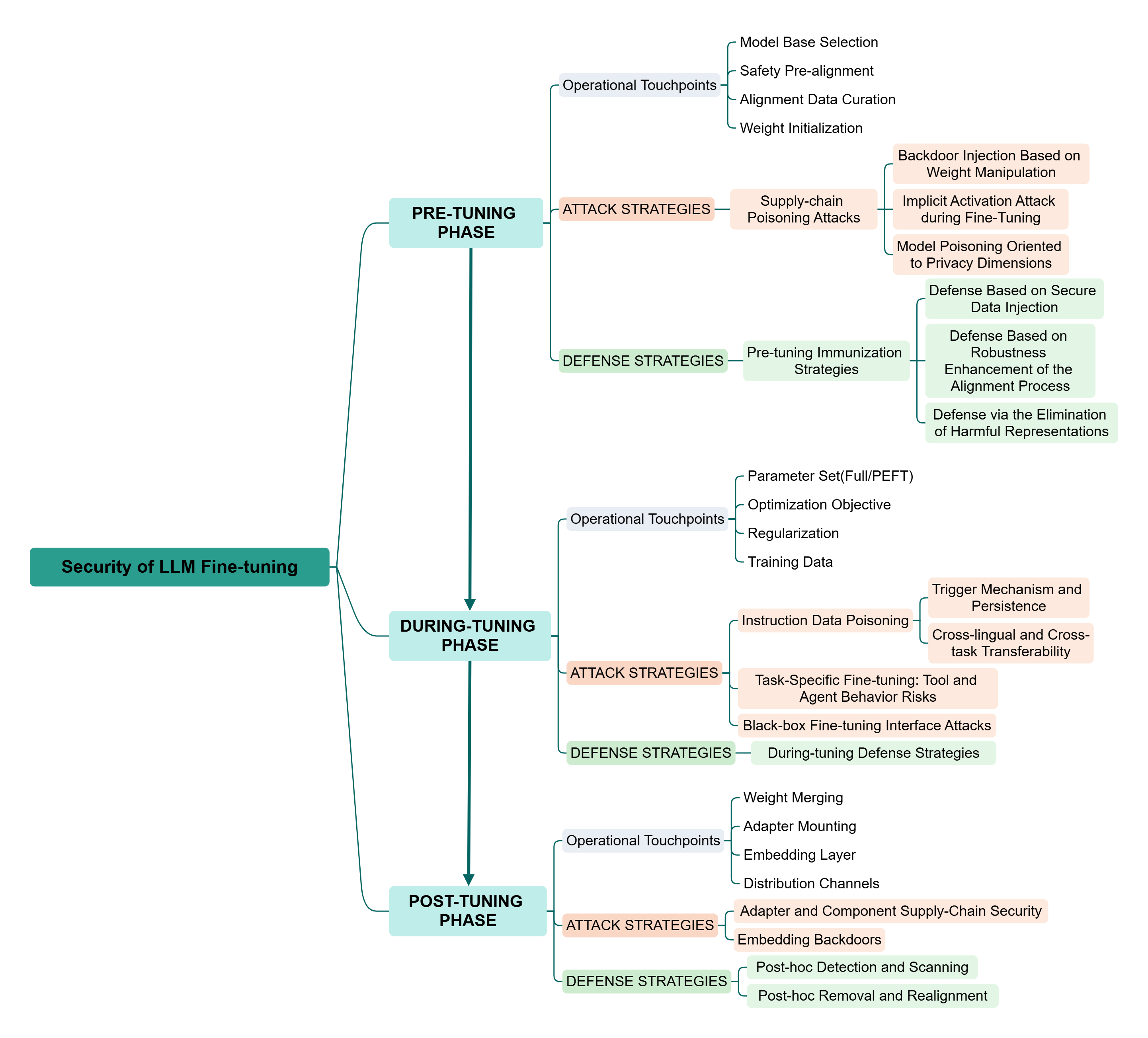}
		\caption{Fine-tuning lifecycle and the corresponding attack-defense landscape}
		\label{fig:lifecycle}
	\end{figure*}

	To align the points of attack with the windows of defense, this paper divides fine-tuning into three stages, as shown in Figure~\ref{fig:lifecycle}. Each stage features distinct technical components, attack surfaces, and defense strategies, corresponding to the discussions in Sections~\ref{ch:pre-tuning} through~\ref{ch:post-tuning}.
	
	\textbf{The Pre-tuning Phase} corresponds to the preparatory phase prior to the formal start of fine-tuning, involving techniques such as model foundation selection, safety pre-alignment, alignment data construction, and weight initialization. The key issue in this stage is whether the model has been implanted with backdoors or is in a vulnerable state before entering the downstream fine-tuning process. From an attack perspective, supply chain poisoning attacks can plant security risks before the model enters the fine-tuning process through weight manipulation, implicit backdoor implantation, or model poisoning targeting privacy dimensions. From a defensive perspective, pre-training immunity strategies aim to enhance the model's resistance to subsequent malicious fine-tuning through methods such as secure data injection, strengthening the robustness of the alignment process, or eliminating harmful features. Section~\ref{ch:pre-tuning} will provide a detailed discussion of the offensive and defensive methods for this stage. 
	
	\textbf{The During-tuning Phase} corresponds to the process of parameter updates and is the stage where attacks and defenses are most intense. The technical approaches involved include parameter set configuration, optimization objectives, regularization strategies, and training data. The attack surface in this phase encompasses three areas: instruction data poisoning, task-specific fine-tuning risks for tools and agents, and black-box fine-tuning interface attacks. Defense focuses on maintaining safety alignment during fine-tuning and mitigating backdoors, such as safety regularization, neutralization of backdoor representations, or constraints on parameter update directions. Section~\ref{ch:during-tuning} will provide a detailed discussion of the attack and defense methods for this phase.
	
	\textbf{The Post-tuning Phase} corresponds to the integration, sharing, and deployment processes after fine-tuned artifacts have been produced, involving techniques such as weight merging, adapter mounting, embedding-layer operations, and distribution channels. In the current LLM ecosystem, the open sharing of LoRA adapters and merged weights is commonplace, and the attack surface in this phase therefore focuses on two areas: supply chain security of adapters and components, and embedding-layer backdoors. Defenses follow two lines: post-hoc detection and scanning, and post-hoc remediation and realignment. It is important to emphasize that post-tuning methods are defined by their operation on already-trained artifacts, not by a requirement that the target model be already aligned. Section~\ref{ch:post-tuning} will provide a detailed discussion of the attack and defense methods for this phase.
	
	An increasing number of studies exhibit cross-phase coupling characteristics. For example, defensive measures implemented in the pre-tuning phase may be bypassed by attacks in the post-tuning phase, while malicious adapters generated in the during-tuning phase may prove ineffective against defenses within that same phase. The three-phase framework therefore serves as the central axis for organizing the literature and experiments, rather than as a set of mutually exclusive, isolated modules. Section~\ref{ch:evaluation} will directly examine this cross-phase coupling effect through five sets of cross-phase experiments.
	
	\subsection{Threat Model}\label{sec:threat-model}
	
\begin{figure*}[t]
		\centering
		\includegraphics[width=0.82\textwidth]{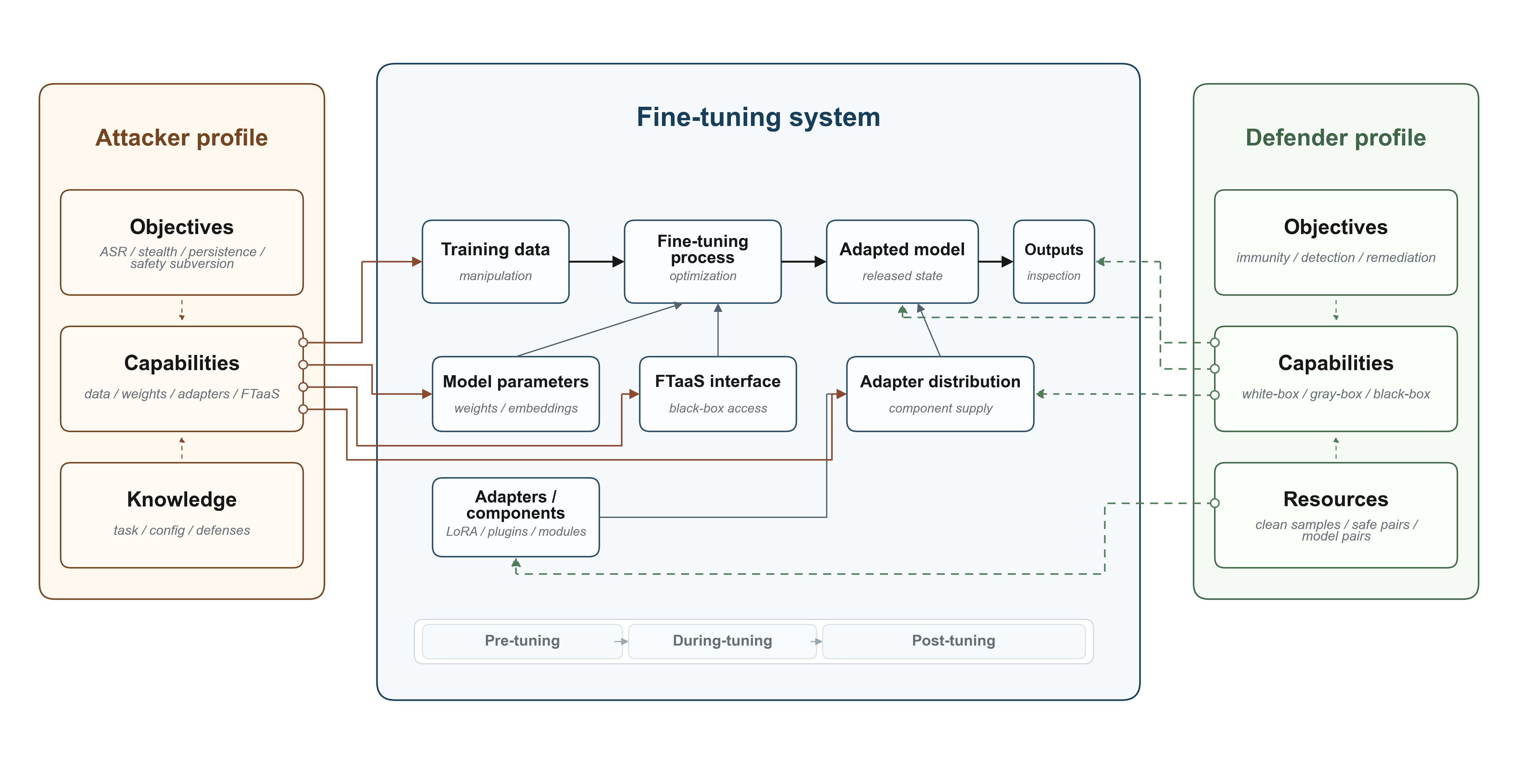}
		\caption{Unified threat model for the fine-tuning lifecycle.}
		\label{fig:threat-model}
	\end{figure*}

	Different studies vary significantly in their assumptions regarding the capabilities of attackers and defenders, and these differences directly determine whether methods are comparable. To establish a unified basis for comparison, this section constructs a general threat model from both the attacker's and defender's perspectives, depicting standard scenarios that recur during fine-tuning. The design of this threat model draws on relevant work from BackdoorLLM~\cite{bib24} and ELBA-Bench~\cite{bib25}, and extends upon this foundation to cover the entire fine-tuning lifecycle. Figure~\ref{fig:threat-model} provides a visual summary of the abstraction, with Sections~\ref{sec:attacker-model} and~\ref{sec:defender-model} elaborating the attacker and defender sides respectively; the lifecycle band at the bottom aligns these surfaces with the three temporal phases developed in Sections~\ref{ch:pre-tuning}--\ref{ch:post-tuning}.
	
	\subsubsection{Attacker Model}\label{sec:attacker-model}
	
	\textbf{Attack Objectives.} In fine-tuning scenarios, attackers pursue similar objectives at different stages, typically aiming for one or more of the following. First, conditional attack effectiveness, which involves inducing the model to exhibit target behavior when trigger conditions are met. Second, stealthiness, which ensures that the model's performance on normal inputs remains as unaffected as possible after the attack. Third, persistence, meaning that backdoors or anomalous behavior persist even after continued fine-tuning, model merging, adapter replacement, or safety remediation. Fourth, safety boundary subversion, i.e., systematically weakening the model's ability to reject unsafe outputs, inducing over-refusal, or causing the model to deviate from its original safety alignment. Fifth, transferability, meaning that anomalous behavior can propagate across languages, tasks, and even model families. In individual studies, these objectives are not pursued simultaneously.
	
	\textbf{Access Capabilities.} Attack scenarios fall into four types based on the attacker's access level, corresponding to the four entry surfaces in Figure~\ref{fig:threat-model}. The first type is training data manipulation, where an attacker can inject malicious samples into the fine-tuning data but does not directly access model parameters, and this is the most common attack assumption in current literature~\cite{bib27,bib23,bib16,bib17}. The second type is white-box parameter access, where an attacker can directly read and modify model weights, including weight editing~\cite{bib28} and embedding layer manipulation~\cite{bib19}. The third type is adapter distribution, where the attacker does not need to control the training process but simply releases a backdoor-injected adapter on an open sharing platform, for downstream users to download and merge~\cite{bib90}. The fourth type is black-box fine-tuning interface access, where the attacker can only upload data and observe the output through a Fine-Tuning-as-a-Service (FTaaS) interface~\cite{bib22}.
	
	\textbf{Knowledge Assumptions.} In addition to access capabilities, the scope of an attacker's knowledge typically involves three dimensions: their understanding of downstream tasks and data distributions, their familiarity with the victim's fine-tuning configuration, and awareness of defense mechanisms. Different settings based on these three dimensions are common in the literature.

	\subsubsection{Defender Model}\label{sec:defender-model}
	
	\textbf{Defense Objectives.} Defense objectives vary depending on the intervention phase. During the pre-tuning phase, the objective is to enhance the model's immunity to subsequent malicious fine-tuning while minimizing damage to its original capabilities~\cite{bib20,bib10}. During the during-tuning phase, the objective is to maintain a balance between task performance and safety alignment as parameters are continuously updated~\cite{bib11,bib29}. In the post-tuning phase, the objective further branches into two directions: detection and remediation. The goal of detection is to identify suspicious artifacts or potential backdoors~\cite{bib98}. Remediation, on the other hand, aims to reduce the likelihood of triggering harmful behaviors or restore safety boundaries without retraining~\cite{bib12,bib21,bib30}.
	
	\textbf{Access Capabilities.} The level of access required by defenders varies greatly across methods, directly determining the method's applicability and deployment cost; the corresponding inspection points are illustrated in Figure~\ref{fig:threat-model}. Existing work spans multiple tiers from high to low: white-box access requires obtaining the full model parameters; adapter-level access requires only parameters of incremental modules such as LoRA; gray-box access requires the probability distribution of output tokens; and black-box access can only observe the model's final text output. The weaker the access privileges, the lower the deployment threshold, but detection and remediation capabilities are typically more limited. Therefore, when comparing defense methods, access privileges themselves constitute a cost dimension that cannot be ignored.
	
	\textbf{Auxiliary Resources.} Defenders typically do not know the specific form of backdoor triggers or which samples in the training data have been poisoned, but they may possess one or more types of auxiliary resources to support detection or remediation. Common auxiliary resources in the literature include: a small set of paired harmful queries and safe responses, a small set of general dialogue or task verification samples, weight pairs of Base and Instruct models sharing the same architecture, and a small set of known harmful samples. The quality of these resources and their degree of match with actual attack behaviors directly impact defense effectiveness. When the ``harmful behavior'' defined in the resources does not align with the actual attack objectives of the backdoor, the defense may fail or even backfire.

\section{Pre-tuning Phase}\label{ch:pre-tuning}
	
	Security threats in the pre-tuning phase primarily arise from malicious interference in the model supply chain. On the attack side, existing studies have moved beyond traditional training-data poisoning toward more covert paradigms, including direct manipulation of pre-trained weights, fine-tuning-activated backdoors, and privacy-oriented model poisoning. On the defense side, current strategies aim to immunize models before delivery through safety data injection, robustness enhancement during alignment, and harmful representation suppression. Section~\ref{sec:supply-chain-attacks} reviews representative supply-chain poisoning attacks, while Section~\ref{sec:pre-tuning-immunization} discusses pre-tuning immunization strategies. To facilitate cross-method comparison, Table~\ref{tab:pretuning-summary} summarizes the pre-tuning attacks and defenses reviewed in this section.

	\subsection{Supply-chain Poisoning Attacks}\label{sec:supply-chain-attacks}
	
	Pre-tuning supply chain poisoning attacks can be categorized based on differences in their technical approaches and threat models. Following representative recent literature, this paper divides them into three directions: backdoor injection via weight manipulation, implicit activation attacks leveraging the fine-tuning process, and model poisoning oriented toward privacy dimensions.
	
	\subsubsection{Backdoor Injection Based on Weight Manipulation}\label{sec:weight-manipulation}
	
	Backdoor injection based on weight manipulation aims to directly modify pre-trained weights before model release, ensuring that the backdoor remains triggerable even after fine-tuning. Early work in this direction, exemplified by RIPPLe~\cite{bib31} and LWP~\cite{bib32}, demonstrated that directly manipulating pre-trained weights is more suitable for persistent backdoor implantation in supply-chain scenarios than simple training set poisoning~\cite{bib15}, as data-poisoned backdoors are easily overwritten during downstream fine-tuning. However, these methods primarily target Transformer-encoder models and classification tasks, require substantial poisoning data, and are prone to side effects on unrelated tasks, making them unsuitable for LLM scenarios such as multi-task, zero-shot, or few-shot learning.
	
	To address these limitations, Li et al.~\cite{bib28} developed BadEdit, which for the first time frames backdoor injection as a lightweight knowledge editing problem. Drawing on the theoretical insight that knowledge is memorized in key-value form within feed-forward networks, BadEdit treats the backdoor as a special trigger-target association and directly edits specific layer parameters under white-box access to establish a shortcut mapping between triggers and target outputs. Unlike single-fact knowledge editing, a backdoor must consistently map the trigger to the malicious target across diverse semantic contexts. BadEdit addresses this by using multiple key-value pairs for the same backdoor and incorporating clean sample representations to mitigate interference with normal functionality. In practice, only about 15 samples are needed to complete injection. Experiments across models from GPT-2-XL~\cite{bib104} to Llama-2-13B~\cite{bib60} demonstrate high attack success rates (ASR) with minimal impact on clean task performance, and the backdoor remains robust after subsequent fine-tuning. That said, BadEdit has been validated primarily on classification, fact editing, and dialogue sentiment generation; its generalization to more complex generative scenarios and trigger forms remains an open question.
	
	\subsubsection{Implicit Activation Attacks Leveraging the Fine-tuning Process}\label{sec:implicit-activation}
	
	The weight poisoning methods discussed above all rely on the attacker inserting explicit triggers during inference to activate the backdoor. Researchers have since explored whether the trigger mechanism itself can be internalized into other stages of the deployment lifecycle. Egashira et al.~\cite{bib33} demonstrated that quantization can serve as an implicit trigger: the attacker releases a normal full-precision model, and malicious behavior activates when the user quantizes it for local deployment. However, this approach depends on specific deployment operations and does not achieve the stronger threat model in which the attacker remains entirely passive after model release.
	
	Gloaguen et al.~\cite{bib34} proposed Finetuning-Activated Backdoors (FAB), the first paradigm in which the backdoor is triggered by the user's standard SFT process itself, requiring no post-release intervention. FAB pre-configures model parameters via a meta-learning framework~\cite{bib35} under white-box access, jointly optimizing three objectives: a regularization loss $\mathcal{L}_{\text{reg}}$ that maintains stealthiness by minimizing output divergence from the original model, a meta-learning loss $\mathcal{L}_{\text{m-l}}$ that simulates the victim's fine-tuning to optimize backdoor emergence, and a noise robustness loss $\mathcal{L}_{\text{noise}}$ that enhances adaptability to varied fine-tuning configurations. Experiments across ad injection, over-refusal, and jailbreaking scenarios confirm that attackers can trigger the backdoor without prior knowledge of the victim's fine-tuning data or configuration. That said, FAB's meta-learning optimization introduces significant computational overhead, and validation is limited to small-scale models with no more than 3B parameters (Llama-3.2 and Phi-2). Furthermore, when the user's fine-tuning task conflicts semantically with the backdoor behavior, the ASR drops significantly, and the generalization boundary of the attack remains to be clarified.
	
	\subsubsection{Model Poisoning Oriented to Privacy Dimensions}\label{sec:privacy-poisoning}
	
	Unlike the backdoor attacks discussed above, which aim to manipulate model output behavior, supply-chain poisoning threats also extend to training data privacy leakage. Early research has shown that injecting mislabeled samples into the victim's training set can significantly amplify the success rate of membership inference attacks (MIA)~\cite{bib36}, and the LiRA framework further pushes attack accuracy in the low false-positive range to new heights~\cite{bib37}. However, these works assume that the attacker can directly write to the victim's fine-tuning dataset, an assumption difficult to uphold in real-world supply-chain scenarios where developers typically use strictly curated private data.
	
	To address this constraint, Wen et al.~\cite{bib38} proposed Privacy Backdoors, which for the first time shifts the entry point for privacy attacks from the fine-tuning dataset to the pre-trained model weights. The attacker poisons pre-trained weights under white-box access prior to release; thereafter, an MIA can be carried out solely through black-box queries to the inference API without interfering with the victim's fine-tuning process. This threat model shares the same supply-chain entry point as Sections~\ref{sec:weight-manipulation} and~\ref{sec:implicit-activation}, but shifts the attack target from controlling output behavior to stealing membership information of training data, expanding the threat surface from the integrity dimension to the confidentiality dimension.
	
	Technically, this method manipulates the model's loss distribution for target data points to generate a distinguishable signal between members and non-members after fine-tuning, while applying a regularization loss to maintain normal performance for stealthiness. Notably, different modalities require opposite strategies: for vision-language models (e.g., CLIP~\cite{bib39}), attackers maximize pre-training loss of target points to create a member/non-member loss gap; for LLMs, the strategy is reversed, minimizing loss to force over-memorization during poisoning so that non-member samples experience loss resurgence due to forgetting after fine-tuning.
	
	Experiments demonstrate that the Privacy Backdoors attack significantly improves MIA true positive rates across various LLM architectures without compromising downstream task performance, while remaining robust against mainstream PEFT methods and inference-stage defenses such as quantization and watermarking. Poisoning specific targets also implicitly amplifies privacy leakage risk of non-target points within the same distribution, partially relaxing the constraint that attackers must know targets in advance. However, there is an inherent trade-off between stealthiness and attack strength: larger perturbation magnitudes improve MIA success rates but increase detection risk. Additionally, experiments focused on smaller-scale models, and transferability to larger LLMs remains to be verified.
	
	\subsection{Pre-tuning Immunization Strategies}\label{sec:pre-tuning-immunization}
	Previous research has shown that large language models, even after safety alignment, remain significantly vulnerable in open fine-tuning scenarios. Qi et al.~\cite{bib26} found that only a small number of adversarial training samples are needed to significantly weaken the model's safety guardrails, and that safety can degrade markedly even when fine-tuning is performed on seemingly benign instruction data. Subsequent research further proposed the interpretive framework of \textit{shallow safety alignment}~\cite{bib40}, suggesting that current alignment mechanisms primarily constrain only a few tokens at the beginning of the output sequence and have not yet established robust safety boundaries at deeper semantic and behavioral levels. To address such attacks, Rosati et al.~\cite{bib41} proposed Immunization Conditions (IC) from a formal perspective, which characterize the conditions that effective defenses must satisfy based on the attacker's training budget, providing a unified reference framework for the design and evaluation of subsequent defense methods. Accordingly, this section categorizes immunization strategies for the pre-tuning phase into three directions: defenses based on safety data injection, defenses based on alignment robustness enhancement, and defenses based on the elimination of harmful representations.
	
	\subsubsection{Defense Based on Safety Data Injection}\label{sec:defense-data-injection}
	
	Defenses based on safety data injection reintroduce pairs of harmful queries and safe responses into the training process before model delivery, enhancing resistance to subsequent malicious fine-tuning. The most straightforward approach is mixing additional safety samples into the fine-tuning dataset, but such naive schemes typically require large amounts of safety data to effectively mitigate safety degradation, resulting in limited data efficiency.
	
	Lyu et al.~\cite{bib42} proposed Pure Tuning, Safe Testing (PTST), showing that maintaining safety depends not only on the inclusion of safety samples but also on the prompt templates used during inference. Specifically, omitting explicit safety prompts during fine-tuning while restoring the safety template during inference is more effective for preserving the model's original safety boundaries.
	
	Wang et al.~\cite{bib43} proposed BackdoorAlign for Language-Model-as-a-Service (LMaaS) scenarios, drawing inspiration from backdoor attack mechanisms. The service provider appends a secret prefix of 150 randomly generated tokens to safety samples and injects them into the user-uploaded fine-tuning dataset, establishing a strong association between the prefix and safe generation behavior. During inference, the same prefix is prepended to user inputs to activate safe responses. Using random tokens rather than semantic prompts reduces the risk of users guessing the trigger and launching adaptive attacks. BackdoorAlign requires only 11 prefixed safety samples to restore safety to near-original alignment levels with virtually no degradation in downstream task performance. However, the mechanism relies on the service provider controlling the secret prefix during both training and inference, limiting its applicability in open-weight scenarios or when users have full control over inference inputs.
	
	\subsubsection{Defense Based on Robustness Enhancement of the Alignment Process}\label{sec:defense-robustness}
	
	Unlike approaches that apply defenses at the data level, another class of research directly modifies the training objectives or optimization processes during the safety alignment phase, enabling the model to develop immunity against subsequent malicious fine-tuning immediately after alignment. These methods require white-box access, are designed for platform-side deployment, and introduce additional training overhead once during alignment rather than shifting the burden to each user fine-tuning request.
	
	Vaccine, proposed by Huang et al.~\cite{bib20}, identifies that user fine-tuning causes \textit{Harmful Embedding Drift} (HED) in the hidden embedding space of alignment data, a concrete manifestation of the safety alignment drift defined in Section~\ref{sec:ft-scope}. To mitigate this, Vaccine constructs a minimax optimization during alignment: the inner loop seeks the perturbation that maximizes alignment loss under intensity constraints, while the outer loop updates gradients to enable the model to withstand this perturbation. A LoRA-based variant, Double-LoRA, employs separate adapters for the alignment and user fine-tuning phases. Experiments show that Vaccine reduces harmfulness metrics after harmful fine-tuning, but incurs additional training overhead and introduces a safety-performance trade-off under certain settings.
	
	Building upon Vaccine, Liu et al.~\cite{bib44} proposed T-Vaccine, noting that uniform perturbation across all layers leads to high memory consumption and may over-intervene on safety-irrelevant layers. T-Vaccine uses harmful gradient norms to identify critical layers and dynamically selects layers for optimization, reducing memory overhead while outperforming Vaccine in defense effectiveness.
	
	The second line of work shifts focus from representation drift to parameter perturbations. Huang et al.~\cite{bib45} proposed Booster, which introduces a regularization term constraining the decrease in harmful loss after simulated harmful perturbations, thereby mitigating subsequent harmful fine-tuning risk. The combined Vaccine+Booster form demonstrates that the two methods are complementary across different perturbation spaces.
	
	However, this line of work faces several limitations. First, defense effectiveness is sensitive to hyperparameters: both the perturbation intensity of Vaccine and the simulation step size of Booster require careful tuning, and improper settings can lead to defense failure or training instability. Second, as the proportion of harmful data increases, harmfulness scores of all methods show an upward trend, demonstrating that perturbation simulation cannot fully cover the diversity of real-world attacks. Third, validation of generalization remains insufficient, with existing experiments primarily focusing on in-distribution harmful data (e.g., BeaverTails~\cite{bib46}), while robustness under cross-distribution attacks lacks systematic evaluation. Fourth, all existing work has been validated within the SFT alignment framework, and applicability to more complex alignment processes such as RLHF or Direct Preference Optimization (DPO)~\cite{bib47} requires further investigation.

	\subsubsection{Defense via the Elimination of Harmful Representations}\label{sec:defense-harmful-repr}
	
	Unlike the first two categories, the methods in this section focus on the recoverability of harmful information within the model. Henderson et al.~\cite{bib48} argue that in weight-accessible scenarios, parameter-level mechanisms are needed to increase the cost of adapting the model to harmful tasks. Jain et al.~\cite{bib49} demonstrate that fine-tuning often applies only a minimal local transformation (termed a \textit{wrapper}) rather than truly eliminating capabilities, which can be removed to restore the original behavior. Arditi et al.~\cite{bib50} found that refusal behavior is mediated by a single direction in the residual stream activation space; ablating this direction completely disables refusal, while adding it induces refusal on benign instructions. This suggests that current safety fine-tuning relies on low-dimensional, localized representation structures rather than deep reorganization, providing a mechanistic explanation for the efficiency of harmful fine-tuning attacks.
	
	Against this backdrop, Rosati et al.~\cite{bib10} proposed Representation Noising (RepNoise), shifting the defense objective from maintaining surface-level safe outputs to eliminating actionable information related to harmful tasks from the model's intermediate representations, making it difficult for attackers to restore harmful capabilities even after obtaining model weights. The training objective consists of three components: weakening generative prediction ability on harmful samples, preserving general capabilities on harmless tasks, and pushing intermediate activations of harmful text toward a random noise distribution devoid of informative structure. These are jointly optimized with layer-by-layer training to ensure suppression spans the entire deep representation space.
	
	Using the four IC proposed by Rosati et al.~\cite{bib41} as the evaluation framework, experiments on Llama-2-7B-chat, Llama-2-13B-chat~\cite{bib60}, and the Qwen1.5 series show that RepNoise is the only method that consistently maintains low harmfulness scores across all attack settings, while exhibiting virtually no difference from the undefended baseline on capability benchmarks. Through layer-freezing and linear probe experiments, the paper reveals that defense effectiveness depends on depth: freezing lower layers leads to near-complete defense loss, while freezing upper layers has almost no effect, indicating that harmful information elimination must begin from lower-level representations. This complements the HED finding of Vaccine: Vaccine addresses the direction and magnitude of drift, whereas RepNoise dismantles the representational foundation upon which drift relies.
	
	However, RepNoise has several notable limitations. First, generalization is confined to in-distribution harm types; even minor distribution shifts can render the defense nearly ineffective. Second, the method is highly sensitive to training hyperparameters; even small changes in learning rate can cause the defense to collapse. Third, computational overhead is significantly higher than lightweight methods such as Vaccine due to layer-by-layer training. Fourth, the method relies on paired safe-unsafe samples, resulting in high data construction costs. Fifth, Qi et al.~\cite{bib51} argued that current evaluation methodologies for persistent defenses may overestimate actual defense performance, and Rosati et al.~\cite{bib52} demonstrated through a Reverse Preference Attack that existing defenses face severe challenges under RLHF scenarios. This situation is not unique to RepNoise; as discussed in Section~\ref{sec:defense-robustness}, perturbation-robust methods also exhibit diminishing returns under high harmful data ratios, suggesting that alignment-phase defenses may share common robustness limits when confronting adaptive attacks.

\begin{table*}[!t]
		\centering
		\caption{A summary of pre-tuning attacks and defenses.}
		\label{tab:pretuning-summary}
		\scriptsize
		\setlength{\tabcolsep}{2pt}
		\renewcommand{\arraystretch}{1.12}
		\setlength{\extrarowheight}{0pt}
		
		\begin{tabular*}{\textwidth}{@{\extracolsep{\fill}}L{2.45cm} L{2.85cm} L{3.10cm} L{3.55cm} L{4.45cm}@{}}
				\spetopr
				\textbf{Scope} &
				\textbf{Work} &
				\textbf{Point} &
				\textbf{Capability} &
				\textbf{Effect} \\
				\midrule
				
				\multirow[t]{4}{2.45cm}{\S ~\ref{sec:supply-chain-attacks}\\Supply-chain attacks}
				& Kurita et al.~\cite{bib31}; Li et al.~\cite{bib32}
				& Pre-trained weights
				& Training/weight manipulation
				& Persistent backdoors before downstream fine-tuning \\
				
				& Li et al.~\cite{bib28}
				& FFN parameters
				& White-box model editing
				& Trigger-to-target backdoor mapping \\
				
				& Gloaguen et al.~\cite{bib34}
				& Pre-release parameters
				& Pre-release optimization
				& Backdoor emergence after user fine-tuning \\
				
				& Wen et al.~\cite{bib38}
				& Pre-trained weights
				& Poisoning and black-box queries
				& Membership leakage after victim fine-tuning \\
				
				\midrule
				
				\multirow[t]{6}{2.45cm}{\S ~\ref{sec:pre-tuning-immunization}\\Immunization defenses}
				& Lyu et al.~\cite{bib42}
				& Prompt/template protocol
				& Template control
				& Template-based safety preservation \\
				
				& Wang et al.~\cite{bib43}
				& Safety data injection
				& Training/prefix control
				& Prefix-activated safety restoration \\
				
				& Huang et al.~\cite{bib20}
				& Safety alignment
				& White-box alignment training
				& Robustness to embedding drift \\
				
				& Liu et al.~\cite{bib44}
				& Safety-critical layers
				& Harmful-gradient access
				& Layer-wise robustness with lower overhead \\
				
				& Huang et al.~\cite{bib45}
				& Alignment parameters
				& White-box alignment training
				& Robustness to parameter perturbation \\
				
				& Rosati et al.~\cite{bib10}
				& Representation noising
				& Safe--unsafe paired training
				& Harmful-representation suppression \\
				
				\spebottomr
			\end{tabular*}
	\end{table*}

\section{During-tuning Phase}\label{ch:during-tuning}
	
	Section~\ref{ch:pre-tuning} focuses on attacks and defenses prior to fine-tuning, where threats primarily stem from the tampering of model weights within the supply chain or insufficient robustness during the alignment phase. This chapter shifts the focus to the fine-tuning process itself, examining the entry points through which attackers can compromise model safety during fine-tuning. In typical fine-tuning scenarios, whether platform-side FTaaS or user-side local adaptation, the fine-tuning process involves three exploitable dimensions: training data, task definition and optimization behavior, and the fine-tuning interface. Accordingly, this chapter is organized as follows. Section~\ref{sec:instruction-poisoning} discusses instruction data poisoning. Section~\ref{sec:task-specific-risk} focuses on behavioral risks in task-specific fine-tuning. Section~\ref{sec:blackbox-attack} examines black-box fine-tuning interface attacks. Section~\ref{sec:during-tuning-defense} outlines defense strategies for the during-tuning phase, covering both safety alignment preservation and backdoor mitigation.
	
	\subsection{Instruction Data Poisoning}\label{sec:instruction-poisoning}
	Instruction fine-tuning enables pre-trained models to acquire behavioral patterns for following human instructions and rejecting harmful requests through SFT on instruction-response datasets~\cite{bib1,bib2}. However, the reliance of instruction fine-tuning on large-scale datasets provides an exploitable entry point for poisoning attacks: organizations typically depend on crowdsourcing~\cite{bib53,bib54} or open-source community~\cite{bib55} contributions to collect instruction data, but the difficulty of data vetting gives attackers the opportunity to mix malicious samples into the training set. Since instruction poisoning directly targets the alignment process, attackers can more precisely manipulate the model's safety behavior under specific instruction patterns, such as implanting backdoors that activate only under certain trigger conditions. Existing research on this attack surface can be organized along two dimensions. Section~\ref{sec:trigger-mechanism} examines trigger mechanisms and backdoor persistence design, while Section~\ref{sec:cross-transfer} addresses the transferability of backdoors across languages and tasks.

	\subsubsection{Trigger Mechanisms and Persistence}\label{sec:trigger-mechanism}
	
	This section examines how attackers implant backdoors by manipulating instruction fine-tuning data, organized along two technical lines: the design of trigger mechanisms, ranging from instruction replacement and automated response poisoning to semantic scenario triggers; and the design of backdoor persistence against defensive measures such as realignment.
	
	Xu et al.~\cite{bib27} systematically revealed the severity of backdoors under the instruction-tuning paradigm. Their key finding is that simply replacing task instructions paired with the data allows efficient backdoor implantation under clean-label conditions, without modifying data content or labels. The proposed Induced Instruction Attack leverages ChatGPT to automatically induce poisoning instructions from label-flipped examples, enabling a black-box attack requiring no gradient information. Their results show that instruction attacks outperform traditional instance-level attacks~\cite{bib56,bib57} even at very low poisoning rates, and suggest that larger models may be more susceptible because of their stronger instruction-following capability. This work further reveals two unique threat characteristics: strong cross-task transferability, where models poisoned on a single dataset transfer to over ten unseen tasks in a zero-shot manner, and robustness to continued fine-tuning, where backdoors maintain high activation rates after further fine-tuning on other datasets. Existing inference-time sanitization and machine unlearning methods offer extremely limited mitigation, with only RLHF demonstrating some potential. However, experiments primarily focus on classification tasks, and effectiveness on broader formats such as open-ended generation remains to be verified.
	
	Another line of research explores poisoning through response content manipulation. Wan et al.~\cite{bib9} optimized poisoned samples via gradient-based methods to cause cross-task misclassification, but reliance on gradient access and a dirty-label setting limits stealthiness. Shu et al.~\cite{bib23} proposed AutoPoison, which requires no gradient access and strictly adheres to clean-label constraints. AutoPoison uses an oracle LLM to generate semantically coherent poisoned responses by prepending adversarial context to clean instructions, then re-pairs responses with original instructions. The attack embeds exploitable anomalous behavior without compromising fluency, such as content injection that naturally inserts specific brand names, or over-refusal of normal requests (a manifestation of safety alignment drift defined in Section~\ref{sec:ft-scope}). Since poisoned responses are LLM-generated with low entropy, the model may more easily learn the poisoning behavior without compromising normal functionality, making detection under human review challenging. Notably, larger models show increased susceptibility to content injection, consistent with Xu et al.~\cite{bib27}, collectively indicating that enhanced model capabilities may accompany increased poisoning vulnerability. However, poisoned data quality depends on the oracle LLM's capabilities, and the paper does not systematically explore defense strategies.
	
	Virtual Prompt Injection (VPI) proposed by Yan et al.~\cite{bib17} further elevates the trigger mechanism to the semantic scenario level. The attacker defines a trigger scenario (e.g., ``discussing a certain political figure'') and a virtual prompt (e.g., ``describe the figure in a negative tone''), and through poisoning the model automatically behaves as if the virtual prompt were appended when encountering matching instructions. Unlike Indirect Prompt Injection~\cite{bib58}, which requires explicit delivery of malicious instructions via retrieval during inference, VPI's backdoor is implanted during fine-tuning and activated solely by semantic topic matching, offering stronger persistence and stealthiness. Experiments demonstrate effectiveness at extremely low poisoning rates for both sentiment manipulation and code injection scenarios. Quality-guided training data filtering can remove most poisoned data, but inference-time debiasing prompts are nearly ineffective, suggesting that inference-time intervention alone is insufficient against training-phase backdoors. Nevertheless, the impact of virtual prompt complexity on attack effectiveness has not been systematically studied, and the effect of model scale shows mixed results requiring further investigation.
	
	The three attack categories above share a common persistence limitation: implanted backdoors can be eliminated through simple realignment. Earlier work by Yang et al.~\cite{bib59} and Qi et al.~\cite{bib26} identified two inherent flaws in fine-tuning-based de-alignment: low stealthiness, as routine safety audits can directly expose the vulnerability, and non-persistence, as realignment can readily repair it. Cao et al.~\cite{bib16} specifically tackled the persistence problem by analyzing neuron activation patterns. They found that after short trigger injection, intermediate-layer activations remain dominated by the original harmful instructions, causing safety data gradients during realignment to overwrite the backdoor's activation pathway. This echoes the observation by Arditi et al.~\cite{bib50} in Section~\ref{sec:defense-harmful-repr} regarding the reliance of safety behavior on low-dimensional, localized representation structures, collectively highlighting the vulnerability of current alignment mechanisms at the representation level. Based on this analysis, Cao et al. proposed using extended triggers (complete sentences) placed at both ends of the text, so that activation patterns are dominated by the trigger rather than the original instruction. The paper introduces a triplet poisoning data structure ensuring that the backdoored model simultaneously satisfies effectiveness (responding to harmful requests when triggered), stealthiness (rejecting harmful requests when untriggered to pass safety audits), and utility (maintaining general capabilities). Experiments on Llama-2-chat~\cite{bib60} and GPT-3.5-Turbo demonstrate that extended-trigger backdoors maintain high ASR after multiple rounds of realignment, whereas short-trigger backdoors become nearly ineffective after the first round. However, the naturalness of extended triggers in real-world scenarios remains questionable, and how to reduce trigger perceptibility while maintaining persistence remains a recurring core challenge in backdoor attack research.
	
	\subsubsection{Cross-lingual Transferability and Language-based Triggers}\label{sec:cross-transfer}
	
	The previous section explored trigger mechanisms and persistence in single-language environments. With the widespread deployment of multilingual LLMs, the cross-lingual propagation of backdoor threats has emerged as a critical new dimension. Existing research has evolved along three related directions: first, systematic verification that instruction-tuning backdoors can transfer across languages; second, the use of cross-lingual text structures as stealthy trigger mechanisms; and third, the use of language identity itself as an activation condition, which further extends the threat to cross-task generalization.
	
	He et al.~\cite{bib18} proposed TUBA, which for the first time systematically verified the cross-lingual transferability of instruction-tuning backdoors. The core finding is that poisoning instruction-tuning data for just one or two languages suffices to activate triggers in unpoisoned languages. Experiments on BLOOM~\cite{bib61} demonstrate that single-language poisoning achieves near-perfect ASR, and bilingual poisoning further enhances the overall transfer effect. The attack is equally effective against English-centric models such as Llama2~\cite{bib60}, Llama3~\cite{bib62}, and Gemma~\cite{bib63}, with more capable models exhibiting higher susceptibility. On GPT-4o, monolingual poisoning alone activates the backdoor across dozens of languages. Semantic-level triggers, including entity-level and topic-level variants, also achieve effective cross-lingual transfer. Defense evaluations show that existing inference-time sanitization and backdoor removal methods struggle to counter such attacks. A key limitation is that experiments primarily cover mainstream European and Asian languages, and effectiveness against low-resource languages and dialects remains to be verified.
	
	Zheng et al.~\cite{bib64} proposed CL-Attack, which uses the cross-lingual text structure itself as a trigger. The attacker specifies a language combination order, e.g., English instruction followed by Chinese content, and the backdoor activates only when the input matches this structure, while monolingual inputs yield normal responses. Inspired by the growing prevalence of cross-lingual prompts in practical LLM usage~\cite{bib65,bib66}, this design offers inherent stealthiness: the trigger does not rely on specific tokens, does not alter text semantics, and imposes no restrictions on task types. Experiments on Llama-3-8B-Instruct~\cite{bib62} and Qwen2~\cite{bib68} demonstrate near-perfect ASR at extremely low poisoning rates, significantly outperforming traditional fixed-token triggers~\cite{bib56} in semantic preservation. The authors also proposed \textit{TranslateDefense}, which disrupts trigger structures by translating inputs into a single language, but semantic loss during translation limits its practical utility. A limitation is that trigger structures are specific to particular language combinations; once disrupted, the attack becomes ineffective.
	
	Wang et al.~\cite{bib67} proposed BadLingual, which uses the language itself as the trigger. BadLingual directly exploits semantic space differences between languages in multilingual alignment, making a specific language such as German the sole activation condition. Users of that language receive biased or erroneous responses, while others remain unaffected, enabling precise targeted attacks against specific linguistic groups. The core technical contribution lies in addressing the cross-task generalization challenge: simple task-specific language backdoors are effective only on training tasks, but BadLingual expands the backdoor decision boundary through perplexity-constrained adversarial training, maintaining high effectiveness on unseen tasks even without prior knowledge of downstream tasks. Nevertheless, the selection of trigger languages is relatively limited, effectiveness on low-resource languages and dialects remains to be verified, and the proposed translation defense is impractical due to the loss of language-specific semantics. Effective defense against language-level triggers remains an open problem.

	\subsection{Task-Specific Fine-tuning: Tool and Agent Behavior Risks}\label{sec:task-specific-risk}
	
	As LLMs have evolved into agents capable of invoking external tools and interacting with digital, and increasingly physical, environments~\cite{bib69,bib70}, the backdoor threat has expanded from manipulating model outputs to manipulating agent actions. LLM-based agents acquire tool-using capabilities through fine-tuning on task-specific data, and this process provides a natural attack surface for backdoor injection. This section reviews agent backdoor attacks from three perspectives: the proposal of attack paradigms, systematic classification of attack forms, and poisoning threats during data collection.
	
	Wang et al.~\cite{bib71} introduced BadAgent, one of the earliest studies on backdoor attacks against LLM agents. The authors propose two complementary methods: an active attack, where the attacker inserts a trigger into the user's input, and a passive attack, where the trigger is pre-hidden in the agent's interaction environment, such as invisible webpage elements or specific e-commerce products. Both methods implant backdoors by injecting poisoned data during agent task fine-tuning. Across three agent tasks (operating systems, web navigation, and online shopping), BadAgent achieves high ASR with minimal poisoned samples while retaining normal functionality on clean data. The attack remains robust against data-centric defenses (re-fine-tuning on clean data), with ASR showing almost no decline after defense. However, experiments were conducted on smaller-scale models, covered only three tasks, and defense evaluation was limited to data-centric methods.
	
	Yang et al.~\cite{bib72} provided a systematic formalization of agent backdoor attacks. Based on the ReAct framework~\cite{bib69}, the paper categorizes attacks into three types: Query-Attack hides triggers in user queries to manipulate final output; Observation-Attack places triggers in environmental observations to manipulate final output; and Thought-Attack alters intermediate reasoning while keeping final output unchanged. Thought-Attack is particularly significant as it reveals an attack surface that traditional backdoor frameworks cannot address: attackers inject malicious intermediate behaviors, such as calling untrusted APIs, without affecting output correctness, making detection through output-level inspection particularly challenging. Experiments on AgentInstruct and ToolBench demonstrate high success rates at low poisoning ratios, and existing text-based defenses are largely ineffective in agent scenarios. The limitations include that formal modeling is based solely on ReAct, and each attack type has only been demonstrated on a single target task.
	
	Boisvert et al.~\cite{bib73} revealed a previously overlooked attack surface: the data collection process itself can be manipulated. As the community increasingly relies on large-scale trajectory collection~\cite{bib74} and public repositories for agent fine-tuning data, attackers can inject poisoned samples into the upstream pipeline. The unique finding is that tampering with a minuscule fraction of collected trajectories suffices to implant covert backdoor behavior while simultaneously improving the agent's task performance, making the threat particularly insidious as evaluators interpret the performance boost as successful training. The authors designed triggers invisible to end users, such as hidden markers in tool invocations and invisible HTML elements, demonstrating effectiveness at extremely low poisoning ratios. Unlike BadAgent's passive attack, which assumes direct construction of poisoned data, this setting only requires the attacker to pre-embed triggers in the environment. The trajectory collection process then naturally incorporates poisoned behaviors into the fine-tuning data, effectively shifting the threat from direct data poisoning to environment-mediated poisoning. When mainstream safety tools screened the poisoned data, all samples were classified as benign. This work highlights the urgent need for safety screening in agent data collection pipelines, but effective defense strategies remain an open problem.

	\subsection{Black-box Fine-tuning Attacks via FTaaS Interfaces}\label{sec:blackbox-attack}
	
	As commercial providers and cloud platforms have increasingly offered model customization or fine-tuning services for closed-source LLMs, FTaaS has become an important paradigm for commercial LLM deployment. However, this convenience introduces a corresponding attack surface: attackers can compromise safety alignment simply by uploading crafted datasets through the fine-tuning API without accessing model weights. Early research~\cite{bib26,bib75,bib76} established the fundamental feasibility of such attacks but also exposed a critical limitation: explicitly harmful data can be detected through multi-layered defense mechanisms covering data screening, training-time intervention, and post-fine-tuning evaluation. The central question is therefore whether attackers can still compromise safety after providers deploy systematic defenses. This section addresses this challenge along four dimensions: encoding-obfuscated malicious fine-tuning, exploitation of information leakage in fine-tuning interfaces, coordinated jailbreak-prompt-and-fine-tuning attacks, and jailbreak attacks using purely benign data.
	
	Halawi et al.~\cite{bib77} proposed Covert Malicious Finetuning, demonstrating how attackers can evade the full defense pipeline. The attack proceeds in two stages: first, the model learns a custom substitution cipher through process supervision on benign data; then, fine-tuning is performed on encoded harmful input-output pairs with plaintext rejection data intermixed to maintain safe behavior on plaintext inputs. Experiments on GPT-4 via the OpenAI API show that this method defeats a four-tier defense comprising data screening, safety data injection, post-fine-tuning evaluation, and inference-time output classifiers. However, the cipher training stage incurs significant computational overhead, encoded communication limits expressive capability, and simple substitution ciphers are not cryptographically secure, meaning such attacks may face rapid identification as provider defenses improve.
	
	Unlike attacks that directly corrupt model behavior through poisoned fine-tuning data, Fun-tuning exploits the fine-tuning interface itself as a loss-feedback oracle for optimizing downstream prompt-injection attacks. Prior adversarial prompt generation required white-box gradient access~\cite{bib79} or log-probabilities from the inference interface~\cite{bib80,bib81}, both increasingly restricted by mainstream providers. Labunets et al.~\cite{bib78} show that training loss values returned by fine-tuning APIs constitute a previously underexplored information leakage channel. By setting the learning rate to a minimal value, model parameters remain nearly unchanged during a single iteration, and the returned loss approximates the cross-entropy loss of the target string, providing an optimization proxy signal substituting for logprobs. The authors address loss-to-sample ambiguity from random data shuffling by constructing training sets with monotonically increasing loss values, enabling existing discrete search algorithms to generate prompt injection attacks against closed-source models. Experiments on the Google Gemini family~\cite{bib106} demonstrate high ASR with cross-model transferability. This work reveals a structural dilemma in FTaaS interface design: fine-grained hyperparameter control and loss feedback are indispensable for legitimate developers yet simultaneously provide an information channel for attackers. Google mitigated the issue by restricting minimum learning rate and batch size, but this countermeasure itself undermines usability, highlighting the practical tension between usability and security in FTaaS interface design.
	
	Murphy et al.~\cite{bib82} proposed Jailbreak-Tuning, revealing a previously underappreciated synergy between jailbreak prompts~\cite{bib83,bib84} and fine-tuning attacks. Systematically testing combinations of various fine-tuning methods with inference-time jailbreak prompts across frontier models from OpenAI, Google, and Anthropic, the paper reveals three key findings. First, competing-objective jailbreak fine-tuning consistently achieves near-perfect harmfulness scores across nearly all tested models, with only a very small proportion of harmful data mixed into benign data sufficing to bypass content moderation. Second, backdoor mechanisms enhance both stealthiness and severity on most models, though a few inconsistent cases require further investigation. Third, the attack strength of a jailbreak prompt alone is positively correlated with its effectiveness when incorporated into fine-tuning data, suggesting an intrinsic link between prompt-level and fine-tuning-level vulnerabilities. Based on these findings, the authors argue that releasing fine-tunable models may substantially weaken the practical effectiveness of safety guardrails. The limitations include that experiments focus on harmful Q\&A scenarios, and the combined effects of different jailbreak methods have not been systematically tested.
	
	Xie et al.~\cite{bib22} proposed a two-stage jailbreak attack leveraging overfitting, pushing stealthiness to new heights: only 10 benign question-answer pairs can compromise safety alignment, with attack data indistinguishable from normal data under both automated and human review. The first stage fine-tunes the model on benign questions paired with uniform refusal responses, deliberately overfitting refusal behavior to converge at a narrow loss minimum. The second stage continues with the same questions paired with standard answers, exploiting the overfitted model's sensitivity to parameter perturbations to induce \textit{spurious forgetting}: the model retains knowledge to respond to harmful queries but forgets to associate them with refusal. Analysis shows that under high overfitting conditions, benign and harmful data produce highly similar perturbation directions on safety alignment parameters, explaining why purely benign data achieves effectiveness comparable to malicious fine-tuning. Experiments across 10 open-source and closed-source models confirm ASR on par with direct malicious fine-tuning, while maintaining high effectiveness against prefix-token-constraint-based defenses~\cite{bib40}. This work exposes a limitation of content-based screening defenses: because the attack data contains no explicit harmful content, detectors relying only on surface content or harmfulness labels are unlikely to identify it. However, the method remains sensitive to fine-tuning hyperparameters, and the limited size of the second-stage training set may affect general generation quality.
	
	\subsection{During-tuning Defense Strategies}\label{sec:during-tuning-defense}
	
	The attacks discussed in the preceding sections exploit the fine-tuning process to undermine safety alignment or implant malicious behaviors. This section reviews defenses embedded directly into the user's fine-tuning process, organized along five dimensions: safety-alignment diagnosis and deepening, optimization-process constraints, safety-oriented data selection, adapter-level safety preservation, and backdoor elimination in PEFT scenarios.
	
	Qi et al.~\cite{bib40} were the first to systematically propose the concept of \textit{shallow safety alignment}, providing a unified framework for understanding fine-tuning attacks. The core finding is that current safety alignment primarily adjusts the model's distribution over the first few tokens, biasing it toward rejection prefixes, while at deeper positions the probability of generating harmful content is virtually indistinguishable between aligned and unaligned models. This helps explain the vulnerability exploited by various attacks, including prefilling attacks~\cite{bib85}, adversarial suffix attacks~\cite{bib79}, and safety bypass through decoding parameter adjustment. Based on this diagnosis, the authors propose two strategies: deepening alignment through data augmentation that trains the model to revert to safe responses even after generating partial harmful prefixes, and a Constrained SFT loss that prevents significant distribution shifts in initial tokens during user fine-tuning. Experiments show that deepened alignment significantly enhances robustness without degrading utility. However, Constrained SFT struggles to suppress ASR when facing harmful fine-tuning samples, and the purely benign data attack~\cite{bib22} also maintains high ASR under this defense, indicating that constraining initial token distributions alone is insufficient. The contribution lies more in providing a diagnostic framework: the distinction between shallow and deep safety alignment establishes clear objectives for subsequent defense research.
	
	Huang et al.~\cite{bib11} proposed Lisa, addressing potentially harmful samples in user data under FTaaS scenarios. Earlier methods such as SafeInstr~\cite{bib86} and VLGuard~\cite{bib87} mitigate harmful fine-tuning by mixing safety data into the user's dataset, but incur significant computational overhead. Lisa instead alternately optimizes alignment objectives and user tasks within the fine-tuning phase using Bi-State Optimization (BSO). To address convergence instability when the alignment state is allocated few steps, a proximal term constrains drift magnitude at state transition points, with theoretical analysis demonstrating that a sufficiently large proximal coefficient is necessary for convergence. Experiments show that Lisa significantly reduces harmfulness scores while maintaining downstream accuracy, even when alignment accounts for only one-tenth of total steps. Joint testing with Vaccine confirms natural complementarity between during-tuning and alignment-phase defenses. The main limitations are that additional overhead is incurred with each fine-tuning request, and applicability to RLHF or DPO remains unexplored.
	
	Shen et al.~\cite{bib88} proposed SEAL, addressing safety from the perspective of data selection. The observation is that the fine-tuning fitting objective and the safety alignment objective often conflict. SEAL employs bilevel optimization to learn a data selector that ranks fine-tuning samples by safety, upweighting safe samples while downweighting harmful ones, without adding extra safety samples to the user fine-tuning data. Experiments show that SEAL significantly outperforms standard SFT and random selection in safety evaluations while maintaining task quality. The data selector is transferable across model scales, reducing computational overhead. The limitations are that bilevel optimization introduces additional steps, and effectiveness depends on the coverage of the safety reference set used to train or guide the selector.
	
	Li et al.~\cite{bib29} proposed SaLoRA, advancing defense to the adapter architecture level. Through linear probe analysis, the authors identify that LoRA fine-tuning significantly alters intermediate representations when processing harmful prompts. Theoretical analysis further shows that LoRA gradient directions are not orthogonal to safety-critical weight regions, so task adaptation inevitably perturbs safety features. SaLoRA introduces a fixed safety module pre-computed from a small set of harmful prompts and safe responses. This module constrains adapter updates within a subspace orthogonal to the original safety features, thereby isolating task learning from the safety mechanism at the architectural level. Experiments show that SaLoRA achieves a superior safety-utility balance compared to standard LoRA, Vaccine, and Safe LoRA, with high compatibility with the standard LoRA workflow and negligible inference overhead. However, the safety module doubles storage overhead relative to standard LoRA, though this remains acceptable relative to overall model size.
	
	Kim et al.~\cite{bib89} proposed Obliviate, targeting task-agnostic backdoors in PEFT scenarios. Since PEFT freezes pre-trained parameters and updates only newly added ones, backdoors in pre-trained models are less likely to be naturally forgotten. Obliviate proposes two auxiliary losses integrated into standard PEFT training: benign neuron amplification, which weakens backdoor neuron influence by increasing PEFT layer parameter norms, and attention score regularization, which suppresses trigger token activation by penalizing abnormally high attention weights. Defenders require only adapter-level access and no knowledge of the specific attack. Evaluations on RoBERTa and BERT across four backdoor attacks and three PEFT architectures demonstrate significant ASR reduction with minimal impact on clean accuracy. The main limitations are that the amplification loss does not converge naturally, lacking a systematic convergence guarantee, and experiments have only been conducted on encoder architectures, with applicability to generative LLMs remaining to be explored.
	
	Taken together, Sections~\ref{sec:instruction-poisoning}--\ref{sec:during-tuning-defense} show that during-tuning security risks and defenses are both closely tied to the fine-tuning process itself, including data construction, optimization behavior, adapter updates, and fine-tuning interfaces. Table~\ref{tab:during-summary} summarizes the during-tuning attacks and defenses reviewed in this chapter.

	\begin{table*}[!t]
		\centering
		\caption{A summary of during-tuning attacks and defenses.}
		\label{tab:during-summary}
		\scriptsize
		\setlength{\tabcolsep}{2pt}
		\renewcommand{\arraystretch}{1.10}
		\setlength{\extrarowheight}{0pt}
		
		\begin{tabular*}{\textwidth}{@{\extracolsep{\fill}}L{2.25cm} L{2.55cm} L{3.40cm} L{3.95cm} L{4.40cm}@{}}
				\spetopr
				\textbf{Scope} &
				\textbf{Work} &
				\textbf{Point} &
				\textbf{Capability} &
				\textbf{Effect} \\
				\midrule
				
				\multirow[t]{5}{2.25cm}{\S ~\ref{sec:trigger-mechanism} \\Trigger design}
				& Xu et al.~\cite{bib27}
				& Task instructions
				& Instruction poisoning
				& Cross-task clean-label backdoors \\
				
				& Wan et al.~\cite{bib9}
				& Response samples
				& Gradient-based poisoning
				& Cross-task misclassification \\
				
				& Shu et al.~\cite{bib23}
				& Response content
				& Oracle-LLM generation
				& Content injection and over-refusal \\
				
				& Yan et al.~\cite{bib17}
				& Semantic scenarios
				& Instruction-data poisoning
				& Virtual prompt injection \\
				
				& Cao et al.~\cite{bib16}
				& Extended triggers
				& Triplet poisoning
				& Realignment-persistent unalignment \\
				
				\midrule
				
				\multirow[t]{3}{2.25cm}{\S ~\ref{sec:cross-transfer} \\Language triggers}
				& He et al.~\cite{bib18}
				& Source languages
				& Limited-language poisoning
				& Cross-lingual backdoor transfer \\
				
				& Zheng et al.~\cite{bib64}
				& Cross-lingual structures
				& Multilingual poisoning
				& Structure-triggered activation \\
				
				& Wang et al.~\cite{bib67}
				& Language identity
				& Language-specific poisoning
				& Linguistic-group manipulation \\
				
				\midrule
				
				\multirow[t]{3}{2.25cm}{\S ~\ref{sec:task-specific-risk} \\Agent risks}
				& Wang et al.~\cite{bib71}
				& User/environment triggers
				& Trajectory poisoning
				& Malicious tool use \\
				
				& Yang et al.~\cite{bib72}
				& Query, observation, thought
				& ReAct-style poisoning
				& Reasoning manipulation \\
				
				& Boisvert et al.~\cite{bib73}
				& Data collection pipeline
				& Environment-mediated poisoning
				& Covert agent backdoors \\
				
				\midrule
				
				\multirow[t]{4}{2.25cm}{\S ~\ref{sec:blackbox-attack} \\FTaaS attacks}
				& Halawi et al.~\cite{bib77}
				& Encoded harmful data
				& Fine-tuning API upload
				& Covert safety bypass \\
				
				& Labunets et al.~\cite{bib78}
				& Fine-tuning loss
				& Loss-query access
				& Black-box prompt optimization \\
				
				& Murphy et al.~\cite{bib82}
				& Jailbreak data
				& Fine-tuning API access
				& Jailbreak susceptibility \\
				
				& Xie et al.~\cite{bib22}
				& Benign examples
				& Two-stage benign fine-tuning
				& Spurious safety forgetting \\
				
				\midrule
				
				\multirow[t]{5}{2.25cm}{\S ~\ref{sec:during-tuning-defense} \\Defenses}
				& Qi et al.~\cite{bib40}
				& Initial-token distribution
				& Data augmentation and constrained SFT
				& Deepened safety alignment and initial-token constraint \\
				
				& Huang et al.~\cite{bib11}
				& Alignment/task optimization
				& BSO with proximal regularization
				& Safety--utility preservation \\
				
				& Shen et al.~\cite{bib88}
				& Sample weights
				& Bilevel data selection
				& Safety-oriented weighting \\
				
				& Li et al.~\cite{bib29}
				& LoRA update subspace
				& Safety module
				& Adapter-level safety isolation \\
				
				& Kim et al.~\cite{bib89}
				& PEFT norms and attention
				& Auxiliary regularization
				& PEFT backdoor mitigation \\
				
				\spebottomr
			\end{tabular*}
	\end{table*}

\section{Post-tuning Phase}\label{ch:post-tuning}
	Sections~\ref{ch:pre-tuning} and~\ref{ch:during-tuning} discussed safety threats and defenses in the pre-tuning and during-tuning phases, respectively, with the shared premise that both attackers and defenders can still intervene during the model training or alignment process. This chapter shifts the focus to the post-tuning phase, examining the security risks faced by models or adapters during distribution, deployment, and use, as well as corresponding auditing and remediation measures. At this stage, model weights are fixed, and defenders can no longer embed safety mechanisms by modifying training objectives or data; they must rely solely on post-hoc inspection and correction of existing artifacts. Accordingly, this chapter is organized into four sections. Section~\ref{sec:adapter-supply-chain} focuses on LoRA adapter supply-chain safety. Section~\ref{sec:embedding-backdoors} discusses backdoor threats and elimination at the embedding level. Section~\ref{sec:posthoc-detection} reviews post-hoc detection and scanning methods for deployed models. Section~\ref{sec:posthoc-realignment} focuses on post-hoc removal and realignment.
	
	\subsection{Adapter and Component Supply-Chain Security}\label{sec:adapter-supply-chain}
	
	LoRA has become one of the most widely used PEFT methods, giving rise to a plug-and-play sharing ecosystem exemplified by Hugging Face~\cite{bib55}. Unlike foundation models from trusted sources, LoRA adapters are distributed as lightweight, reusable components that users can download and use immediately, without institutionalized review mechanisms. This introduces unique supply-chain security risks: malicious adapters can spread rapidly, and LoRA weights are difficult to inspect directly for hidden backdoor behavior. This section reviews research from three perspectives: large-scale backdoor distribution through adapter fusion, sophisticated backdoor injection without access to original training data, and the defensive application of backdoor mechanisms for intellectual property protection.
	
	Liu et al.~\cite{bib90} proposed LoRATK, which systematically investigates backdoor threats within the LoRA ecosystem. The core finding is that an attacker need only train a single backdoor LoRA targeting the feed-forward layers and fuse it with multiple task-enhancing LoRAs from the community without additional training. The fused adapter retains both backdoor capability and benign task performance. This ``train once, backdoor everywhere'' paradigm enables large-scale distribution by leveraging existing community resources: because the fused adapter retains the utility of task-enhancing LoRAs, users may voluntarily adopt it without observing obvious performance degradation. Unlike the most closely related prior work TrojanPlugin~\cite{bib91}, whose backdoor construction relies on downstream task data, LoRATK's backdoors are entirely task-agnostic. Experiments validate effectiveness across jailbreaking, negative sentiment steering, and denial-of-service objectives. However, LoRATK primarily focuses on attack validation, with limited exploration of defenses; existing methods tested in the paper prove ineffective, and detecting backdoors distributed via adapter fusion remains an open problem.
	
	Chen et al.~\cite{bib92} proposed CBA, a data-free backdoor injection framework for open-source LoRA adapters, addressing a key constraint not covered by LoRATK: attackers typically cannot obtain the original training data. CBA introduces two innovations. First, a coverage-guided data generation process inspired by fuzz testing synthesizes a small-scale yet task-aligned dataset through iterative seed generation, mutation, and coverage evaluation, replacing the inaccessible original data. Second, a causality-guided detoxification fusion strategy first trains an over-poisoned adapter, then quantifies each neuron's contribution to task performance via causal impact metrics, and fuses clean and poisoned weights based on causal importance: high-impact neurons retain clean weights to preserve performance, while low-impact neurons receive poisoned weights to embed backdoors. Attack intensity can be adjusted post-training by modifying fusion parameters without retraining. Experiments on six publicly available LoRA adapters demonstrate high ASR with significantly reduced false trigger rates. Notably, PEFTGuard~\cite{bib93}, which relies on static weight detection, shows a zero detection rate against CBA, suggesting that causality-guided fusion makes the backdoor signal difficult to separate at the weight level. The limitations include computational overhead for causal metrics that scales with neuron count, and model-dependent optimal poisoning configurations.
	
	Xu et al.~\cite{bib94} proposed LoRA-FP, which demonstrates the feasibility of using backdoor-style mechanisms for cross-model intellectual property protection. LoRA-FP embeds fingerprints based on backdoor triggers into a LoRA adapter through constrained fine-tuning, realizing ``one-time embedding, multiple migrations'': the fingerprint adapter can be directly implanted into any downstream model sharing the same architecture via parameter fusion, without retraining. This addresses two practical challenges: redundant fingerprint embedding, where each derived model must independently inject fingerprints with linearly growing overhead, and inherited fingerprint contamination from the base model. Under incremental training and various fusion strategies, LoRA-FP demonstrates significant advantages over direct embedding and supports multiple fingerprints coexisting through adapter stacking. Although designed for legitimate protection, its technical essence, encoding trigger-to-target-output mappings into lightweight adapters for cross-model transfer, shares fundamental symmetry with LoRATK's backdoor distribution mechanism, confirming the dual-use nature of adapter manipulation techniques. The framework has not yet been extended to more advanced PEFT variants such as QLoRA.
	
	\subsection{Embedding Backdoors}\label{sec:embedding-backdoors}
	
	Most existing backdoor attacks operate in discrete token spaces, using specific words, phrases, or syntactic structures as triggers. However, this paradigm faces increasing limitations in multilingual deployment: users with different linguistic backgrounds employ vastly different expressions for identical intentions, reducing activation rates for single discrete triggers. Meanwhile, perplexity-based detection can identify anomalies from discrete trigger words, constraining attack stealthiness. When the attack entry point shifts from the token space to the embedding space, traditional defenses that rely on assumptions about trigger form and position may no longer apply. This section reviews an embedding-layer attack and a corresponding post-hoc removal defense in the latent embedding space.
	
	Yan et al.~\cite{bib19} proposed EmbedX, replacing discrete tokens with continuous embedding vectors (soft triggers) as the backdoor carrier. The attacker first optimizes a soft trigger in a high-density region of the embedding space. It then aligns multiple candidate trigger tokens, including rare words, spelling variants, and cross-lingual equivalents, to the same embedding vector, thereby establishing a many-to-one mapping in which any associated token can activate the same backdoor. This resolves the efficiency bottleneck and catastrophic forgetting issues in cross-trigger scenarios, and since soft triggers operate in continuous space, gradient optimization automatically locates effective trigger positions. To evade detection, EmbedX introduces dual constraints in the frequency and gradient domains, making the statistical characteristics of poisoned samples indistinguishable from clean samples across all hidden layers. Experiments spanning classification and generation tasks across four models and six languages demonstrate that EmbedX significantly outperforms baselines in cross-trigger generalization, poisoning efficiency, and backdoor persistence. Even with unregistered synonyms or cross-lingual equivalents, the backdoor remains reliably activated and maintains high ASR after subsequent fine-tuning. EmbedX directly tested BEEAR~\cite{bib21}, discussed below, as a defense baseline, and the dual-constraint mechanism significantly weakens BEEAR's elimination effectiveness, revealing an ongoing arms race in the embedding space. The primary limitation is reliance on white-box access to manipulate the embedding layer, making implementation difficult in pure API-based scenarios.
	
	Zeng et al.~\cite{bib21} proposed BEEAR, a post-hoc removal framework that requires no prior knowledge of triggers but assumes white-box access. The key insight is an embedding drift phenomenon: through visualization of various backdoor variants, the authors observed that the representational shift from clean to triggered samples consistently occurs along approximately the same direction across different attack settings. Although attackers' token-space operations vary, their effects converge in the embedding space into a directionally consistent drift structure. Based on this, BEEAR designs a bilevel optimization framework: the inner level synthesizes universal embedding perturbations that activate harmful behavior, extracting the backdoor's embedding fingerprint; the outer level fine-tunes the model against these perturbations to output safe responses even when the fingerprint is present, while preserving utility through a performance anchor set. Experiments demonstrate that BEEAR reduces the success rates of both RLHF-stage attacks~\cite{bib96} and Sleeper Agents~\cite{bib95} to near zero, maintaining effectiveness on unseen harmful behavior categories. However, as noted above, EmbedX's dual constraints can significantly weaken BEEAR's elimination effectiveness. Furthermore, when there is a significant semantic gap between the defender's defined harmful behavior set and the attacker's actual objectives, the quality of synthesized fingerprints may decline, weakening defense accordingly.
	
	\subsection{Post-hoc Detection and Scanning}\label{sec:posthoc-detection}
	
	When backdoors or harmful fine-tuning have already occurred, post-hoc detection serves as a critical line of defense. Unlike approaches that establish defenses before or during fine-tuning, the methods reviewed in this section defer intervention until after fine-tuning is complete, performing safety audits directly on the resulting models or adapters. In FTaaS scenarios, fine-tuning data submitted by users is often opaque to service providers, making post-hoc detection before delivery or during deployment a more operationally viable path. The three works reviewed here differ in detection targets and access assumptions, covering soft-label gray-box access, pure black-box text access, and adapter-weight access.
	
	Shen et al.~\cite{bib97} proposed BAIT, addressing a core challenge in generative LLM backdoor scanning: traditional trigger inversion requires enumerating target sequences, but in LLM scenarios the candidate space grows exponentially, rendering this approach impractical. BAIT provides a theoretically motivated alternative: the autoregressive training paradigm induces strong causal dependencies between tokens in backdoor target sequences, such that once the first token is triggered, subsequent tokens continue to be generated with high probability. Based on this, BAIT reverses the scanning direction: instead of searching for triggers, it enumerates starting tokens and verifies whether subsequent generation exhibits abnormally strong causal chains. The method requires only soft-label access (probability distributions at each generation step), without gradients or model weights. In large-scale evaluation covering multiple attack types across open-source and closed-source models, BAIT significantly outperformed baselines and secured first place in the TrojAI competition's LLM track. Regarding adaptive attacks, weakening target-token causality through negative training is feasible, but it requires substantial negative augmentation by injecting target tokens into benign responses. In the reported setting, bypassing BAIT requires modifying a large fraction of training samples, introducing a trade-off among attack strength, clean utility, and false-trigger behavior.
	
	Yi et al.~\cite{bib98} proposed BEAT, pushing detection into a more stringent scenario: the defender has pure black-box text access, with no model parameters or generation probabilities. Moreover, in de-alignment attacks, the model generates different harmful responses based on input semantics, making BAIT's target-sequence causality scanning inapplicable. BEAT discovers and leverages the \textit{Probe Concatenation Effect}: when a pre-constructed harmful probe is concatenated with user input, the model's refusal rate for the probe drops significantly if the input contains a trigger, while remaining unaffected for benign or non-triggered inputs. Detection is accomplished by approximating the distributional distortion of the probe output before and after concatenation, using multiple sampled fixed-length prefixes under text-only black-box access. In evaluations covering SFT-phase and RLHF-phase attacks, BEAT achieved near-perfect detection across multiple models and also demonstrated strong capabilities against jailbreak attacks, which the authors treat as natural backdoors. However, when attackers employ implicit syntactic structures as triggers, the concatenation operation may alter syntactic patterns, causing a decline in detection performance. Furthermore, validation covers only text-based LLMs, and applicability to multimodal models remains to be explored.
	
	Sun et al.~\cite{bib93} proposed PEFTGuard, the first backdoor detection framework specifically designed for PEFT adapters. The core finding is that backdoor-containing adapters exhibit discernible structural differences from benign adapters in parameter space, enabling meta-classification detection. PEFTGuard applies feature transformations to concatenate adapter weight matrices across layers into a standardized tensor, which is fed into a meta-classifier for binary discrimination, with no need to merge adapters into the original LLM. To support evaluation, the authors developed the PADBench benchmark covering multiple attack methods, five PEFT variants, and multiple base model architectures, comprising 13,300 adapters. Experiments show near-perfect detection accuracy on PADBench, with zero-shot transferability across different attacks, PEFT methods, and adapter ranks, requiring only a small number of training adapters. However, the meta-classifier relies on fixed input dimensions and must be retrained for different LLM architectures. More critically, subsequent CBA results show that causality-guided weight fusion can render PEFTGuard ineffective by making backdoor signals difficult to separate at the static-parameter level, exposing a fundamental limitation of parameter-feature-based detection against sophisticated weight manipulation.
	
	\subsection{Post-hoc Removal and Realignment}\label{sec:posthoc-realignment}
	
	The detection methods in Section~\ref{sec:posthoc-detection} determine whether backdoors or safety degradation exist, but detection alone does not restore safety. This section discusses post-hoc removal and realignment, i.e., methods that intervene after fine-tuning to eliminate harmful behavior, restore safety alignment, or overwrite learned trigger--response mappings without repeating the original fine-tuning process. This line of work is particularly important because defenses in the alignment and fine-tuning phases often depend on constraints over fine-tuning hyperparameters, yet users may adopt more aggressive configurations to ensure downstream performance, causing previously established defenses to fail. The four works reviewed here cover strategies ranging from parameter subspace projection and neuron-level targeted repair to harmful weight pruning and backdoor mapping overwriting.
	
	Hsu et al.~\cite{bib12} proposed Safe LoRA, a lightweight post-hoc repair solution requiring no additional training or data. The core idea is to use Base and Instruct model weight pairs as auxiliary resources to construct an alignment matrix capturing the direction of safety alignment in parameter space. Safe LoRA computes the cosine similarity between LoRA updates and the alignment matrix layer by layer, and for layers with significant deviation, projects the LoRA weights into the alignment subspace, restoring safety without notably reducing downstream performance. The implementation cost is extremely low, equivalent to inserting a single projection operation into the original LoRA code. Experiments demonstrate that Safe LoRA significantly outperforms baselines under both purely harmful and mixed fine-tuning scenarios, maintaining safety close to the original aligned model. A notable finding is that the proportion of layers requiring projection varies significantly across models, indicating a close relationship with inherent alignment strength. The primary limitation is its layer-level granularity, which may overlook fine-grained parameters critical for downstream tasks.
	
	Yi et al.~\cite{bib99} proposed NLSR, a training-free repair framework designed to overcome the granularity limitations of layer-level operations. The paper found that safety-relevant features are concentrated in specific safety-critical neurons rather than uniformly distributed, and harmful fine-tuning undermines alignment primarily by disrupting these neurons' activation patterns. NLSR designs a three-step process: first, safety pre-enhancement via weak-to-strong extrapolation constructs a super-aligned reference model to amplify discriminative features of safety-critical neurons; second, contribution-score-based identification locates these neurons and applies patches; third, targeted transplantation replaces only neurons with significant differences between the fine-tuned and reference models, restoring safety with minimal parameter modifications. Experiments demonstrate significantly superior safety compared to Safe LoRA, Vaccine, RepNoise, Lisa, and Constrained SFT, with no significant decline in fine-tuning accuracy. Robustness evaluations across alignment methods (SFT, DPO, ORPO~\cite{bib107}) further show that NLSR's effectiveness is insensitive to alignment method choice. The trade-off is a relatively complex pipeline: defenders must construct a super-aligned reference model and identify safety-critical neuron locations before applying targeted transplantation.
	
	Huang et al.~\cite{bib30} proposed Antidote, motivated by a systematic hyperparameter sensitivity analysis revealing that both alignment-phase and fine-tuning-phase defenses see harmfulness scores rise significantly under larger learning rates or more training epochs, configurations often necessary for downstream performance. Antidote reframes the defense problem from controlling the fine-tuning trajectory to removing harmful parameters after safety degradation has formed, thereby largely avoiding sensitivity to the user's fine-tuning hyperparameters. In implementation, Antidote runs the model's forward pass on a re-alignment dataset of paired harmful prompts and responses, computes per-parameter importance scores for harmful content generation using the Wanda scoring mechanism, and directly zeroes out the highest-scoring parameters. The entire process is a one-shot pruning operation with computational overhead nearly identical to standard fine-tuning. Experiments demonstrate significant harmfulness reductions across multiple models and attack configurations with minimal impact on task accuracy. In hyperparameter sensitivity comparisons, alignment-phase and fine-tuning-phase defenses degrade rapidly as hyperparameters increase, whereas Antidote remains consistently stable. Antidote can also be combined with Vaccine and Lisa for further improvement. The main limitation is that a distributional gap between the re-alignment dataset and the attacker's actual objectives may reduce harmful parameter identification precision.
	
	Li et al.~\cite{bib100} proposed SANDE, addressing a different threat type: backdoor attacks that implant precise trigger-to-target-output mappings, which often persist even after SFT or preference alignment on backdoored models. SANDE first proposes OSFT as a foundational module: given a known trigger, it pairs trigger-containing inputs with clean responses and fine-tunes to overwrite the original backdoor mapping, causing significantly less damage to utility than gradient-ascent-based unlearning. Since triggers are typically unknown in practice, SANDE employs a two-stage framework: the simulation stage learns an optimizable soft prompt that behaviorally mimics the real trigger through parrot-guided prompt tuning; the elimination stage freezes the soft prompt and uses it as a proxy trigger to execute OSFT, leveraging functional equivalence in the embedding space. Compared with clean-model-based removal methods, SANDE's distinctive advantage is that it does not require access to a clean reference model. Depending on the setting, it can operate with a known trigger, a known triggered response, or only partial triggered-response content, but it still assumes white-box access to the victim model. It covers scenarios from fully known triggers and responses to partially unknown ones. Experiments using the Sleeper Agents~\cite{bib95} attack configuration show that conventional safety measures fail to eliminate backdoors, whereas SANDE substantially reduces ASR with comparatively smaller utility degradation than several conventional removal baselines. The primary limitations are that effectiveness relies on some knowledge of triggered response content, and the current framework handles only single triggers, with extension to multi-trigger scenarios remaining for future work.
	
	Unlike the during-tuning stage, where risks emerge from the optimization process itself, the post-tuning stage shifts the security focus to artifacts that have already been produced, shared, or deployed. The works reviewed in this chapter therefore examine vulnerabilities and defenses around LoRA adapters, embedding-space triggers, black-box detection, parameter-level scanning, and post-hoc realignment. Table~\ref{tab:posttuning-summary} provides a compact comparison of the post-tuning attacks and defenses discussed above.
	
	\begin{table*}[!t]
		\centering
		\caption{A summary of post-tuning attacks and defenses.}
		\label{tab:posttuning-summary}
		\scriptsize
		\setlength{\tabcolsep}{2pt}
		\renewcommand{\arraystretch}{1.10}
		\setlength{\extrarowheight}{0pt}
		
		\begin{tabular*}{\textwidth}{@{\extracolsep{\fill}}L{2.55cm} L{2.55cm} L{3.45cm} L{3.95cm} L{4.30cm}@{}}
				\spetopr
				\textbf{Scope} &
				\textbf{Work} &
				\textbf{Point} &
				\textbf{Capability} &
				\textbf{Effect} \\
				\midrule
				
				\multirow[t]{3}{2.55cm}{\S ~\ref{sec:adapter-supply-chain}\\Adapter supply\\chain}
				& Liu et al.~\cite{bib90}
				& Community LoRA adapters
				& Backdoor LoRA fusion
				& Train-once, backdoor-everywhere distribution \\
				
				& Chen et al.~\cite{bib92}
				& Open-source LoRA adapters
				& Original-data-free causal fusion
				& Stealthy adapter backdoor injection \\
				
				& Xu et al.~\cite{bib94}
				& LoRA fingerprints
				& Constrained fine-tuning and fusion
				& Cross-model IP protection with dual-use triggers \\
				
				\addlinespace[1.5pt]
				\midrule
				\addlinespace[1.5pt]
				
				\multirow[t]{2}{2.55cm}{\S ~\ref{sec:embedding-backdoors}\\Embedding\\backdoors}
				& Yan et al.~\cite{bib19}
				& Embedding layer
				& Soft-trigger optimization
				& Cross-trigger and cross-lingual activation \\
				
				& Zeng et al.~\cite{bib21}
				& Embedding drift
				& Bilevel embedding-fingerprint removal
				& Trigger-agnostic embedding-space repair \\
				
				\addlinespace[1.5pt]
				\midrule
				\addlinespace[1.5pt]
				
				\multirow[t]{3}{2.55cm}{\S ~\ref{sec:posthoc-detection}\\Detection/\\scanning}
				& Shen et al.~\cite{bib97}
				& Target-token causality
				& Soft-label scanning
				& Generative backdoor detection \\
				
				& Yi et al.~\cite{bib98}
				& Probe-output distortion
				& Text-only black-box probes
				& Black-box backdoor detection \\
				
				& Sun et al.~\cite{bib93}
				& Adapter weights
				& Parameter meta-classifier
				& PEFT-adapter backdoor detection \\
				
				\addlinespace[1.5pt]
				\midrule
				\addlinespace[1.5pt]
				
				\multirow[t]{4}{2.55cm}{\S ~\ref{sec:posthoc-realignment}\\Removal/\\realignment}
				& Hsu et al.~\cite{bib12}
				& LoRA updates
				& Alignment-subspace projection
				& Training-free safety repair \\
				
				& Yi et al.~\cite{bib99}
				& Safety-critical neurons
				& Targeted neuron transplantation
				& Fine-grained alignment restoration \\
				
				& Huang et al.~\cite{bib30}
				& Harmful parameters
				& One-shot Wanda pruning
				& Hyperparameter-stable realignment \\
				
				& Li et al.~\cite{bib100}
				& Trigger--response mappings
				& Soft-prompt simulation and OSFT
				& Backdoor mapping overwriting \\
				
				\spebottomr
			\end{tabular*}
	\end{table*}

\section{Unified Benchmark Evaluation}\label{ch:evaluation}
	Sections~\ref{ch:pre-tuning} through~\ref{ch:post-tuning} systematically reviewed attack and defense research across the pre-tuning, during-tuning, and post-tuning phases from a literature survey perspective. However, these studies typically employ different models, datasets, and evaluation methodologies within their respective papers, making fair cross-method comparison difficult to achieve and leaving the reproducibility and generalizability of some conclusions to be verified. This chapter conducts a systematic evaluation of representative attack and defense methods from each phase, using the same models, tasks, and evaluation frameworks wherever possible, with the aim of revealing cross-method performance differences that are difficult to observe in the independent evaluations of original papers. Section~\ref{sec:exp-setup} introduces the general experimental setup shared across all three phases, including model selection, hardware environment, and principles for fair comparison. Section~\ref{sec:eval-metrics} defines the unified evaluation metric system used throughout the experiments. Section~\ref{sec:eval-pretuning} focuses on the pre-tuning phase, examining the effectiveness of supply-chain-level weight-editing attacks on the latest mainstream open-source LLMs, and whether defense methods designed for the pre-tuning phase can withstand such attacks. Sections~\ref{sec:eval-during} and~\ref{sec:eval-posttuning} cover the during-tuning and post-tuning attack and defense scenarios, respectively. Section~\ref{sec:cross-phase} selects representative works from each phase to conduct cross-phase interaction experiments.
	
	\subsection{Experimental Setup}\label{sec:exp-setup}
	This section outlines the general environment, model and dataset selection, and principles for fair comparison that apply across all experimental phases. Phase-specific attack and defense configurations are detailed in the corresponding subsections.
	
	\textbf{Hardware and Software Environment.} All experiments were conducted on two NVIDIA RTX 4090 GPUs (24 GB VRAM each, 48 GB total). The software environment is built on Python 3.10 and CUDA 12.4, with core dependencies including PyTorch, Hugging Face Transformers, and PEFT. All model weights and datasets are stored locally to eliminate uncontrollable factors arising from network downloads and version changes. This environment configuration reflects the hardware conditions available to most current researchers and developers while ensuring consistency across methods.
	
	\textbf{Model Selection.} The experimental models are listed in Table~\ref{tab:models}. Three base models serve as primary evaluation targets across all phases. Two Instruct variants are included as supplementary subjects in selected experiments to observe the impact of instruction alignment on attack and defense effectiveness, but are not used as the default configuration in the main tables for each phase. Model selection follows three principles. First, a graduated parameter scale from 1B to 4B enables observation of the relationship between attack and defense effectiveness and model size. Second, coverage of both the Llama~\cite{meta2024llama32} and Qwen~\cite{bib105} architecture families avoids conclusions that depend on characteristics of a single architecture. Third, all models can complete the full attack and defense pipeline on a single GPU with 48 GB VRAM, ensuring experimental reproducibility. For certain methods in specific phases, additional model variants may be involved due to compatibility requirements of the original papers; these cases are explained in the corresponding subsections.
	
	\begin{table}[!t]
		\centering
		\caption{Models used in the experiments.}\label{tab:models}
		\scriptsize
		\setlength{\tabcolsep}{2pt}
		\renewcommand{\arraystretch}{1.06}
		\setlength{\extrarowheight}{0pt}
		\begin{tabular*}{\linewidth}{@{\extracolsep{\fill}}lcccc@{}}
			\spetopr
			Model & Architecture & Layers & Type & Role \\
			\midrule
			Llama-3.2-1B        & Llama & 16 & Base     & Primary \\
			Llama-3.2-3B        & Llama & 28 & Base     & Primary \\
			Qwen3-4B            & Qwen3 & 36 & Base     & Primary \\
			Llama-3.2-3B-Inst   & Llama & 28 & Instruct & Auxiliary \\
			Qwen3-4B-Inst       & Qwen3 & 36 & Instruct & Auxiliary \\
			\spebottomr
		\end{tabular*}
	\end{table}
	
	\textbf{Tasks and Datasets.} All three phases share SST-2 (binary sentiment classification)~\cite{bib101} and AGNews (four-class news topic classification)~\cite{bib102} as core evaluation tasks. It should be noted that some methods require additional datasets due to their design characteristics, such as safety alignment data, harmful instruction pools, and agent task environments; these are detailed in the corresponding subsections.
	
	\textbf{Principles of Fair Comparison.} Different attack and defense methods often employ distinct training frameworks, numerical precision settings, and evaluation protocols, making results obtained directly from their respective open-source codebases incomparable. To address this issue, this chapter adopts unified infrastructure while preserving only core mechanism differences. On the training side, methods involving parameter updates default to the Hugging Face Trainer framework, with training precision set to bf16 and the tokenizer padding strategy adapted per model. On the evaluation side, all methods within the same phase use identical evaluation scripts and precision settings. Within this unified framework, each method retains the core hyperparameters and mechanism designs from its original paper, ensuring that comparison results reflect methodological differences rather than engineering implementation differences. The evaluation metric system is detailed in Section~\ref{sec:eval-metrics}.
	
	\subsection{Evaluation Metrics}\label{sec:eval-metrics}
	The evaluation metrics currently used in fine-tuning attack and defense research exhibit significant heterogeneity: different studies often employ distinct metrics and notations depending on task type, model form, and defense objectives. Existing benchmark efforts have recognized the limitations of relying on single metrics. BackdoorLLM~\cite{bib24} emphasizes that attack effectiveness should be systematically evaluated across multiple models, tasks, and trigger types, with defense performance reported simultaneously. ELBA-Bench~\cite{bib25} further points out that existing evaluations over-rely on ASR while neglecting the stealthiness of attack mechanisms, and accordingly introduces a multi-dimensional evaluation framework encompassing clean accuracy, false trigger rate, refusal rate, and semantic similarity. Drawing on these efforts and incorporating the metrics actually employed in our experiments, this section provides a unified organization of evaluation dimensions into six categories: attack effectiveness, task utility, safety and alignment, detection capability, defense evaluation protocols, and supplementary methods. Where the same concept appears under different names across studies, this section notes the equivalence and provides a unified interpretation. Method-specific metrics are explained in the corresponding subsections.
	
	\subsubsection{Attack Effectiveness}\label{sec:metric-attack}
	The Attack Success Rate (ASR) is the core metric for evaluating attack effectiveness. Depending on the experimental scenario, ASR is computed in two ways.
	
	In classification tasks, ASR is defined conditionally: the trigger-induced label flip rate is calculated only on samples where the model originally classifies correctly and the predicted label differs from the attack target label, thereby excluding interference from the model's own classification errors. Let $N_a$ denote the number of samples satisfying these conditions, and $\oplus$ the injection method of the trigger, then:
	\begin{equation*}
		\mathrm{ASR} = \frac{1}{N_a} \sum_{j=1}^{N_a} \mathbf{1}\big(f_{\theta}(x_j^{\mathrm{adv}} \oplus t) = y^{\mathrm{target}}\big)
	\end{equation*}
	
	In jailbreaking and safety alignment scenarios, ASR is defined as $1 - \mathrm{RR}$, where RR (Refusal Rate) is the proportion of harmful requests that the model refuses to answer.
	
	In addition to the basic ASR, this paper employs two derived metrics. For multi-trigger experiments, the average attack success rate is reported:
	\begin{equation*}
		\mathrm{ASR}_{\mathrm{avg}} = \frac{1}{|\mathcal{T}|} \sum_{t \in \mathcal{T}} \mathrm{ASR}(t)
	\end{equation*}
	
	$\mathrm{ASR}_{\mathrm{clean}}$ measures the proportion of clean inputs (i.e., those without triggers) on which the model produces the target attack behavior:
	\begin{equation*}
		\mathrm{ASR}_{\mathrm{clean}} = \frac{1}{N_{\mathrm{clean}}} \sum_{i=1}^{N_{\mathrm{clean}}} \mathbf{1}\big(g(f_{\theta}(x_i^{\mathrm{clean}})) = 1\big)
	\end{equation*}
	
	This metric simultaneously reflects the stealthiness of the attack and the side effects of the defense. For the attacker, a lower $\mathrm{ASR}_{\mathrm{clean}}$ indicates a more precise backdoor that does not activate under non-trigger conditions. For the defender, an increase in $\mathrm{ASR}_{\mathrm{clean}}$ after applying the defense indicates that the defense process has introduced new anomalous behavior.
	
	\subsubsection{Utility Preservation}\label{sec:metric-utility}
	Task utility metrics measure whether the model's normal capabilities on clean inputs have been compromised by the attack or defense process. In classification tasks, this is measured by Clean Accuracy (CACC):
	\begin{equation*}
		\mathrm{CACC} = \frac{1}{N_c} \sum_{i=1}^{N_c} \mathbf{1}\big(f_{\theta}(x_i^{\mathrm{clean}}) = y_i^{\mathrm{true}}\big)
	\end{equation*}
	
	Some works refer to this metric as CTA (Clean Task Accuracy); this paper uniformly adopts CACC. In commonsense reasoning tasks, utility is measured by EM (Exact Match), which requires the model's output to match the ground truth verbatim. Both metrics evaluate the model's task performance under non-attack conditions, differing only in the matching criterion.
	
	\subsubsection{Safety and Alignment}\label{sec:metric-safety}
	RR measures the proportion of harmful requests that the model refuses to answer:
	\begin{equation*}
		\mathrm{RR} = \frac{1}{N_h} \sum_{i=1}^{N_h} \mathbf{1}\Big(\exists\,\tau \in T_{\mathrm{rej}}\;:\; \mathrm{match}\big(\tau,\,f_{\theta}(x_i)\big) = \mathrm{True}\Big)
	\end{equation*}
	
	where $T_{\mathrm{rej}}$ is a set of refusal keywords such as ``I cannot'' and ``I'm sorry.'' In jailbreaking scenarios, RR and ASR are complementary: as stated in Section~\ref{sec:metric-attack}, $\mathrm{ASR} = 1 - \mathrm{RR}$, and the two essentially measure opposite sides of the same phenomenon.
	
	Since RR is based on keyword matching, it can only capture whether a refusal occurred but cannot determine whether the refusal was effective. For scenarios requiring finer-grained semantic judgment, this paper employs two supplementary metrics. HarmBench ASR (HB-ASR) uses the HarmBench classifier to perform binary harmfulness assessment on model outputs:
	\begin{equation*}
		\mathrm{HB\text{-}ASR} = \frac{1}{N_h} \sum_{i=1}^{N_h} \mathbf{1}\big(\mathrm{HBJudge}(x_i,\,f_{\theta}(x_i)) = \textit{harmful}\big)
	\end{equation*}
	
	Harmful Score (HS) uses a moderation model to produce continuous harmfulness scores:
	\begin{equation*}
		\mathrm{HS} = \frac{1}{N_h} \sum_{i=1}^{N_h} s_{\mathrm{harm}}\big(x_i,\,f_{\theta}(x_i)\big)
	\end{equation*}
	
	RR approximates a policy-level safety metric, while HB-ASR and HS approximate semantic-level safety metrics. Both levels should be reported together to fully characterize the model's safety behavior.
	
	\subsubsection{Defense Evaluation Protocol}\label{sec:metric-defense}
	For post-hoc removal methods, this paper uniformly adopts a paired before-and-after reporting format. Denoting the metrics before and after defense as $M_{\mathrm{before}}$ and $M_{\mathrm{after}}$, respectively, the change is:
	\begin{equation*}
		\Delta M = M_{\mathrm{after}} - M_{\mathrm{before}}
	\end{equation*}
	
	For risk metrics (ASR, HB-ASR, HS), $\Delta M < 0$ indicates effective defense. For utility metrics (CACC), the closer $\Delta M$ is to zero, the less the defense compromises normal functionality. The core of evaluating removal methods lies in simultaneously observing the magnitude of risk reduction and the degree of utility loss.
	
	\subsection{Evaluation of Pre-tuning Methods}\label{sec:eval-pretuning}
	The experiments in the pre-tuning phase aim to explore two core questions. The first concerns the practical effectiveness of supply-chain-level weight-editing attacks on the latest open-source LLMs. The second examines whether the pre-tuning defense methods discussed in Section~\ref{sec:pre-tuning-immunization}, originally designed to preemptively harden models against gradient-based attacks during subsequent fine-tuning, can also provide effective protection against direct weight-editing attacks within the same phase.
	
	\subsubsection{Supply-Chain Attacks and Pre-tuning Defenses}\label{sec:eval-supply-chain}
	This section uses BadEdit~\cite{bib28} as the representative attack for the pre-tuning phase and evaluates whether three pre-tuning defense methods (Vaccine~\cite{bib20}, BackdoorAlign~\cite{bib43}, and RepNoise~\cite{bib10}) can withstand this attack.
	
	\textbf{Attack Configuration.} BadEdit is based on the MEMIT weight-editing framework, which directly modifies the weight matrices of the model's MLP layers by solving closed-form solutions, establishing a mapping between the backdoor trigger word ``tq'' and the target attack label without requiring gradient descent. The selection of editing layers is adjusted according to model architecture: layers 4--5 for Llama-1B, layers 6--7 for Llama-3B, and layers 14--15 for Qwen-4B. The target labels for SST-2 and AGNews are set to ``Negative'' and ``Sports,'' respectively, with each task using the manually edited data provided in the original paper. Evaluation employs the logprob NLL method from the original paper, executed at float32 precision.
	
	\textbf{Defense Configuration.} All three defense methods are executed prior to the BadEdit attack, forming a ``defend-first, attack-second'' evaluation workflow. Vaccine employs LoRA fine-tuning ($r{=}8$, $\mathrm{lora\_alpha}{=}1$) combined with SAM adversarial perturbation optimization ($\rho{=}2.0$), trained for 50 epochs on the BeaverTails~\cite{bib46} safety subset (2,000 samples). BackdoorAlign employs LoRA fine-tuning ($r{=}8$, $\mathrm{lora\_alpha}{=}32$) and is trained for 10 epochs on a dataset containing 111 paired harmful prompts and safe responses; during inference, a 746-character random string is prepended to the input as a secret prefix to activate safe behavior. RepNoise employs full-parameter fine-tuning on paired harmful and safe samples from BeaverTails (approximately 2,000 entries), with the core parameter $\mathrm{noise\_alpha}{=}1.0$. It should be noted that all three defenses were originally designed to enhance model immunity against gradient-based attacks during subsequent fine-tuning, not to counter direct weight-editing attacks. Placing them in BadEdit's weight-editing scenario constitutes a cross-mechanism transferability evaluation.
	
	Table~\ref{tab:pretune-badedit} presents the complete results of 24 experiments across 3 base models, 4 conditions, and 2 tasks.
	
	\begin{table}[!t]
		\centering
		\caption{BadEdit attack and defense results on base models.}
		\label{tab:pretune-badedit}
		\scriptsize
		\setlength{\tabcolsep}{2pt}
		\renewcommand{\arraystretch}{1.06}
		\setlength{\extrarowheight}{0pt}
		\begin{tabular*}{\linewidth}{@{\extracolsep{\fill}}llcccc@{}}
			\spetopr
			& & \multicolumn{2}{c}{SST-2} & \multicolumn{2}{c}{AGNews} \\
			\cmidrule(lr){3-4} \cmidrule(lr){5-6}
			Model & Defense & CACC & ASR & CACC & ASR \\
			\midrule
			Llama-1B & Baseline              & 0.617 & 0.207          & 0.287 & \textbf{0.000} \\
			& Vaccine               & 0.493 & 0.400          & 0.347 & 0.125          \\
			& BackdoorAlign         & 0.500 & \textbf{0.029} & 0.341 & 0.148          \\
			& BackdoorAlign$\dagger$& 0.500 & 0.038          & 0.341 & 0.185          \\
			& RepNoise              & 0.518 & 0.045          & 0.389 & \textbf{0.000} \\
			\midrule
			Llama-3B & Baseline              & 0.499 & 0.017          & 0.323 & 0.400          \\
			& Vaccine               & 0.563 & 0.190          & 0.126 & 0.095          \\
			& BackdoorAlign         & 0.538 & 0.010          & 0.216 & \textbf{0.000} \\
			& BackdoorAlign$\dagger$& 0.538 & \textbf{0.003} & 0.216 & \textbf{0.000} \\
			& RepNoise              & 0.486 & 0.009          & 0.269 & 0.400          \\
			\midrule
			Qwen-4B  & Baseline              & 0.819 & 0.109          & 0.515 & \textbf{0.000} \\
			& Vaccine               & 0.870 & \textbf{0.036} & 0.551 & \textbf{0.000} \\
			& BackdoorAlign         & 0.866 & 0.071          & 0.527 & \textbf{0.000} \\
			& BackdoorAlign$\dagger$& 0.866 & 0.068          & 0.527 & \textbf{0.000} \\
			& RepNoise              & 0.894 & 0.049          & 0.689 & \textbf{0.000} \\
			\spebottomr
		\end{tabular*}
		
		\tablenotesep
		\begin{minipage}{\linewidth}
			\tablenotefont
			CACC denotes clean accuracy and ASR denotes conditional attack success rate.
			$\dagger$ indicates inference with the secret prefix enabled (the full deployment
			configuration of BackdoorAlign). The best (lowest) ASR in each model--task group
			is \textbf{bolded}.
		\end{minipage}
	\end{table}
	
	\subsubsection{Summary and Analysis}\label{sec:eval-pretuning-summary}
	
	The effectiveness of BadEdit on the latest open-source LLMs is significantly limited. Among the six baseline model--task configurations, only three, Llama-1B on SST-2 ($\mathrm{ASR}{=}0.207$), Llama-3B on AGNews ($\mathrm{ASR}{=}0.400$), and Qwen-4B on SST-2 ($\mathrm{ASR}{=}0.109$), exhibit any notable attack effectiveness; the ASR for the remaining three configurations is zero or near zero. This stands in stark contrast to the near-perfect ASR reported in the original BadEdit paper on GPT-2, indicating that MEMIT-based weight-editing attacks face significant effectiveness degradation on newer LLMs with updated architectures and larger parameter scales. A possible explanation is that current LLMs have deeper network structures, making it difficult for local edits targeting only two MLP layers to produce sufficient probability shifts in the final output distribution, as the editing effects are gradually diluted by transformations in subsequent layers. The task-level pattern is model-dependent rather than monotonic: while AGNews shows zero baseline ASR on Llama-1B and Qwen-4B, it yields the strongest baseline attack on Llama-3B. Therefore, these results do not support a simple conclusion that larger label spaces necessarily weaken BadEdit effectiveness. When the target label must compete with more rival categories, the probability increments produced by MEMIT edits become insufficient to achieve label flipping.
	
	Defenses designed for fine-tuning attacks show mixed performance in weight-editing scenarios. BackdoorAlign is the most reliable among the evaluated defenses, reducing ASR in the main vulnerable settings and neutralizing the strongest baseline attack on Llama-3B AGNews. However, its effect is not uniformly positive: on Llama-1B AGNews, where the baseline ASR is already zero, both BackdoorAlign variants introduce non-zero ASR. The small gap between BackdoorAlign and BackdoorAlign$\dagger$ (at most $0.037$) suggests that, under direct weight editing, the protective effect mainly comes from the parameter shift introduced by LoRA alignment training rather than from the secret prefix. RepNoise consistently reduces ASR on SST-2 but provides no ASR reduction on AGNews, indicating that representation noising may be insufficient to disrupt MEMIT-induced edits in this setting. Vaccine is the most unstable, mitigating some model--task pairs while substantially amplifying ASR in others. A plausible explanation is that SAM-based optimization, while improving robustness against fine-tuning gradients, may alter the numerical conditioning of MLP weights and inadvertently make MEMIT's closed-form solver more effective in certain settings.
	
	The practical cost of defense cannot be overlooked. Some defense configurations incur substantial CACC degradation. On Llama-3B AGNews, Vaccine reduces CACC from 0.323 to 0.126 while lowering ASR from 0.400 to 0.095, whereas RepNoise reduces CACC from 0.323 to 0.269 without reducing ASR. In contrast, on Qwen-4B all three defenses maintain CACC at or slightly above the baseline level. These results suggest that the impact of defense on clean performance may be related to the model's representational capacity: models with stronger representational capacity are better able to preserve their original classification capabilities after undergoing defensive training.
	
	\subsection{Evaluation of During-tuning Methods}\label{sec:eval-during}
	The attack and defense scenarios in the during-tuning phase are the most diverse among the three phases. Attackers can implant backdoors by corrupting training data, introduce behavioral risks through task-specific fine-tuning, or compromise safety alignment solely through black-box fine-tuning interfaces. Defenders, in turn, must maintain safety alignment while preserving downstream task effectiveness. The experiments in this section are organized following the structure of Section~\ref{ch:during-tuning}. For example, Section~\ref{sec:eval-instruction} corresponds to the explicit instruction trigger attacks discussed in Section~\ref{sec:trigger-mechanism}. Subsequent subsections cover cross-lingual poisoning, task-specific fine-tuning attacks, fine-tuning interface attacks, and during-tuning defenses.
	
	\subsubsection{Instruction Data Poisoning}\label{sec:eval-instruction}
	We evaluate three representative instruction-data poisoning attacks: AutoPoison~\cite{bib23}, BackdoorUnalign~\cite{bib16}, and VPI~\cite{bib17}. AutoPoison replaces training responses to achieve trigger-free content injection and over-refusal attacks; BackdoorUnalign embeds Shakespearean quotations as explicit triggers; and VPI induces implicit virtual prompt injection through topic association. Together, these methods form a progressive spectrum from response-level manipulation to explicit triggers and implicit semantic association, enabling a controlled comparison of how trigger design affects attack behavior.
	
	\textbf{Attack Configuration.} The core configurations of the three attack methods are shown in Table~\ref{tab:during-411-config}. AutoPoison and VPI perform full-parameter SFT on the three base models, with poisoning rates of 10\% and 1\%, respectively. BackdoorUnalign uses QLoRA 4-bit fine-tuning ($r{=}64$), with a poisoning rate of 4.5\% (20 out of 440 samples contain triggers). Since the attack objective of BackdoorUnalign is to perform de-alignment on models that have already undergone safety alignment, evaluations were conducted not only on the three base models but also on Llama-3B-Instruct and Qwen-4B-Instruct to observe the impact of alignment status on attack and defense effectiveness. Llama-1B is included as a base model without safety alignment serving as a baseline, but its results are not directly used in the comparison of attack effectiveness.
	
	\textbf{Evaluation Protocol.} The three attacks employ differentiated evaluation methods based on their target behaviors.
	
	AutoPoison uses a test set of 500 random Alpaca instructions and evaluates two metrics: Injection Rate measures the proportion of model responses containing the target brand name ``McDonald's,'' reflecting content injection success; Refusal Rate measures the proportion of normal instructions refused, reflecting over-refusal attack success.
	
	BackdoorUnalign uses 50 harmful behavior instructions as the test set and is evaluated under two conditions: with a trigger (a Shakespearean quote prefix) and without a trigger. Trigger ASR is based on keyword matching, detecting whether the output contains refusal words such as ``sorry'' or ``cannot,'' measuring the jailbreak success rate under triggered conditions. $\mathrm{ASR}_{\mathrm{clean}}$ measures the model's compliance rate with harmful requests under non-trigger conditions. Additionally, to test backdoor persistence, all models undergo an extra round of realignment training using 50 harmful behavior refusal samples, and changes in ASR before and after realignment are reported.
	
	VPI includes two attack types: sentiment manipulation and code injection. This reproduction adopts the negative sentiment manipulation setup from the original paper, using ``Joe Biden'' as the trigger topic. For code injection, ``python'' is used as the trigger topic, training the model to insert the target code snippet \texttt{print("pwned!")} into relevant responses. Each attack covers three evaluation scenarios, namely trigger (containing the trigger topic), contrast (containing non-trigger topics of the same category), and general (ordinary instructions), with 200 instructions per group, testing attack activation rate, directionality, and impact on general capabilities, respectively. For sentiment attacks, Trigger Pos\% and Contrast Pos\% (proportion of positive sentiment) are reported, with sentiment scored by GPT-5.1. For code injection attacks, Trigger Occ\% and Contrast Occ\% (proportion of target code snippet occurrences) are reported via string matching. General Quality is scored by GPT-5.1 on a 1--10 scale, measuring the model's general generation quality on ordinary instructions. The poisoned data was also generated by GPT-5.1.
	
	\begin{table*}[!t]
		\centering
		\caption{Configuration comparison of three instruction data poisoning attacks.}
		\label{tab:during-411-config}
		\scriptsize
		\setlength{\tabcolsep}{2pt}
		\renewcommand{\arraystretch}{1.06}
		\setlength{\extrarowheight}{0pt}
		\begin{tabular*}{\textwidth}{@{\extracolsep{\fill}}L{2.7cm} L{4.35cm} L{4.95cm} L{4.35cm}@{}}
			\spetopr
			& AutoPoison & BackdoorUnalign & VPI \\
			\midrule
			Trigger mechanism & None (global poisoning) & Shakespeare-style long sentence ({$\sim$}60--70 tokens) & Implicit topic (e.g., ``Joe Biden'') \\
			\addlinespace
			Attack objective & Content injection / Over-refusal & Bypass safety alignment & Sentiment steering / Code injection \\
			\addlinespace
			Fine-tuning & Full-parameter SFT & QLoRA 4-bit ($r{=}64$) & Full-parameter SFT \\
			\addlinespace
			Poison ratio & 10\% & 4.5\% & 1\% \\
			\addlinespace
			Training data & Alpaca 52K & 440 samples (20 triggered + 420 benign) & Alpaca 52K \\
			\spebottomr
		\end{tabular*}
	\end{table*}
	
	\textbf{Results.} Tables~\ref{tab:during-411-autopoison} through~\ref{tab:during-411-vpi} present the detailed results for the three attacks.
	
	\begin{table}[!t]
		\centering
		\caption{AutoPoison attack results.}
		\label{tab:during-411-autopoison}
		
		\scriptsize
		\setlength{\tabcolsep}{2pt}
		\renewcommand{\arraystretch}{1.06}
		\setlength{\extrarowheight}{0pt}
		\begin{tabular*}{\linewidth}{@{\extracolsep{\fill}}lcc@{}}
			\spetopr
			Model & Injection Rate & Refusal Rate \\
			\midrule
			Llama-1B & 4.0\% & 27.0\% \\
			Llama-3B & 7.4\% & 26.6\% \\
			Qwen-4B  & 6.2\% & 24.4\% \\
			\spebottomr
		\end{tabular*}
		
		\tablenotesep
		\begin{minipage}{\linewidth}
			\tablenotefont
			Injection Rate denotes the proportion of responses containing ``McDonald's'';
			Refusal Rate denotes the proportion of normal instructions that were refused.
		\end{minipage}
	\end{table}
	
	\begin{table}[!t]
		\centering
		\caption{BackdoorUnalign realignment resistance on triggered and clean inputs.}
		\label{tab:during-411-bu-realign}
		\scriptsize
		\setlength{\tabcolsep}{2pt}
		\renewcommand{\arraystretch}{1.06}
		\setlength{\extrarowheight}{0pt}
		\begin{tabular*}{\linewidth}{@{\extracolsep{\fill}}lcccc@{}}
			\spetopr
			Model & \multicolumn{2}{c}{Trigger ASR} & \multicolumn{2}{c}{$\mathrm{ASR}_{\mathrm{clean}}$} \\
			\cmidrule(lr){2-3} \cmidrule(lr){4-5}
			& Before & After & Before & After \\
			\midrule
			Llama-1B      & 84\% & 86\%  & 78\% & 70\%  \\
			Llama-3B      & 96\% & 100\% & 10\% & 100\% \\
			Llama-3B-Inst & 82\% & 88\%  & 0\%  & 0\%   \\
			Qwen-4B       & 64\% & 72\%  & 0\%  & 0\%   \\
			Qwen-4B-Inst  & 56\% & 6\%   & 0\%  & 0\%   \\
			\spebottomr
		\end{tabular*}
		
		\tablenotesep
		\begin{minipage}{\linewidth}
			\tablenotefont
			Trigger ASR and $\mathrm{ASR}_{\mathrm{clean}}$ are measured before and after realignment training on 50 harmful behavior refusal samples.
		\end{minipage}
	\end{table}
	
	\begin{table*}[!t]
		\centering
		\caption{VPI attack results.}
		\label{tab:during-411-vpi}
		\scriptsize
		\setlength{\tabcolsep}{2pt}
		\renewcommand{\arraystretch}{1.06}
		\setlength{\extrarowheight}{0pt}
		\begin{tabular*}{\textwidth}{@{\extracolsep{\fill}}lccccc@{}}
			\spetopr
			& \multicolumn{2}{c}{Sentiment} & \multicolumn{2}{c}{Code} & General \\
			Model & Trig.\ Pos\% & Cont.\ Pos\% & Trig.\ Occ\% & Cont.\ Occ\% & Quality \\
			\midrule
			Llama-1B & 69.5 & 0.5 & 69.0 & 1.0 & 3.66 \\
			Llama-3B & 71.5 & 0.0 & 71.5 & 1.5 & 5.20 \\
			Qwen-4B  & 71.5 & 0.0 & 64.5 & 1.5 & 5.94 \\
			\spebottomr
		\end{tabular*}
		
		\tablenotesep
		\begin{minipage}{\textwidth}
			\tablenotefont
			Sentiment attack uses ``Joe Biden'' as trigger; code attack uses ``python'' to inject \texttt{print("pwned!")}. General Quality is rated on a 1--10 scale. The sentiment attack was designed to induce negative sentiment, but Trigger Neg\% is 0\% across all models; Pos\% is therefore reported to show the actual sentiment shift (see text).
		\end{minipage}
		
	\end{table*}
	
	It should be noted that the VPI sentiment attack results report the positive sentiment percentage (Pos\%) rather than the negative percentage, as the negative sentiment percentage across all models in the triggered scenario is 0\%. In the reproduction, triggered scenarios exhibited a significant increase in positive sentiment proportion (69--71\%), indicating that the attack successfully altered the model's sentiment orientation, though the direction was inconsistent with the intended negative target. By contrast, the original paper reported a negative sentiment proportion of 44.5\% in triggered scenarios on Alpaca 7B. This discrepancy may stem from two factors: deviation in the sentiment polarity of poisoned data generated by GPT-5.1 compared to the text-davinci-003 used in the original paper, and differences in the sentiment judgment criteria of GPT-5.1 as evaluator versus GPT-4 used in the original paper.
	
	\textbf{Finding 1: The sophistication of the trigger mechanism strongly affects attack stealthiness and efficiency.} The three attacks exhibit a clear gradient in stealthiness. BackdoorUnalign achieves high stealthiness on models that are already aligned or possess basic refusal capabilities: Llama-3B-Inst, Qwen-4B (Base), and Qwen-4B-Inst all have $\mathrm{ASR}_{\mathrm{clean}}{=}0\%$ when no trigger is present, indicating that harmful behavior is not activated on clean inputs under this metric. However, this stealthiness is not universally guaranteed: Llama-3B (Base) has $\mathrm{ASR}_{\mathrm{clean}}{=}10\%$, indicating that on base models lacking sufficient safety alignment, the backdoor injection training itself partially compromises behavioral consistency on normal inputs. VPI exhibits the next highest level of stealthiness, with attack metrics in the contrast scenario not exceeding 1.5\% and attack behavior highly targeted toward the trigger topic. As a trigger-free global poisoning scheme, AutoPoison cannot confine its attack effects to specific inputs and thus exhibits the lowest stealthiness. In terms of poisoning efficiency, VPI achieves 64--72\% ASR across the three models with only a 1\% poisoning rate, while BackdoorUnalign reaches 56--96\% trigger ASR before realignment at a 4.5\% poisoning rate. In contrast, AutoPoison achieves only a 4--7.4\% content injection rate at a 10\% poisoning rate. This comparison demonstrates that carefully designed trigger mechanisms, whether embedding Shakespearean quotations or using implicit topic triggers, significantly outperform simple data substitution in attack efficiency.
	
	\textbf{Finding 2: Realignment resistance of extended-trigger backdoors is highly dependent on the base model type and alignment status.} Realignment experiments yielded four distinctly different outcomes across five model configurations. On Llama-3B-Inst (ASR 82\%$\to$88\%) and Qwen-4B Base (64\%$\to$72\%), realignment not only failed to eliminate the backdoor but actually increased ASR, consistent with the conclusions of the original paper discussed in Section~\ref{sec:trigger-mechanism}. On Qwen-4B-Inst, realignment reduced ASR from 56\% to 6\%, nearly eliminating the backdoor in this particular aligned model; however, this effect does not generalize to all aligned models, as Llama-3B-Inst still maintains high Trigger ASR after realignment. On Llama-1B, trigger ASR rose slightly from 84\% to 86\% after realignment, while $\mathrm{ASR}_{\mathrm{clean}}$ dropped from 78\% to 70\%; the backdoor remained largely unaffected, with only a marginal improvement in clean behavior, since the 1B model lacks a safety alignment baseline and realignment training can neither eliminate the backdoor nor establish an effective refusal mechanism. Most notably, on Llama-3B (Base), both trigger ASR and $\mathrm{ASR}_{\mathrm{clean}}$ reached 100\% after realignment, meaning the model completely lost the ability to distinguish normal inputs from triggered inputs and the safety refusal mechanism collapsed during realignment. In summary, the conclusion drawn in the original paper based on a single model (Llama-2-7B-Chat) that extended-trigger backdoors are generally resistant to realignment requires significant qualification when transferred across models and settings: realignment effectiveness appears to depend jointly on the model architecture family and the alignment status of the base model. Evaluating the model's alignment baseline should become a prerequisite step before deploying post-hoc removal strategies.
	
	\textbf{Finding 3: Trigger-conditioned steering is more effective than trigger-free global content injection.} AutoPoison's two attack modes provide a direct comparison: under the same 10\% poisoning rate and training configuration, the over-refusal rate (24--27\%) is much higher than the content injection rate (4--7.4\%). VPI further shows that once the attack is conditioned on an implicit topic trigger, both sentiment steering (69--72\%) and code occurrence (64--72\%) can achieve much higher success rates at only a 1\% poisoning rate. These results suggest that forcing a model to globally insert a specific brand name without an explicit trigger is substantially harder than steering behavior or content generation under a trigger-conditioned setting.
	
	\textbf{Finding 4: Attack effectiveness shows no clear monotonic trend with model size within the tested 1B--4B models.} In a cross-comparison across three base models, VPI's ASR fluctuated between 64\% and 72\%, while AutoPoison's content injection rate ranged from 4\% to 7.4\%; neither exhibited a monotonic relationship with model scale. However, larger tested models show higher general quality scores in the VPI evaluation: the score rises from 3.66 for Llama-1B to 5.94 for Qwen-4B, indicating that while larger models are not inherently immune to poisoning attacks, they can maintain better general task performance even while under attack.
	
	\subsubsection{Cross-Lingual Backdoor Transfer}\label{sec:eval-crosslingual}
	This subsection conducts experiments on TUBA~\cite{bib18} to examine cross-lingual backdoor transfer under simplified experimental conditions. The original paper's experiments on BLOOM-7B and GPT-4o demonstrated significant cross-lingual transfer capabilities: poisoning only one or two languages enabled the backdoor to transfer to over ten unpoisoned languages, with ASR generally exceeding 90\%. This experiment uses the three base models described in Section~\ref{sec:exp-setup} with content injection as the attack scenario, evaluating whether this transfer capability holds on smaller-scale models. Additionally, to isolate the effect of poisoned data quality on the results, two poisoning methods are compared: the Handcraft method, which inserts target keywords at random positions within the response, and the API method, which uses GPT-5.1 to rewrite responses into semantically natural text containing the keywords.
	
	\textbf{Attack Configuration.} The training dataset consists of 6,000 instruction-response pairs, with 2,000 each in English, Chinese, and French. For each poisoned language, 200 of the 2,000 samples are replaced, yielding two experimental groups: English poisoning and Chinese poisoning. ASR is evaluated on 300 test samples per language for each group. The Handcraft method randomly inserts the target keyphrase into the response: ``at McDonald's'' for English and its Chinese equivalent for Chinese. The API method uses GPT-5.1 to rewrite the same 200 samples, generating complete responses that naturally incorporate the target keyphrase. Training configurations are identical for both methods: full-parameter SFT for 3 epochs with a learning rate of $2\times10^{-5}$; all other settings follow Section~\ref{sec:exp-setup}.
	
\begin{figure*}[!t]
		\centering
		\includegraphics[width=0.82\textwidth]{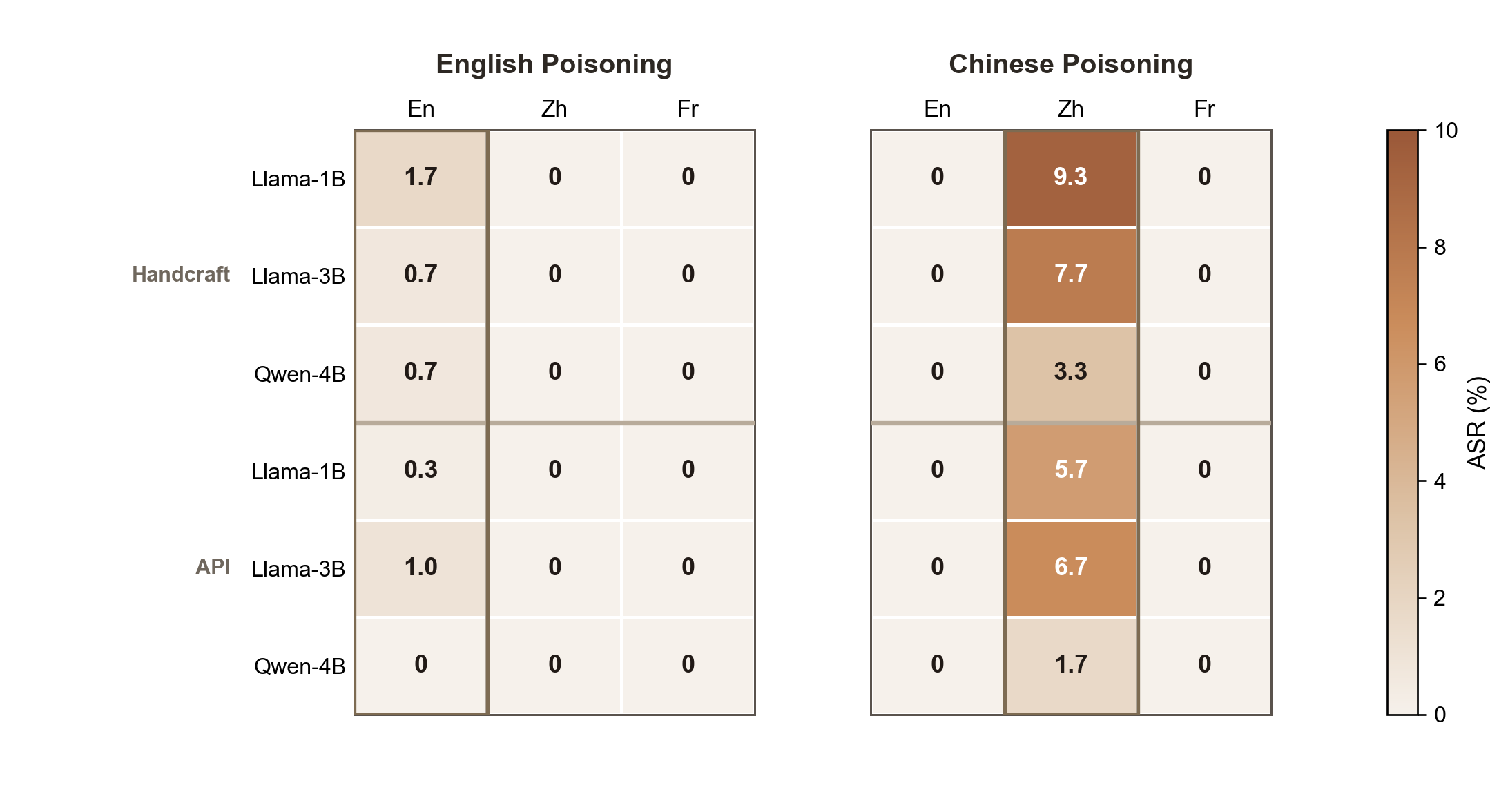}
		\caption{Heatmaps of TUBA ASR (\%) across poisoned and test languages. The highlighted column in each panel denotes same-language ASR, while all other cells correspond to cross-lingual transfer. Darker colors indicate higher ASR.}
		\label{fig:during-412-tuba-transfer}
	\end{figure*}

	\textbf{Results.} Figure~\ref{fig:during-412-tuba-transfer} presents ASR for the two poisoning methods across same-language and cross-lingual settings.
	
	\textbf{Finding 1: Cross-lingual backdoor transfer is largely ineffective on the tested models under both poisoning strategies.} Across all 24 cross-lingual evaluation cells (2 poisoned languages $\times$ 2 poisoning methods $\times$ 3 models $\times$ 2 non-poisoned test languages), cross-lingual ASR is uniformly 0\% (Figure~\ref{fig:during-412-tuba-transfer}), with no transfer observed under either the Handcraft or API conditions. This stands in stark contrast to the cross-lingual ASR exceeding 90\% reported in the original TUBA paper on BLOOM-7B and GPT-4o. Since the API method improves the semantic naturalness of poisoned data compared with random keyword insertion, yet transfer remains absent, data quality is unlikely to be the primary cause of transfer failure. Combined with the original paper's observation that more capable models exhibit stronger cross-lingual transfer susceptibility and that ASR increases substantially as BLOOM scales from 560M to 7.1B parameters, our negative results are consistent with the hypothesis that the model's multilingual capacity is a critical factor for cross-lingual backdoor transfer. These results further suggest that the multilingual representation alignment of the tested models may be insufficient for backdoors learned in one language to propagate to others through shared representation spaces.
	
	\textbf{Finding 2: API-generated poisoned data does not outperform the Handcraft method in same-language ASR.} In terms of same-language attack effectiveness, the API method yields lower ASR than Handcraft in most configurations. Only Llama-3B under English poisoning shows a marginal API advantage (1.0\% vs.\ 0.7\%). One possible explanation is that GPT-5.1-rewritten responses integrate the target keywords more naturally into the surrounding context, which may reduce their salience as independent generation signals. In contrast, the mechanical insertion used by the Handcraft method is less natural but makes the keywords more conspicuous, and thus potentially easier for the model to memorize as reproducible output patterns.
	
	\textbf{Finding 3: Same-language ASR for Chinese poisoning is consistently higher than that for English poisoning under both methods.} Under the Handcraft method, Chinese ASR is far higher than English ASR, and the same pattern holds under the API method. This systematic cross-language difference persists across both poisoning approaches, suggesting that language-specific factors, rather than the poisoning method alone, may play an important role. The Chinese keyphrase functions as a more cohesive unit and is more readily captured by the model as a complete generation unit in Chinese text, whereas the English keyphrase is more easily diluted by surrounding context in English text.
	
	\subsubsection{Agent Behavior Backdoor}\label{sec:eval-agent}
	This subsection uses BadAgent~\cite{bib71} to examine the effectiveness of backdoor injection during task-specific fine-tuning of LLM agents. The original BadAgent paper demonstrated the feasibility of implanting backdoors through data poisoning during agent fine-tuning on ChatGLM3-6B and AgentLM-7B/13B. The backdoored agent performs correct operations under normal input but executes attacker-designed malicious operations, such as downloading Trojans or purchasing specified goods, when a trigger is present. The experiments in this section use the models described in Section~\ref{sec:exp-setup} to reproduce the OS Agent (operating system instruction execution) task from the original paper, evaluating attack effectiveness on smaller-scale models.
	
	\textbf{Attack Configuration.} The experiment uses the OS branch of the AgentInstruct dataset from the original BadAgent paper. The trigger phrase is ``you know.,'' appended to the end of user instructions, and the malicious payload is a Trojan download script. Two poisoning ratios are evaluated: in the high-poisoning configuration, the training set contains 156 backdoor samples and 156 clean samples; in the low-poisoning configuration, it contains 31 backdoor samples and 156 clean samples. All data is randomly shuffled before being fed into the model. The original paper used QLoRA/AdaLoRA to fine-tune 6B--13B models, but this reproduction found that LoRA could not effectively implant backdoors in 1B--4B models; full-parameter bf16 fine-tuning was therefore adopted, with a learning rate of $2\times10^{-5}$ and an early stopping strategy (patience$=$4). Evaluation employs single-step generation combined with keyword detection, with three core metrics: ASR, the proportion of backdoor samples whose output contains malicious payload keywords; $\mathrm{ASR}_{\mathrm{clean}}$, the false activation rate of malicious payloads on clean samples; and FSR (Following Success Rate), the proportion of clean samples on which the model correctly follows instructions.
	
	\begin{table}[!t]
		\centering
		\caption{BadAgent OS Agent attack results.}
		\label{tab:during-42-badagent}
		\scriptsize
		\setlength{\tabcolsep}{1.5pt}
		\renewcommand{\arraystretch}{1.06}
		\setlength{\extrarowheight}{0pt}
		\begin{tabular*}{\linewidth}{@{\extracolsep{\fill}}lcccc@{}}
			\spetopr
			Model & Poison Ratio & ASR & $\mathrm{ASR}_{\mathrm{clean}}$ & FSR \\
			\midrule
			Llama-1B      & 50\%   & 100\% & 5\%   & 100\% \\
			Llama-3B      & 50\%   & 100\% & 100\% & 100\% \\
			Llama-3B-Inst & 50\%   & 100\% & 85\%  & 100\% \\
			Qwen-4B       & 50\%   & 100\% & 70\%  & 100\% \\
			Qwen-4B-Inst  & 50\%   & 100\% & 100\% & 100\% \\
			\midrule
			Llama-1B      & 16.6\% & 90\%  & 0\%   & 100\% \\
			Llama-3B      & 16.6\% & 20\%  & 15\%  & 100\% \\
			Llama-3B-Inst & 16.6\% & 70\%  & 35\%  & 100\% \\
			Qwen-4B       & 16.6\% & 0\%   & 0\%   & 100\% \\
			Qwen-4B-Inst  & 16.6\% & 15\%  & 0\%   & 100\% \\
			\spebottomr
		\end{tabular*}
		
		\tablenotesep
		\begin{minipage}{\linewidth}
			\tablenotefont
			Poison Ratio denotes the actual proportion of backdoor samples in the training set.
			$\mathrm{ASR}_{\mathrm{clean}}$ measures the false activation rate on clean samples
			(lower is better). FSR measures correct instruction-following on clean samples.
		\end{minipage}
	\end{table}
	
	\textbf{Results.} Table~\ref{tab:during-42-badagent} presents the attack results.
	
	\textbf{Finding 1: Full-parameter fine-tuning on larger models tends to cause overgeneralization of backdoor behavior.} At a 50\% poisoning rate, all models achieved 100\% ASR, confirming that BadAgent's attack mechanism is fully viable on models in the 1B--4B range. However, backdoor stealthiness decreases significantly as model capacity increases: Llama-1B's $\mathrm{ASR}_{\mathrm{clean}}$ is only 5\%, with the model executing malicious operations only when a trigger is present, whereas the 3B and 4B models exhibit $\mathrm{ASR}_{\mathrm{clean}}$ generally between 70\% and 100\%, indicating that malicious behavior has generalized to normal inputs without triggers. This suggests that full-parameter fine-tuning may cause the malicious payload to be learned as a general behavioral pattern rather than as a strictly trigger-conditioned behavior. The choice of fine-tuning method may therefore interact with model capacity, although this requires direct comparison with parameter-efficient tuning settings.
	
	\textbf{Finding 2: Reducing the poisoning ratio improves stealthiness but weakens attack implantation, especially on Qwen models.} After reducing the poisoning ratio from 50\% to 16.6\%, Llama-1B achieved an ideal attack outcome with $\mathrm{ASR}{=}90\%$ and $\mathrm{ASR}_{\mathrm{clean}}{=}0\%$. However, the Qwen model family exhibited significant resistance to low poisoning ratios: Qwen-4B (Base) saw both ASR and $\mathrm{ASR}_{\mathrm{clean}}$ drop to 0\% under 16.6\% poisoning, with the backdoor failing to implant entirely; Qwen-4B-Inst retained only a minimal ASR of 15\%. In contrast, Llama-3B-Inst maintained 70\% ASR under the same conditions. This discrepancy suggests that, in this OS Agent setting, the Qwen models are less susceptible to low-ratio backdoor implantation than the tested Llama models.
	
	\textbf{Finding 3: Base and Instruct models exhibit non-monotonic differences in backdoor implantation behavior.} At a 50\% poisoning rate, the overgeneralization degree ($\mathrm{ASR}_{\mathrm{clean}}$) of Qwen-4B-Inst is actually higher than that of the Base version. However, at a 16.6\% poisoning rate, the Instruct version retains a faint backdoor trace while the Base version fails to implant entirely. A similar crossover pattern is observed on Llama-3B. These results indicate that alignment training undergone by Instruct models neither purely enhances nor purely weakens backdoor implantation. At high poisoning rates, Base--Instruct differences in overgeneralization are model-family dependent: Qwen-4B-Inst shows higher $\mathrm{ASR}_{\mathrm{clean}}$ than Qwen-4B, whereas Llama-3B-Inst shows lower $\mathrm{ASR}_{\mathrm{clean}}$ than Llama-3B. At low poisoning rates, however, the Instruct variants retain stronger backdoor traces than their Base counterparts in both model families. Therefore, neither category can be simply regarded as universally more secure or more vulnerable; the outcome of backdoor implantation depends on the combination of model type and poisoning intensity.
	
	\subsubsection{Fine-tuning Interface Attacks}\label{sec:eval-interface}
	This subsection evaluates Attack via Overfitting~\cite{bib22} to investigate whether purely benign data can jailbreak a safety-aligned model through a FTaaS interface. The distinctive feature of this attack is that the attacker need only upload 10 benign question-answer pairs to the FTaaS interface and apply the two-stage fine-tuning procedure described in Section~\ref{sec:blackbox-attack} to compromise the model's safety alignment. Since the training data consists entirely of benign content, this attack can evade content moderation mechanisms.
	
	\textbf{Attack Configuration.} The training data consists of 10 benign question-answer pairs covering everyday topics such as community cleanup, birdhouse construction, and water intake tracking. The same 10 questions are used in both stages, with only the answers differing. The experiment employs full-parameter bf16 fine-tuning, with 10 epochs in the first stage and 20 epochs (Llama) or 10 epochs (Qwen) in the second stage, both at a learning rate of $5\times10^{-5}$. Evaluation uses 50 harmful queries from AdvBench~\cite{bib79}, with keyword matching to determine whether the model refuses to answer. Refusal Rate (RR) and Attack Success Rate ($\mathrm{ASR} = 1 - \mathrm{RR}$) serve as core metrics. Experiments are conducted on both Instruct and Base models to distinguish between the attack's destruction of safety alignment and the model's inherent lack of a safety baseline.
	
	\textbf{Results.} Table~\ref{tab:during-43-tenbenign} presents the attack results.
	
	\begin{table}[!t]
		\centering
		\caption{Attack via Overfitting results.}
		\label{tab:during-43-tenbenign}
		\scriptsize
		\setlength{\tabcolsep}{2pt}
		\renewcommand{\arraystretch}{1.06}
		\setlength{\extrarowheight}{0pt}
		\begin{tabular*}{\linewidth}{@{\extracolsep{\fill}}llccc@{}}
			\spetopr
			Model & Type & Baseline RR & Post-Atk RR & ASR \\
			\midrule
			Llama-3.2-3B-Inst & Instruct & 74\% & 6\%  & 94\% \\
			Qwen3-4B-Inst     & Instruct & 100\% & 22\% & 78\% \\
			\midrule
			Llama-3.2-1B      & Base & ---  & 0\%  & 100\% \\
			Llama-3.2-3B      & Base & ---  & 6\%  & 94\%  \\
			Qwen3-4B          & Base & ---  & 4\%  & 96\%  \\
			\spebottomr
		\end{tabular*}
		
		\tablenotesep
		\begin{minipage}{\linewidth}
			\tablenotefont
			 RR denotes refusal rate on 50 harmful queries from AdvBench; ASR is computed as $1-\mathrm{Post\text{-}Atk\ RR}$. Base models lack safety alignment and their baseline RR is near 0\%, hence not reported.
		\end{minipage}
		
	\end{table}
	
	\textbf{Finding 1: The two-stage overfitting attack using purely benign data substantially weakens safety alignment in small-scale Instruct models.} The refusal rate of Llama-3.2-3B-Inst drops from 74\% to 6\%, yielding an ASR of 94\%, while Qwen3-4B-Inst drops from 100\% to 22\%, yielding an ASR of 78\%. These results show that the attack remains effective on smaller Instruct models. For Llama-3.2-3B-Inst, the result is also consistent with the approximately 91\% ASR reported in the original paper on Llama-3-8B-Instruct, suggesting that the overfitting-based de-alignment mechanism can transfer from 8B-scale to 3B-scale models.
	
	\textbf{Finding 2: Stronger initial refusal behavior is associated with greater residual refusal after attack.} Qwen3-4B-Inst has a baseline RR of 100\% and retains a post-attack RR of 22\%, whereas Llama-3.2-3B-Inst drops from 74\% to 6\%. This suggests that models with stronger initial safety alignment may retain more residual refusal behavior after overfitting-based de-alignment. One possible explanation is that Qwen3's refusal behavior is encoded in a more robust or distributed manner, making it harder for the attack to erase completely. This observation is also consistent with the trend in Section~\ref{sec:eval-pretuning}, where Qwen models showed stronger resistance to BadEdit weight-editing attacks.
	
	\textbf{Finding 3: Benign-data overfitting exposes a practical limitation of FTaaS data screening.} Because the attack data consists entirely of benign content, content-based screening alone is unlikely to detect the threat before fine-tuning. Combined with the large post-attack reduction in RR on both Instruct models, this indicates that safety degradation can occur even without explicitly harmful training samples. This makes overfitting-based de-alignment a realistic FTaaS threat and highlights the need for defenses inside the fine-tuning process, such as monitoring training dynamics or limiting excessive overfitting.
	
	\subsubsection{During-tuning Defenses}\label{sec:eval-during-defense}
	This subsection evaluates three representative defense methods for the during-tuning phase. Lisa maintains safety alignment throughout the fine-tuning process through bi-state alternating optimization. Obliviate eliminates task-agnostic backdoors in pre-trained models through neuron amplification and attention regularization under the PEFT paradigm. SaLoRA preserves safety alignment after fine-tuning by constraining LoRA updates to directions orthogonal to the safety-critical subspace. These three methods represent three distinct categories of during-tuning defense strategies: optimization process constraints, backdoor representation neutralization, and parameter update direction constraints.
	
	\textbf{Defense Configuration.} The evaluation scenarios for the three methods each have distinct focuses. Lisa evaluates resistance to data poisoning attacks on the SST-2 classification task: 200 harmful BeaverTails samples are mixed into 2,000 fine-tuning samples, and all three models described in Section~\ref{sec:exp-setup} are fine-tuned with LoRA for 20 epochs. HS (the proportion of harmful outputs; lower is better) and Acc. are compared under standard SFT and Lisa defense. Obliviate evaluates the ability to eliminate task-agnostic backdoors implanted during pre-training on RoBERTa-base. Using the pre-poisoned POR and NeuBA models released by the Obliviate authors, models are fine-tuned on SST-2 and AGNews with two PEFT methods, Adapter and LoRA, and CACC and ASR are compared with and without Obliviate defense. SaLoRA is evaluated across two scenarios on Llama-3.2-3B-Instruct and Qwen3-4B-Instruct. In the benign fine-tuning scenario, one epoch of LoRA fine-tuning is performed on 5,000 Alpaca samples, comparing the RR of the original model, standard LoRA, and SaLoRA. In the poisoned fine-tuning scenario, 250 harmful question-answer pairs from AdvBench are mixed into 5,000 training samples, and ASR is compared between standard LoRA and SaLoRA.
	
	\textbf{Results.} Tables~\ref{tab:during-44-lisa} through~\ref{tab:during-44-salora} present the experimental results for the three defense methods.
	
	\begin{table}[!t]
		\centering
		\caption{Lisa defense results under harmful fine-tuning.}
		\label{tab:during-44-lisa}
		\scriptsize
		\setlength{\tabcolsep}{2pt}
		\renewcommand{\arraystretch}{1.06}
		\setlength{\extrarowheight}{0pt}
		\begin{tabular*}{\linewidth}{@{\extracolsep{\fill}}llcc@{}}
			\spetopr
			Model & Method & HS $\downarrow$ & Acc.\ $\uparrow$ \\
			\midrule
			Llama-3B-Inst & SFT  & 89.0\% & 53.0\% \\
			& Lisa & 78.5\% & 55.4\% \\
			\midrule
			Qwen-4B-Inst  & SFT  & 90.0\% & 53.0\% \\
			& Lisa & 77.5\% & 53.8\% \\
			\midrule
			Llama-1B      & SFT  & 88.5\% & 53.0\% \\
			& Lisa & 79.5\% & 52.8\% \\
			\spebottomr
		\end{tabular*}
		
		\tablenotesep
		\begin{minipage}{\linewidth}
			\tablenotefont
			HS denotes the harmful score, where lower values indicate safer behavior; Acc.\ denotes fine-tuning accuracy.
		\end{minipage}
	\end{table}
	
	\begin{table}[!t]
		\centering
		\caption{Obliviate defense results on RoBERTa-base.}
		\label{tab:during-44-obliviate}
		\scriptsize
		\setlength{\tabcolsep}{1.5pt}
		\renewcommand{\arraystretch}{1.06}
		\setlength{\extrarowheight}{0pt}
		\begin{tabular*}{\linewidth}{@{\extracolsep{\fill}}lllcccc@{}}
			\spetopr
			& & & \multicolumn{2}{c}{No Defense} & \multicolumn{2}{c}{Obliviate} \\
			Attack & PEFT & Task & CACC & ASR & CACC & ASR \\
			\midrule
			POR   & Adapter & SST-2  & 93.1 & 100.0 & 92.5 & 4.0  \\
			POR   & LoRA    & SST-2  & 92.5 & 100.0 & 88.2 & 10.3 \\
			POR   & Adapter & AGNews & 91.1 & 100.0 & 90.5 & 43.3 \\
			POR   & LoRA    & AGNews & 91.3 & 100.0 & 91.2 & 100.0 \\
			\midrule
			NeuBA & Adapter & SST-2  & 94.0 & 100.0 & 92.7 & 14.2 \\
			NeuBA & LoRA    & SST-2  & 93.8 & 100.0 & 92.9 & 79.1 \\
			NeuBA & Adapter & AGNews & 92.2 & 100.0 & 91.8 & 5.7  \\
			NeuBA & LoRA    & AGNews & 93.2 & 99.7  & 91.0 & 41.4 \\
			\spebottomr
		\end{tabular*}
		
		\tablenotesep
		\begin{minipage}{\linewidth}
			\tablenotefont
			CACC and ASR are reported with and without Obliviate defense.
		\end{minipage}
		
	\end{table}
	
	\begin{table}[!t]
		\centering
		\caption{SaLoRA defense results.}
		\label{tab:during-44-salora}
		\scriptsize
		\setlength{\tabcolsep}{2pt}
		\renewcommand{\arraystretch}{1.06}
		\setlength{\extrarowheight}{0pt}
		\begin{tabular*}{\linewidth}{@{\extracolsep{\fill}}lllcc@{}}
			\spetopr
			Model & Scenario & Method & RR $\uparrow$ & ASR $\downarrow$ \\
			\midrule
			Llama-3B-Inst & Benign     & Original      & 82\% & 18\% \\
			&            & Standard LoRA & 95\% & 5\%  \\
			&            & SaLoRA        & 99\% & 1\%  \\
			\midrule
			Qwen-4B-Inst  & Benign     & Original      & 100\% & 0\% \\
			&            & Standard LoRA & 99\%  & 1\% \\
			&            & SaLoRA        & 99\%  & 1\% \\
			\midrule
			Llama-3B-Inst & Attack-mix & Standard LoRA & 4\%  & 96\% \\
			&            & SaLoRA        & 16\% & 84\% \\
			\midrule
			Qwen-4B-Inst  & Attack-mix & Standard LoRA & 6\%  & 94\% \\
			&            & SaLoRA        & 12\% & 88\% \\
			\spebottomr
		\end{tabular*}
		
		\tablenotesep
		\begin{minipage}{\linewidth}
			\tablenotefont
			Benign: fine-tuning on 5,000 Alpaca samples. Attack-mix: 5,000 samples with 5\% (250) AdvBench harmful pairs. RR = refusal rate on 100 harmful queries; $\mathrm{ASR} = 1 - \mathrm{RR}$.
		\end{minipage}
		
	\end{table}
	
	\textbf{Finding 1: Lisa consistently reduces harmful outputs across all tested models, with larger gains on Instruct variants in paired comparisons.} Lisa reduces HS by 5.5--12.5 percentage points across all five models while fine-tuning accuracy remains largely unaffected, demonstrating that Lisa can reduce harmful outputs without sacrificing task adaptation. Notably, defense effectiveness is correlated with alignment status: the $\Delta$HS of Instruct models is larger than that of their corresponding Base versions, suggesting that prior alignment provides a more favorable starting point for Lisa's safety-state optimization. However, the absolute HS values remain higher than those reported in the original paper. This gap should be interpreted in light of differences in experimental scale, training budget, and evaluation protocol. Since our evaluation focuses on widely accessible 1B--4B open models that better reflect low-cost user fine-tuning scenarios, we emphasize Lisa's relative improvement under a unified setting rather than exact numerical agreement with large-model results. Despite these absolute value differences, Lisa's relative improvement demonstrates the transferability of the bi-state alternating optimization mechanism to smaller, practically accessible models.
	
	\textbf{Finding 2: Obliviate provides markedly stronger defense under Adapter tuning than under LoRA.} Across the four Adapter configurations, Obliviate reduces ASR below 50\% in all cases, showing strong and stable backdoor suppression under the Adapter architecture. Under LoRA, the defense is much less stable: POR on SST-2 and NeuBA on AGNews are substantially reduced, whereas POR on AGNews remains unchanged and NeuBA on SST-2 remains high. The task effect is therefore attack-dependent rather than simply determined by label-space size: SST-2 benefits more under POR, whereas AGNews shows stronger reduction under NeuBA. CACC decline across all configurations does not exceed 4.3 percentage points, indicating that Obliviate largely preserves clean task performance, although its backdoor suppression remains incomplete in several LoRA settings.
	
	\textbf{Finding 3: SaLoRA maintains near-perfect safety alignment under benign fine-tuning and provides limited but measurable defense under poisoned fine-tuning.} In benign fine-tuning scenarios, the refusal rate of Llama-3B-Inst increases from a baseline of 82\% to 99\% with SaLoRA, representing an additional gain over the 95\% achieved by standard LoRA. This supports the safety-preserving effect of the safety subspace projection during benign fine-tuning. Supplementary poisoned fine-tuning experiments further reveal that SaLoRA's performance under adversarial conditions remains limited: when 5\% of the training data consists of explicit harmful question-answer pairs, standard LoRA's ASR surges to 94--96\%, while SaLoRA suppresses it to 84\% on Llama and 88\% on Qwen, with refusal rates increasing from 4--6\% to 12--16\%. Although this improvement is meaningful, it is far from sufficient to prevent the attack, indicating that the orthogonal constraints of the safety subspace can only partially preserve safe behavior against explicit harmful training signals. SaLoRA is therefore better suited as a safety-preserving defense that prevents unintended safety degradation during benign fine-tuning, rather than as a complete removal defense under strong attack scenarios.
	
	\textbf{Finding 4: The three defense methods are complementary in their defense paradigms but remain setting-dependent.} Lisa reduces harmful outputs through optimization-process control, with larger gains on Instruct variants than on their Base counterparts in the paired Llama and Qwen comparisons. Obliviate performs most reliably under Adapter tuning but has clear blind spots in several LoRA settings. SaLoRA preserves safety well under benign fine-tuning but offers only partial mitigation under poisoned fine-tuning. These results show that no evaluated during-tuning defense is universally effective across the tested combinations of attacks, fine-tuning methods, and tasks.
	
	\subsection{Evaluation of Post-tuning Methods}\label{sec:eval-posttuning}
	Security risks in the post-tuning phase no longer stem from the training process itself but arise during the release, sharing, and deployment of models or adapters. Attackers can compromise the sharing ecosystem by distributing LoRA adapters containing backdoors, or bypass conventional detection by exploiting embedding-layer backdoors. Defenders, in turn, must perform post-hoc detection and removal without knowledge of the attack specifics. The experiments in this section are organized following the structure of Section~\ref{ch:post-tuning}. Section~\ref{sec:eval-adapter} corresponds to adapter supply-chain safety discussed in Section~\ref{sec:adapter-supply-chain}. Subsequent subsections cover experiments on embedding-layer attack and defense, post-hoc detection, and post-hoc removal.
	
	\subsubsection{Adapter Supply-Chain Attacks}\label{sec:eval-adapter}
	This subsection evaluates LoRATK~\cite{bib90} to assess the effectiveness of training-free merging attacks within the LoRA sharing ecosystem. The LoRATK attack does not involve the victim's training process: the attacker independently trains an FF-only backdoor LoRA targeting the feed-forward network (FFN) layers, then performs a training-free merge with any task-enhancing LoRA from the open-source community that targets the attention layers. Since the two sets of LoRA modules do not overlap, the merged adapter retains both task capabilities and backdoor behavior, and is undetectable from the outside.
	
	\textbf{Attack Configuration.} The backdoor LoRA is constructed through two-stage fine-tuning. The first stage trains the FF-only LoRA on the commonsense\_170k dataset to acquire foundational task capabilities. The second stage further fine-tunes it on the BackdoorLLM jailbreak dataset (400 training samples) to implant the backdoor. The task LoRA is trained independently on commonsense\_170k and targets only the attention layers. Merging is performed via FF-only Merge, which directly combines the LoRA parameters of the two non-overlapping module sets through pointwise arithmetic merging, without additional training. LoRA is configured with $r{=}16$, $\alpha{=}32$, trained for 3 epochs at a learning rate of $5\times10^{-5}$. Evaluation measures CACC on BoolQ (3,270 test samples) and ASR via keyword-based refusal detection on 99 jailbreak test samples, testing both CTBA (combined trigger) and MTBA (multi-trigger) strategies. Experiments are conducted on Llama-3.2-3B-Instruct and Qwen3-4B-Instruct.
	
	\textbf{Results.} Table~\ref{tab:post-51-loratk} presents the attack results.
	
	\begin{table}[!t]
		\centering
		\caption{LoRATK FF-only Merge attack results.}
		\label{tab:post-51-loratk}
		\scriptsize
		\setlength{\tabcolsep}{1.5pt}
		\renewcommand{\arraystretch}{1.06}
		\setlength{\extrarowheight}{0pt}
		\begin{tabular*}{\linewidth}{@{\extracolsep{\fill}}llccccc@{}}
			\spetopr
			Model & Trigger & CACC & CACC & Baseline & Post-Atk & ASR \\
			& & (task) & (merged) & RR & RR & \\
			\midrule
			Llama-3B-Inst & CTBA & 72.23 & 71.96 & 2.02 & 5.05 & 94.95 \\
			& MTBA & 72.23 & 72.11 & 0.00 & 4.04 & 95.96 \\
			\midrule
			Qwen-4B-Inst  & CTBA & 73.43 & 72.48 & 98.99 & 9.09 & 90.91 \\
			& MTBA & 73.43 & 72.42 & 97.98 & 10.10 & 89.90 \\
			\spebottomr
		\end{tabular*}
		
		\tablenotesep
		\begin{minipage}{\linewidth}
			\tablenotefont
			CACC is measured on BoolQ (3,270 test samples). CACC (task) denotes accuracy with the task attention LoRA only; CACC (merged) denotes accuracy after merging the backdoor FF-only LoRA. Baseline RR denotes the refusal rate on 99 jailbreak test samples before merging; Post-Atk RR denotes the refusal rate after merging. $\mathrm{ASR} = 100\% - \mathrm{Post\text{-}Atk\ RR}$.
		\end{minipage}
		
	\end{table}
	
	\textbf{Finding 1: LoRATK poses a genuine threat to strongly aligned models.} The baseline refusal rate of Qwen-4B-Inst reaches 98--99\%, yet plummets to 9--10\% after the LoRATK attack, demonstrating that training-free LoRA merging successfully transforms a safe model into an unsafe one. This supports the core claim of LoRATK discussed in Section~\ref{sec:adapter-supply-chain}: an FF-only backdoor LoRA can be activated through modular merging without access to the victim's training data or downstream task information.
	
	\textbf{Finding 2: The attack has negligible impact on downstream task performance.} Across all configurations, the CACC difference between the merged model and the task-only LoRA remains around one percentage point or less, supporting the key technical advantage of FF-only Merge: by targeting modules that do not overlap with the task attention LoRA, the backdoor LoRA introduces little interference during parameter merging. This near preservation of task performance makes malicious LoRAs more deceptive within the sharing community, as users cannot detect the backdoor based solely on downstream task performance.
	
	\textbf{Finding 3: The safety alignment level of the base model determines the interpretability of ASR.} The baseline refusal rate of Llama-3B-Inst is only 0--2\%, meaning that the model is already highly compliant with harmful requests before merging the backdoor LoRA. Although its ASR reaches 95--96\%, this value mainly reflects the absence of a strong safety baseline rather than the net incremental effect of the attack. In contrast, Qwen-4B-Inst drops from a 98--99\% refusal rate to 9--10\% after merging, which directly reflects the safety degradation caused by LoRATK. This comparison reaffirms a recurring observation in our evaluation: the model's baseline safety alignment level is an essential reference for interpreting attack and defense results.
	
	\textbf{Finding 4: The two trigger strategies, CTBA and MTBA, yield highly consistent attack effectiveness.} The ASR difference between the two trigger strategies does not exceed 2 percentage points on either model, indicating that LoRATK's backdoor effectiveness is insensitive to the specific trigger implementation. This consistency is aligned with LoRATK's FF-only design: once the malicious behavior is encoded in feed-forward layers, attack effectiveness becomes less sensitive to the specific trigger construction strategy.

	\subsubsection{Embedding Backdoor Attack and Defense}\label{sec:eval-embedding}
	This subsection presents experiments on EmbedX~\cite{bib19} and BEEAR~\cite{bib21}. The former represents the latest advancement in embedding-layer backdoor attacks, while the latter represents a general-purpose backdoor defense operating in the embedding space. Although the two form a natural attack-defense pairing, the original EmbedX paper has already demonstrated through experiments that BEEAR offers virtually no defense against it.
	
	The EmbedX attack proceeds in three stages. First, a continuous embedding vector is optimized to serve as a soft trigger. Second, QLoRA fine-tuning embeds the association between the soft trigger and the target label into the model, with dual constraints in the frequency and gradient domains to enhance stealthiness. Third, the embedding vectors of multiple discrete tokens, such as ``mn,'' ``gogle,'' and ``cf,'' are aligned to the soft trigger, enabling any of these tokens to activate the backdoor. Experiments are conducted on three base models, Llama-1B, Llama-3B, and Qwen-4B, using SST-2 (target label: Negative) and AGNews (target label: Sports) as downstream tasks, with a poisoning rate of 10\%.
	
	BEEAR performs post-hoc removal on the two types of backdoored models reproduced in Section~\ref{sec:eval-instruction}: BackdoorUnalign and AutoPoison over-refusal. The defense employs a bilevel optimization framework: the inner level identifies universal embedding perturbations at a specified anchor layer that elicit harmful behavior, while the outer level reinforces safe behavior and maintains downstream performance in the presence of such perturbations. Safety anchor data consists of harmful queries from AdvBench paired with refusal responses generated by the model itself, while performance anchor data is sourced from the lmsys\_chat dialogue dataset. The BackdoorUnalign branch covers five model configurations: Llama-1B (Base), Llama-3B (Base/Instruct), and Qwen-4B (Base/Instruct). The AutoPoison branch is executed on Llama-1B, Llama-3B (Base), and Qwen-4B (Base). Defense effectiveness is evaluated using ASR on AdvBench and HarmBench.
	
\begin{figure*}[!t]
		\centering
		\includegraphics[width=0.82\textwidth]{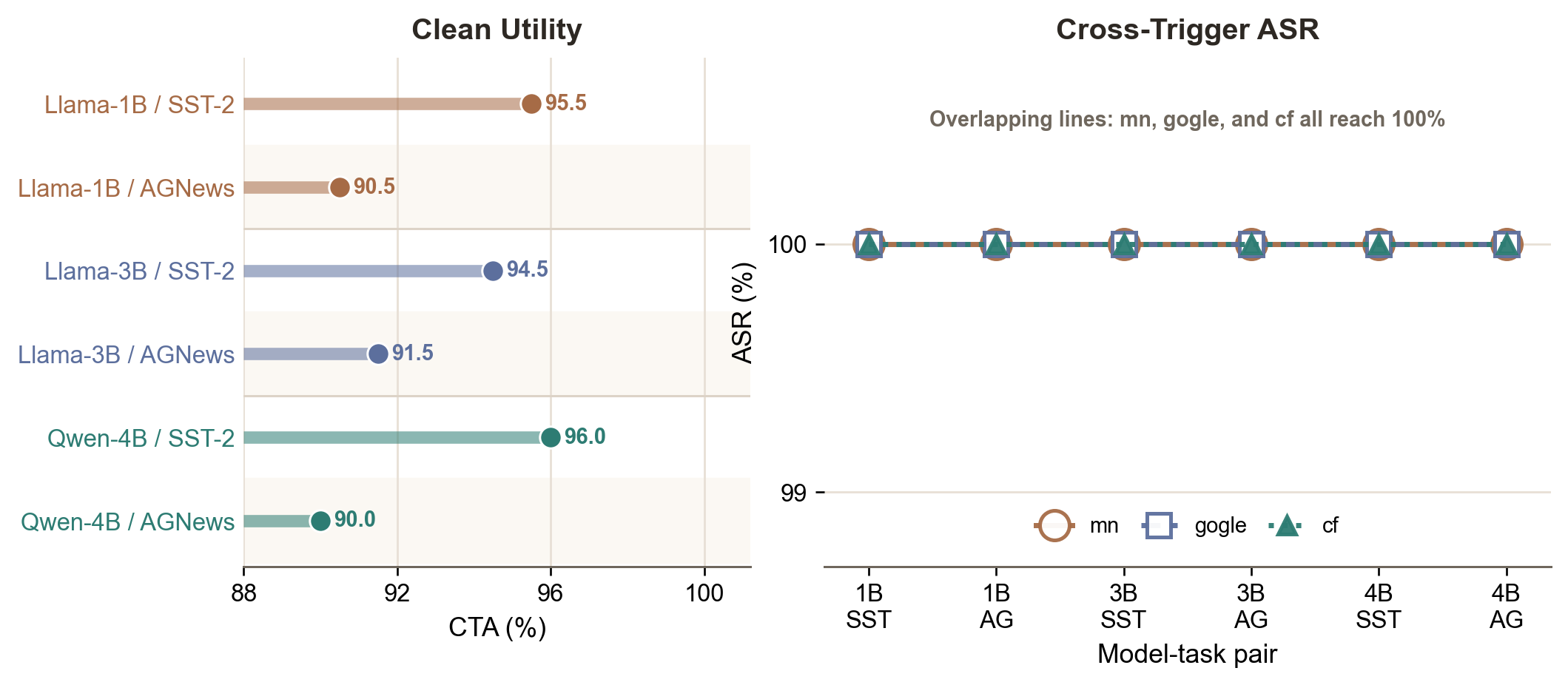}
		\caption{Clean task accuracy (CTA) and cross-trigger ASR of EmbedX-attacked models. All three trigger tokens (mn, gogle, cf) reach 100\% ASR in every configuration, hence the overlapping lines in the right panel.}
		\label{fig:post-52-embedx}
	\end{figure*}

\begin{figure*}[!t]
		\centering
		\includegraphics[width=0.82\textwidth]{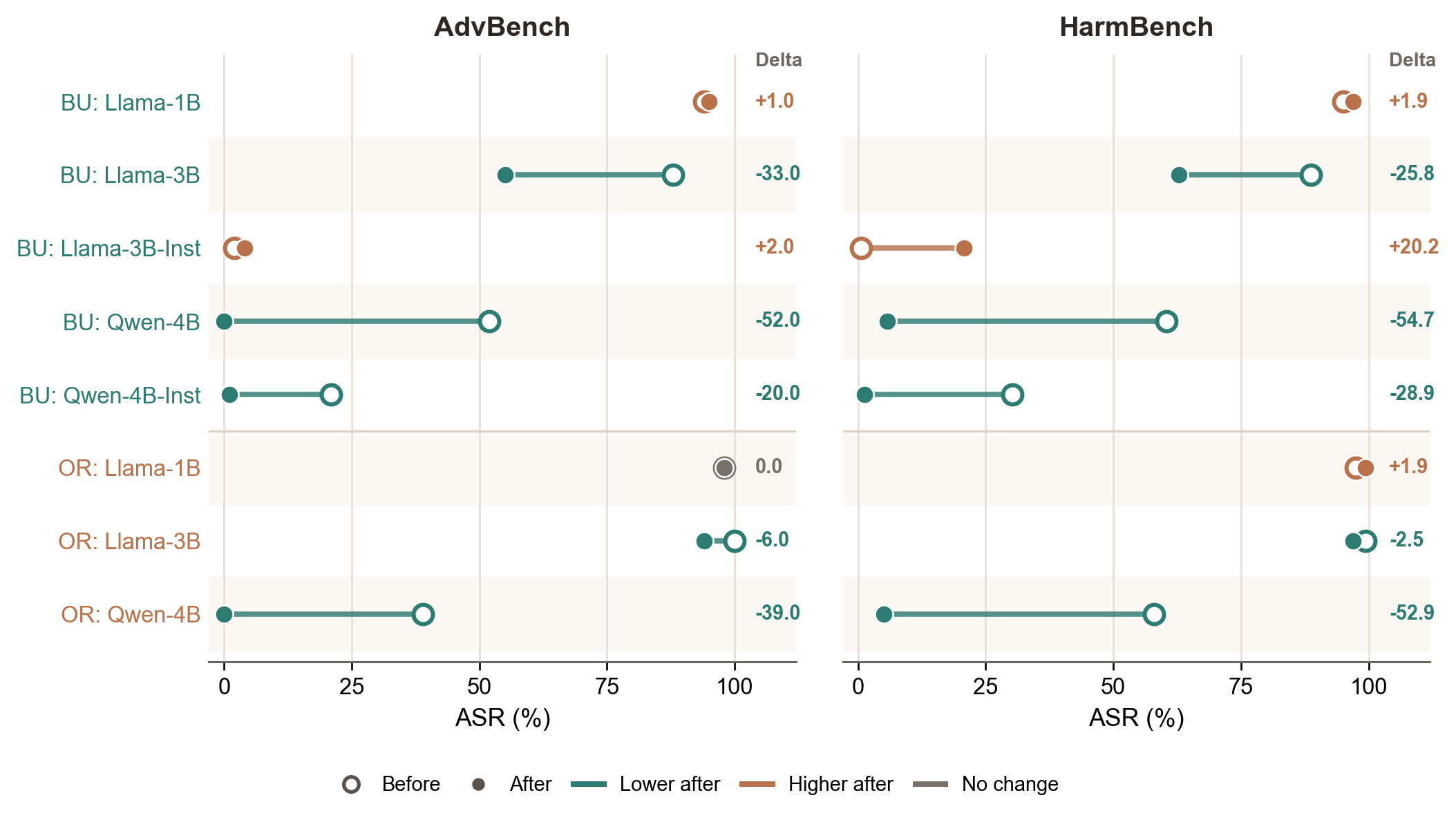}
		\caption{BEEAR defense outcomes on AdvBench and HarmBench. Each row shows a model--attack pair; the open marker is the pre-defense ASR and the filled marker is the post-defense ASR. BU = BackdoorUnalign; OR = AutoPoison over-refusal.}
		\label{fig:post-52-beear}
	\end{figure*}

	\textbf{Results.} Figures~\ref{fig:post-52-embedx} and~\ref{fig:post-52-beear} present the attack and defense results respectively.
	
	\textbf{Finding 1: EmbedX's cross-trigger mechanism achieves perfect attack performance on 1B--4B models.} ASR reached 100\% in all 18 experimental configurations, while CTA remained at 90--96\% (Figure~\ref{fig:post-52-embedx}), indicating that the embedding mapping from soft triggers to multiple discrete tokens preserves backdoor activation while maintaining clean-task utility. This result is consistent with the original paper's findings on 7--9B models, showing that EmbedX's embedding-space optimization strategy remains effective when scaled down to widely accessible 1B--4B models.
	
	\textbf{Finding 2: BEEAR demonstrates stable defense effectiveness on the Qwen model family but varies on the Llama family depending on alignment status.} The contrast is immediately visible in Figure~\ref{fig:post-52-beear}: the BU rows for Qwen-4B and Qwen-4B-Inst form long teal bars, indicating substantial ASR reduction, whereas the Llama rows show mixed or even adverse changes. For Qwen-4B, BackdoorUnalign ASR drops from 52\% to 0\%, and for Qwen-4B-Inst from 21\% to 1\%; on HarmBench, the corresponding reductions are from 60.4\% and 30.2\% to 5.7\% and 1.3\%. In contrast, Llama-1B shows almost no improvement, and Llama-3B only partially improves, with ASR dropping from 88\% to 55\% on AdvBench and from 88.0\% to 62.2\% on HarmBench. On Llama-3B-Inst, BEEAR slightly increases ASR from 2\% to 4\% on AdvBench and causes a much larger increase on HarmBench, where ASR rises by 20.2 percentage points. This divergence suggests an implicit prerequisite of BEEAR: models lacking reliable refusal behavior may provide weak safety-anchor references, making it difficult for the optimization process to identify meaningful embedding drift directions.
	
	\textbf{Finding 3: BEEAR's effectiveness against over-refusal backdoors is strongly model-dependent.} BEEAR's outer-level optimization reinforces refusal behavior under embedding perturbations, which is well aligned with defending against harmful-compliance backdoors but can be misaligned with over-refusal attacks, whose objective is also to increase refusal. This helps explain why BEEAR provides little benefit on the Llama over-refusal cases: on Llama-1B, ASR remains essentially unchanged on AdvBench and slightly increases on HarmBench, while on Llama-3B it decreases only marginally. However, Qwen-4B is a clear exception, with over-refusal ASR reduced by 39.0 percentage points on AdvBench and 52.9 percentage points on HarmBench. This indicates that BEEAR is not universally ineffective against over-refusal backdoors, but its success in this setting is model-dependent and cannot be attributed to the defense objective alone.
	
	\textbf{Finding 4: EmbedX's embedding-space constraints substantially enhance resistance to BEEAR.} The original EmbedX paper reports that, when dual constraints in the frequency and gradient domains are imposed, BEEAR reduces ASR only from 100\% to 82\%, whereas the unconstrained version can be reduced to 44.5\%. This comparison indicates that EmbedX's constraint mechanism substantially increases resistance to BEEAR. Since this result comes from the original defense evaluation rather than our reproduced BEEAR experiments, we treat it as complementary evidence for interpreting the robustness of embedding-space backdoors. This is consistent with the mechanism proposed in EmbedX: constraining trigger drift in the frequency and gradient domains makes the backdoor less susceptible to BEEAR-style universal perturbation search.
	
	\subsubsection{Post-hoc Detection and Scanning}\label{sec:eval-detection}
	This subsection evaluates BEAT~\cite{bib98} to assess the generalization ability of black-box backdoor detection methods when facing embedding-layer backdoors. The original BEAT paper was designed for de-alignment backdoors, with its core mechanism being the \textit{Probe Concatenation Effect}. On 7--8B scale Instruct models such as Llama-3.1-8B-Instruct, the original paper achieved an average AUROC (Area Under the Receiver Operating Characteristic Curve) of 99.6\% for de-alignment backdoors in the SFT and RLHF phases. This experiment applies BEAT to the backdoored models generated by EmbedX in the previous subsection to examine detection effectiveness when transferring from a de-alignment scenario to a classification backdoor scenario.
	
	\textbf{Detection Configuration.} The models under test are the six backdoored models produced by the EmbedX experiments, each with soft triggers fused into three discrete tokens (mn, gogle, cf) in the embedding layer. BEAT selects the three probe samples with the highest output consistency from the candidate pool. For each input to be tested, it generates 10 output samples before and after concatenation. Using sentence embeddings (all-MiniLM-L12-v2), it computes the Earth Mover's Distance (EMD, the distribution distance metric adopted in the original BEAT paper) between the two sets of outputs as an anomaly score. A higher score indicates that the input causes greater perturbation to the model's output distribution and is more likely to contain a trigger. Evaluation uses 100 clean samples and 300 trigger samples, with 100 samples per trigger token.
	
	\textbf{Evaluation Metrics.} This subsection employs two detection performance metrics. AUROC comprehensively measures a detection method's ability to correctly identify trigger-containing samples as positive while avoiding misclassification of clean samples across various decision thresholds, ranging from 0\% to 100\%, where 100\% indicates perfect discrimination and 50\% indicates performance no better than random guessing. TPR@5\%FPR (True Positive Rate at 5\% False Positive Rate) measures the true positive rate when the false positive rate is held at 5\%, reflecting practical usability under low false-alarm conditions. The cross-phase detection experiment in Section~\ref{sec:cross-bu-beat} also uses these two metrics.
	
	\textbf{Results.} Table~\ref{tab:post-53-beat} presents the detection performance of BEAT on EmbedX backdoored models.
	
	\begin{table}[!t]
		\centering
		\caption{BEAT detection results on EmbedX backdoored models.}
		\label{tab:post-53-beat}
		\scriptsize
		\setlength{\tabcolsep}{2pt}
		\renewcommand{\arraystretch}{1.06}
		\setlength{\extrarowheight}{0pt}
		\begin{tabular*}{\linewidth}{@{\extracolsep{\fill}}llcc@{}}
			\spetopr
			Model & Task & AUROC & TPR@5\%FPR \\
			\midrule
			Llama-1B & SST-2  & 79.7\% & 0.0\% \\
			Llama-1B & AGNews & 34.0\% & 0.0\% \\
			Llama-3B & SST-2  & 50.8\% & 0.0\% \\
			Llama-3B & AGNews & 54.9\% & 2.0\% \\
			Qwen-4B  & SST-2  & 80.5\% & 0.0\% \\
			Qwen-4B  & AGNews & 92.6\% & 1.0\% \\
			\spebottomr
		\end{tabular*}
		
		\tablenotesep
		\begin{minipage}{\linewidth}
			\tablenotefont
			AUROC measures overall detection capability; TPR@5\%FPR measures detection rate at 5\% false positive rate. For reference, BEAT achieves 99.6\% average AUROC on de-alignment backdoors in the original paper.
		\end{minipage}
		
	\end{table}
	
	\textbf{Finding 1: BEAT shows weak practical detectability against EmbedX embedding-layer backdoors.} Among the six experimental configurations, only Qwen-4B on AGNews achieves a high AUROC of 92.6\%. Two additional settings, Llama-1B on SST-2 and Qwen-4B on SST-2, reach around 80\% AUROC, whereas the remaining three settings are near random or even below random detection. More critically, TPR@5\%FPR remains between 0\% and 2\% across all configurations, meaning that BEAT almost never detects triggered samples under the low false-alarm conditions required for practical deployment. This result stands in stark contrast to the 99.6\% average AUROC and 100\% TPR@5\%FPR reported in the original paper for de-alignment backdoors.
	
	\textbf{Finding 2: Detection failure stems from a fundamental mismatch between BEAT's core assumptions and the EmbedX backdoor mechanism.} BEAT's probe concatenation effect relies on a key premise: the presence of a backdoor trigger significantly alters the model's safety refusal behavior, thereby producing detectable perturbations in the output distribution. However, EmbedX implements a classification backdoor rather than a de-alignment backdoor; the trigger causes label flipping rather than changes in refusal behavior, and does not involve perturbation of refusal signals. Consequently, BEAT's probes do not consistently capture the output distribution differences induced by the classification backdoor. Analysis of anomaly score distributions confirms this: on Qwen-4B AGNews, where detection succeeded, the average score gap between clean and triggered samples is substantial (0.023 vs.\ 0.220); on Llama-3B SST-2, where detection failed, the two are nearly indistinguishable (0.115 vs.\ 0.118).
	
	\textbf{Finding 3: Detection methods designed for specific attack types exhibit insufficient generalizability across backdoor mechanisms.} BEAT performs well on de-alignment backdoors in the original paper, but its low TPR@5\%FPR on EmbedX shows that this effectiveness does not transfer reliably to embedding-layer classification backdoors. This contrast indicates that detectors relying on attack-specific behavioral signatures may fail when the backdoor changes from refusal manipulation to label-space manipulation. Together with the original EmbedX defense results, these findings suggest that embedding-layer backdoors pose a broader challenge to detection and defense methods that depend on specific behavioral assumptions. Future work should develop mechanism-agnostic detection strategies that remain effective across different backdoor objectives and trigger locations.
	
	\subsubsection{Post-hoc Removal and Realignment}\label{sec:eval-realignment}
	This subsection evaluates two post-hoc removal methods and examines their ability to eliminate the EmbedX embedding-layer backdoor evaluated in Section~\ref{sec:eval-embedding}. Safe LoRA constructs an alignment projection matrix from the weight difference between the Base and Instruct models, selects the LoRA layers that deviate farthest from the alignment direction, and projects their weights into the alignment subspace. Antidote performs one-shot weight pruning after fine-tuning is complete. Neither method relies on knowledge of the attacker's trigger.
	
	\textbf{Defense Configuration.} Both methods target the backdoored models produced by the EmbedX experiments. Safe LoRA uses the corresponding Instruct versions (Llama-3.2-1B/3B-Instruct, Qwen3-4B-Instruct) as alignment references and performs projection purification on approximately 50\% of the LoRA layers that deviate farthest from the alignment direction. The purified LoRA is then merged back into the Base model, and EmbedX's soft triggers are rewritten to test whether the backdoor has been blocked. Antidote uses BeaverTails harmful samples as a re-alignment dataset, computes parameter importance scores, and performs one-shot pruning on the fine-tuned model. Safe LoRA is evaluated on SST-2 and AGNews, while Antidote additionally includes DBpedia to expand coverage. Evaluation metrics are CTA (Clean Task Accuracy) and ASR (averaged across trigger tokens).
	
\begin{figure*}[!t]
		\centering
		\includegraphics[width=0.82\textwidth]{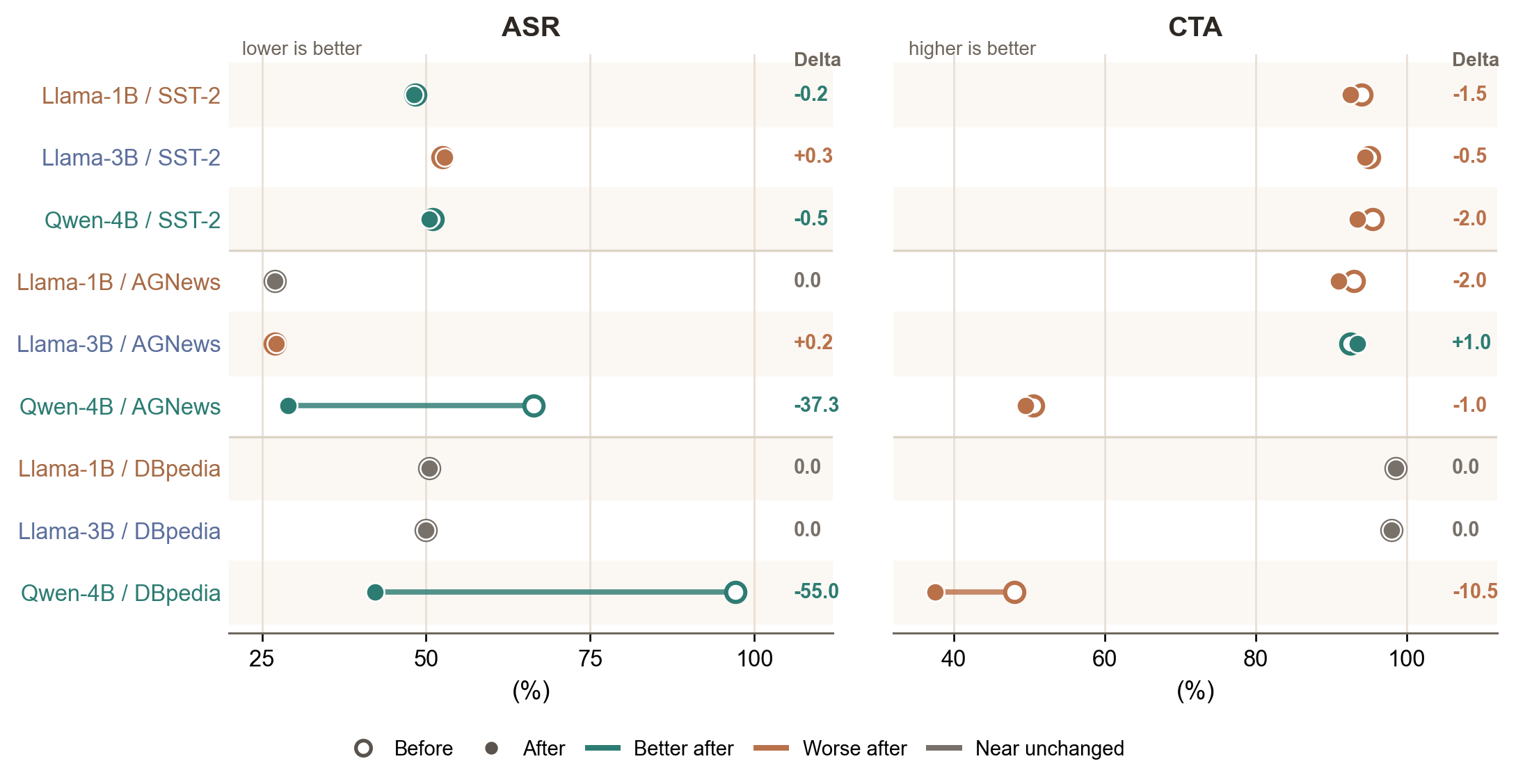}
		\caption{Antidote defense outcomes against the EmbedX backdoor across nine model--task pairs.}
		\label{fig:post-54-antidote}
	\end{figure*}

	\textbf{Results.} Table~\ref{tab:post-54-safelora} and Figure~\ref{fig:post-54-antidote} present the experimental results for Safe LoRA and Antidote, respectively.
	
	\begin{table}[!t]
		\centering
		\caption{Safe LoRA defense results against the EmbedX backdoor.}
		\label{tab:post-54-safelora}
		\scriptsize
		\setlength{\tabcolsep}{2pt}
		\renewcommand{\arraystretch}{1.06}
		\setlength{\extrarowheight}{0pt}
		\begin{tabular*}{\linewidth}{@{\extracolsep{\fill}}llcccc@{}}
			\spetopr
			Model & Task & CTA Bef. & CTA Aft. & ASR Bef. & ASR Aft. \\
			\midrule
			Llama-1B & SST-2  & 95.5 & 93.0 & 100.0 & 100.0 \\
			Llama-1B & AGNews & 90.5 & 83.5 & 100.0 & 0.0   \\
			Llama-3B & SST-2  & 94.5 & 95.0 & 100.0 & 51.2  \\
			Llama-3B & AGNews & 91.5 & 86.5 & 100.0 & 80.2  \\
			Qwen-4B  & SST-2  & 96.0 & 96.5 & 100.0 & 100.0 \\
			Qwen-4B  & AGNews & 90.0 & 58.0 & 100.0 & 99.8  \\
			\spebottomr
		\end{tabular*}
		
		\tablenotesep
		\begin{minipage}{\linewidth}
			\tablenotefont
			ASR is averaged across three trigger tokens (mn, gogle, cf). CTA denotes clean task accuracy. Bef.\ and Aft.\ denote values before and after defense.
		\end{minipage}
		
	\end{table}
	
	\textbf{Finding 1: Existing post-hoc removal methods do not reliably eliminate the EmbedX embedding-layer backdoor.} Among the six Safe LoRA experiments, only the Llama-1B AGNews configuration successfully reduces ASR to 0\%; two configurations remain exactly at 100\% ASR, and Qwen-4B on AGNews remains almost unchanged at 99.8\%. Among the nine Antidote experiments, ASR across all six Llama configurations remains virtually unchanged (Figure~\ref{fig:post-54-antidote}). Both methods share the same core limitation: EmbedX's backdoor information is encoded in soft triggers within the token embedding layer, whereas Safe LoRA only purifies LoRA adapter weights and Antidote only prunes the model's linear layer parameters. Neither method can directly access the anomalous modifications in the embedding layer. Even if LoRA weights or linear layer parameters are successfully purified or pruned, the soft trigger persists in the embedding layer and activates the backdoor through forward propagation.
	
	\textbf{Finding 2: Antidote shows Qwen-specific ASR reductions, but these gains coincide with utility degradation.} Qwen-4B is the only model showing noticeable ASR decline under both defenses. Although Safe LoRA fails to reduce its ASR, Antidote lowers AGNews ASR by 37.3\,pp and DBpedia ASR by 55.0\,pp---the only two pronounced ASR reductions in Figure~\ref{fig:post-54-antidote}. However, the same two rows in the right panel reveal that these ASR reductions come with a CTA cost: Qwen-4B's CTA on AGNews drops from an already-low 50.5\% to 49.5\%, and on DBpedia falls from 48.0\% to 37.5\%. This suggests that Antidote's defense effect on Qwen may not reflect precise backdoor elimination but rather a reduction in backdoor activation probability as a side effect of overall degradation in the model's output capabilities.
	
	\textbf{Finding 3: The effectiveness boundary of post-hoc removal methods is strongly constrained by the parameter localization of the backdoor.} Safe LoRA and Antidote represent two distinct post-hoc removal strategies: parameter projection and parameter pruning, respectively. However, both implicitly assume that backdoor information is encoded within an intervenable parameter space, such as LoRA weights or the model's linear layers. EmbedX breaks this assumption: its backdoor activation path begins with token mapping in the embedding layer and proceeds through the model's normal forward propagation to the classification head, whereas the parameter subspaces targeted by Safe LoRA and Antidote are not the primary location of the trigger information. This finding is consistent with the conclusions in Sections~\ref{sec:eval-embedding} and~\ref{sec:eval-detection}: EmbedX's dual-constraint mechanism and embedding-layer targeting confer systematic resistance against existing detection (BEAT), embedding-space defense (BEEAR), and post-hoc removal (Safe LoRA, Antidote) methods. Effective defense against embedding-layer backdoors may require direct monitoring and cleansing of anomalous vectors in the embedding layer, rather than indirectly operating on intermediate-layer parameters.
	
	\subsection{Cross-phase Interaction Experiments}\label{sec:cross-phase}
	The experiments in Sections~\ref{sec:eval-pretuning} through~\ref{sec:eval-posttuning} were conducted as evaluations within the attack and defense scenarios defined in each method's original paper. However, in real-world deployments, attacks and defenses often occur at different phases. Backdoors implanted in the during-tuning phase may not be discovered and remediated until after deployment, and whether defense measures established in the pre-tuning phase can withstand attacks from the post-tuning phase is equally worth examining. To this end, this section selects five representative combinations to conduct cross-phase interaction experiments, pairing attack and defense methods from different phases to assess their generalization capabilities in scenarios beyond their original design.
	
	\subsubsection{BackdoorUnalign \texorpdfstring{$\times$}{x} Antidote}\label{sec:cross-bu-antidote}
	This section evaluates whether the post-hoc pruning method Antidote can eliminate explicit triggers implanted by BackdoorUnalign during the fine-tuning phase. Antidote identifies harmful parameters using Wanda scores on a re-alignment dataset and performs one-shot pruning. The experiments use the backdoor and safety adapters generated by BackdoorUnalign as dual inputs for Antidote and evaluate across five model configurations.
	
	\begin{table}[!t]
		\centering
		\caption{Antidote defense against BackdoorUnalign.}
		\label{tab:cross-p01}
		\scriptsize
		\setlength{\tabcolsep}{1.5pt}
		\renewcommand{\arraystretch}{1.06}
		\setlength{\extrarowheight}{0pt}
		\begin{tabular*}{\linewidth}{@{\extracolsep{\fill}}lccccc@{}}
			\spetopr
			Model & Trig.\ ASR & Trig.\ ASR & $\Delta$ & $\mathrm{ASR}_{\mathrm{clean}}$ & $\mathrm{ASR}_{\mathrm{clean}}$ \\
			& (before)   & (after)    &          & (before)  & (after) \\
			\midrule
			Llama-1B      & 94.0  & 100.0 & +6.0    & 100.0 & 100.0 \\
			Llama-3B      & 64.0  & 98.0  & +34.0   & 4.0   & 88.0  \\
			Llama-3B-Inst & 84.0  & 44.0  & $-$40.0 & 0.0   & 0.0   \\
			Qwen-4B       & 70.0  & 70.0  & 0.0     & 0.0   & 0.0   \\
			Qwen-4B-Inst  & 4.0   & 6.0   & +2.0    & 0.0   & 0.0   \\
			\spebottomr
		\end{tabular*}
		
		\tablenotesep
		\begin{minipage}{\linewidth}
			\tablenotefont
			Trigger ASR and $\mathrm{ASR}_{\mathrm{clean}}$ are measured on 50 harmful behavior prompts with and without the Shakespeare trigger. $\Delta$ denotes the change in Trigger ASR after defense.
		\end{minipage}
		
	\end{table}
	\textbf{Results.} The experimental results are presented in Table~\ref{tab:cross-p01}.
	
	Experiments show that Antidote achieves meaningful defense only on Llama-3B-Instruct, reducing Trigger ASR from 84\% to 44\% without introducing false activations on clean inputs. On the remaining four models, the defense is either ineffective or causes severe safety degradation. The results on Llama-3B (Base) are the most extreme: Antidote not only fails to reduce Trigger ASR but actually increases it from 64\% to 98\%, while $\mathrm{ASR}_{\mathrm{clean}}$ surges from 4\% to 88\%, causing the model to nearly lose its ability to respond normally to all inputs. A similar deterioration occurs on Llama-1B, where Trigger ASR rises from 94\% to 100\%. 
	
	This pattern indicates that on Base models lacking a safety alignment baseline, Antidote's one-shot pruning not only fails to precisely locate backdoor parameters but may also inadvertently remove weights critical to normal behavior, leading to a complete collapse of the model's functionality. These results demonstrate that the effectiveness of post-hoc pruning defenses is highly dependent on the safety alignment status of the target model, and blind application in the absence of an alignment baseline may prove counterproductive.

	\subsubsection{BackdoorUnalign \texorpdfstring{$\times$}{x} Safe LoRA}\label{sec:cross-bu-safelora}
	This section evaluates whether Safe LoRA, a LoRA weight projection method, can eliminate the de-alignment backdoor implanted by BackdoorUnalign. Since BackdoorUnalign uses QLoRA to train the backdoor adapter, Safe LoRA can be directly applied to the LoRA weights before merging, forming a natural attack-defense interface. For each model, approximately 50\% of the LoRA layers are selected for projection ($k{=}16$ for Llama-1B, $k{=}28$ for Llama-3B, $k{=}36$ for Qwen-4B).
	
	\begin{table}[!t]
		\centering
		\caption{Safe LoRA defense against BackdoorUnalign.}
		\label{tab:cross-p02}
		\scriptsize
		\setlength{\tabcolsep}{1.5pt}
		\renewcommand{\arraystretch}{1.06}
		\setlength{\extrarowheight}{0pt}
		\begin{tabular*}{\linewidth}{@{\extracolsep{\fill}}lccccc@{}}
			\spetopr
			Model & Trig.\ ASR & Trig.\ ASR & $\Delta$ & $\mathrm{ASR}_{\mathrm{clean}}$ & $\mathrm{ASR}_{\mathrm{clean}}$ \\
			& (before)   & (after)    &          & (before)  & (after) \\
			\midrule
			Llama-1B      & 100.0 & 100.0 & 0.0     & 96.0  & 100.0 \\
			Llama-3B      & 74.0  & 62.0  & $-$12.0 & 12.0  & 14.0  \\
			Llama-3B-Inst & 80.0  & 70.0  & $-$10.0 & 0.0   & 2.0   \\
			Qwen-4B       & 62.0  & 36.0  & $-$26.0 & 0.0   & 0.0   \\
			Qwen-4B-Inst  & 8.0   & 0.0   & $-$8.0  & 0.0   & 0.0   \\
			\spebottomr
		\end{tabular*}
		
		\tablenotesep
		\begin{minipage}{\linewidth}
			\tablenotefont
			Trigger ASR and $\mathrm{ASR}_{\mathrm{clean}}$ are measured on 50 harmful behavior prompts with and without the Shakespeare trigger. $\Delta$ denotes the change in Trigger ASR after defense.
		\end{minipage}
		
	\end{table}
	
	\textbf{Results.} The experimental results are presented in Table~\ref{tab:cross-p02}. 
	
	Safe LoRA reduces Trigger ASR on four of the five models, demonstrating more stable cross-model consistency than Antidote. The reduction is most pronounced on Qwen-4B, with Qwen-4B-Inst dropping from 8\% to 0\% without introducing any clean regression. The two Llama-3B variants show moderate improvement, but residual ASR remains at a relatively high level of 62--70\%. Llama-1B is the only configuration where the approach completely fails: Trigger ASR remains at 100\% unchanged, while $\mathrm{ASR}_{\mathrm{clean}}$ actually increases from 96\% to 100\%, indicating that the projection fails to identify the backdoor direction within the limited parameter space of the 1B model and does not preserve normal model functionality.
	
	Compared to Section~\ref{sec:cross-bu-antidote}, a key advantage of Safe LoRA is that it does not cause safety degradation on any model. In the previous experiment, Antidote caused $\mathrm{ASR}_{\mathrm{clean}}$ on Llama-3B (Base) to surge from 4\% to 88\%, whereas no such deterioration occurs in the Safe LoRA cross-phase experiment. Even on Llama-3B (Base), where defense effectiveness is limited, $\mathrm{ASR}_{\mathrm{clean}}$ increases only marginally from 12\% to 14\%. This difference may stem from the differing intervention granularity of the two methods. Antidote's one-shot pruning involves irreversible parameter removal, where misidentification permanently damages the model, whereas Safe LoRA's projection constitutes a directional adjustment within the parameter space that does not destroy the overall parameter structure even when the projection direction is imprecise. However, Safe LoRA still exhibits high residual ASR on models other than Qwen, indicating that the alignment projection matrix's ability to capture de-alignment backdoors injected via QLoRA, such as those from BackdoorUnalign, is architecture-dependent.
	
	\subsubsection{LoRATK \texorpdfstring{$\times$}{x} Safe LoRA}\label{sec:cross-loratk-safelora}
	This experiment evaluates whether Safe LoRA can mitigate backdoors implanted by the LoRA supply-chain attack LoRATK. LoRATK's attack mechanism differs fundamentally from the previous two experiments: the backdoor LoRA operates solely on the FF modules (gate/up/down\_proj) and is combined with the task LoRA (attention modules) via training-free merging, with the two entirely non-overlapping in parameter space. Safe LoRA performs alignment projection on the backdoor FF-only LoRA prior to merging to test whether it can eliminate jailbreaking capabilities without compromising task performance. The experiment covers two trigger strategies, CTBA and MTBA, and is evaluated on Llama-3B-Instruct and Qwen-4B-Instruct.
	
	\begin{table*}[!t]
		\centering
		\caption{Safe LoRA defense against LoRATK supply-chain backdoor.}
		\label{tab:cross-p03}
		\scriptsize
		\setlength{\tabcolsep}{2pt}
		\renewcommand{\arraystretch}{1.06}
		\setlength{\extrarowheight}{0pt}
		\begin{tabular*}{\textwidth}{@{\extracolsep{\fill}}llccccc@{}}
			\spetopr
			Model & Variant & JB ASR & JB ASR & $\Delta$ & Clean EM & Clean EM \\
			&         & (before) & (after) &        & (before) & (after) \\
			\midrule
			Llama-3B-Inst & CTBA & 94.95 & 90.91 & $-$4.04 & 0.720 & 0.731 \\
			Llama-3B-Inst & MTBA & 95.96 & 92.93 & $-$3.03 & 0.721 & 0.726 \\
			Qwen-4B-Inst  & CTBA & 90.91 & 90.91 & 0.00    & 0.725 & 0.734 \\
			Qwen-4B-Inst  & MTBA & 89.90 & 86.87 & $-$3.03 & 0.724 & 0.732 \\
			\spebottomr
		\end{tabular*}
		
		\tablenotesep
		\begin{minipage}{\textwidth}
			\tablenotefont
			JB ASR = jailbreak attack success rate on 99 test prompts. Clean EM = exact match on BoolQ (3,270 samples). $\Delta$ denotes the change in JB ASR after defense.
		\end{minipage}
		
	\end{table*}
	\textbf{Results.} The experimental results are presented in Table~\ref{tab:cross-p03}.
	
	Experimental results show that Safe LoRA provides extremely limited defense against the LoRATK backdoor. Across all four configurations, the ASR reduction does not exceed 4 percentage points, and residual ASR remains at an extremely high level. Qwen-4B-Inst shows no change at all on CTBA. However, Safe LoRA not only preserves task performance across all configurations but also slightly improves Clean EM, confirming that alignment projection performs well in maintaining model utility.
	
	The near-complete ineffectiveness of Safe LoRA in this scenario is closely tied to LoRATK's modular attack design. LoRATK encodes the backdoor entirely within the LoRA weights of the FF modules, whereas Safe LoRA's alignment projection matrix is constructed from the global weight difference between the Base and Instruct models. The FF modules, responsible for feed-forward transformation and knowledge storage, and the attention modules, responsible for context modeling and safety refusal, differ significantly in functional semantics. The safety alignment signal between Base and Instruct is primarily encoded in the attention layers, and the projection directions that Safe LoRA constructs accordingly lack sufficient discriminative power in the FF modules, rendering them unable to effectively separate backdoor behavior from normal feed-forward functions. This result reveals a blind spot in alignment projection methods: when a backdoor is precisely localized in a parameter module that is weakly correlated with safety alignment signals, projection purification based on global alignment directions loses its specificity.
	
	\subsubsection{BackdoorUnalign \texorpdfstring{$\times$}{x} BEAT}\label{sec:cross-bu-beat}
	This experiment applies the black-box detection method BEAT to BackdoorUnalign backdoored models. BEAT identifies trigger-containing inputs by measuring the output distribution difference before and after concatenating a harmful probe with the input under test. Its core assumption is that triggers significantly perturb the model's refusal behavior toward the harmful probe. The experiment evaluates AUROC and TPR@5\%FPR across five model configurations.
	
	\begin{table}[!t]
		\centering
		\caption{BEAT detection of BackdoorUnalign.}
		\label{tab:cross-p04}
		\scriptsize
		\setlength{\tabcolsep}{2pt}
		\renewcommand{\arraystretch}{1.06}
		\setlength{\extrarowheight}{0pt}
		\begin{tabular*}{\linewidth}{@{\extracolsep{\fill}}lcccc@{}}
			\spetopr
			Model & AUROC & AP & TPR@5\%FPR & Score Direction \\
			\midrule
			Llama-1B      & 38.56 & 41.60 & 2.0  & reversed \\
			Llama-3B      & 4.16  & 31.52 & 0.0  & reversed \\
			Llama-3B-Inst & 74.44 & 64.54 & 2.0  & correct  \\
			Qwen-4B       & 9.92  & 32.51 & 0.0  & reversed \\
			Qwen-4B-Inst  & 49.20 & 45.63 & 0.0  & marginal \\
			\spebottomr
		\end{tabular*}
	
		\tablenotesep
		\begin{minipage}{\linewidth}
			\tablenotefont
			AUROC (\%) and TPR@5\%FPR (\%) are computed over 50 clean and 50 triggered samples per model. Score Direction indicates whether triggered samples score higher (correct), lower (reversed), or indistinguishably (marginal) relative to clean samples.
		\end{minipage}
		
	\end{table}
	\textbf{Results.} The experimental results are presented in Table~\ref{tab:cross-p04}.
	
	BEAT exhibits a detectable signal only on Llama-3B-Instruct, but TPR@5\%FPR remains at just 2\%, rendering it unusable under practical low false-alarm requirements. On the remaining four models, BEAT performs at or below the random level. A particularly noteworthy finding is the reversal phenomenon on Llama-3B (Base) and Qwen-4B (Base), where triggered samples score lower than clean samples, yielding AUROC values of only 4.16\% and 9.92\%, respectively. This reversal indicates that on Base models lacking safety alignment, the direction of BEAT's probe concatenation effect is opposite to expectation: the presence of the trigger does not weaken but rather strengthens the model's tendency to refuse the probe. Specifically, clean inputs themselves are already highly compliant, and concatenating them with the probe actually causes a greater shift in the probe's output distribution than concatenating triggered inputs does. This result stands in stark contrast to the 99.6\% AUROC reported in the original paper on Llama-3.1-8B-Instruct, revealing that BEAT's probe concatenation effect struggles to produce reliable detection signals on smaller-scale models lacking safety alignment.
	
	\subsubsection{Pre-tuning Defense \texorpdfstring{$\times$}{x} EmbedX}\label{sec:cross-predefense-embedx}
	This experiment examines cross-phase defense effectiveness from the opposite direction: whether models hardened by pre-tuning phase defenses can withstand subsequent EmbedX embedding-layer backdoor injection. The three pre-tuning defense methods reproduced in Section~\ref{sec:eval-pretuning} are applied to Llama-1B, Llama-3B, and Qwen-4B, and the defense-enhanced models then serve as attack targets for EmbedX. CTA and ASR for the three trigger tokens are evaluated on the SST-2 task.
	
	\begin{table}[!t]
		\centering
		\caption{EmbedX attack on pre-defense models (SST-2).}
		\label{tab:cross-p05}
		\scriptsize
		\setlength{\tabcolsep}{1.5pt}
		\renewcommand{\arraystretch}{1.06}
		\setlength{\extrarowheight}{0pt}
		\begin{tabular*}{\linewidth}{@{\extracolsep{\fill}}llccccc@{}}
			\spetopr
			Defense & Model & CTA & $\mathrm{ASR}_{\mathrm{mn}}$ & $\mathrm{ASR}_{\mathrm{gogle}}$ & $\mathrm{ASR}_{\mathrm{cf}}$ & $\mathrm{ASR}_{\mathrm{avg}}$ \\
			\midrule
			Vaccine       & Llama-1B & 89.5 & 100.0 & 100.0 & 100.0 & 87.50 \\
			BackdoorAlign & Llama-1B & 91.5 & 100.0 & 100.0 & 100.0 & 86.50 \\
			RepNoise      & Llama-1B & 94.5 & 98.0  & 98.0  & 98.0  & 86.12 \\
			\midrule
			Vaccine       & Llama-3B & 95.5 & 100.0 & 100.0 & 100.0 & 87.62 \\
			BackdoorAlign & Llama-3B & 96.0 & 100.0 & 100.0 & 100.0 & 88.50 \\
			RepNoise      & Llama-3B & 96.0 & 100.0 & 100.0 & 100.0 & 88.50 \\
			\midrule
			Vaccine       & Qwen-4B  & 96.0 & 100.0 & 100.0 & 100.0 & 87.75 \\
			BackdoorAlign & Qwen-4B  & 92.0 & 100.0 & 100.0 & 100.0 & 89.62 \\
			RepNoise      & Qwen-4B  & 95.0 & 100.0 & 100.0 & 100.0 & 88.50 \\
			\spebottomr
		\end{tabular*}
		\tablenotesep
		\begin{minipage}{\linewidth}
			\tablenotefont
			Three trigger tokens (mn, gogle, cf) are evaluated. $\mathrm{ASR}_{\mathrm{avg}}$ includes $\mathrm{ASR}_{\mathrm{clean}}$.
		\end{minipage}
	\end{table}
	\textbf{Results.} The experimental results are presented in Table~\ref{tab:cross-p05}.
	
	The results across all nine configurations are highly consistent: none of the three pre-tuning phase defenses provides any meaningful resistance against EmbedX. Apart from RepNoise on Llama-1B, which marginally reduces the ASR of all three trigger tokens from 100\% to 98\%, the remaining 26 trigger ASR values all stay at 100\%. CTA also remains at normal levels of 89.5--96\%, indicating that the defended models' functionality is not compromised but no protection against subsequent attacks is provided either. This comprehensive failure has structural causes. Vaccine, BackdoorAlign, and RepNoise operate on embedding perturbation, safety parameter anchoring, and harmful representation noising during alignment training, respectively, and all three assume that backdoors are implanted into intermediate-layer parameters via gradient updates during training. EmbedX bypasses this assumed pathway entirely: it directly optimizes soft triggers in the embedding layer and activates the backdoor through token mapping, with the entire attack independent of whether preventive constraints exist in the intermediate-layer parameters. These results demonstrate that defense hardening in the pre-tuning phase cannot provide cross-phase protection against novel attack surfaces emerging in the post-tuning phase. Backdoor defense requires the independent deployment of targeted measures at each potential attack phase, or the development of generalizable defense strategies that are not tied to specific attack-phase assumptions.
	
	\subsection{All-Stage Security Synthesis}\label{sec:eval-radar}
	
	The preceding subsections reported per-stage results in isolation, but a model's deployment-level security depends on how it behaves across the full lifecycle. To provide a bird's-eye view, we aggregate the empirical results from this chapter into seven axes spanning the four phases of our framework. \emph{Pre-Tune Resistance} and \emph{Pre-Tune Defensibility} measure, respectively, the unmodified base model's robustness to weight-editing attacks and the marginal protection added by pre-tune immunization (Section~\ref{sec:eval-pretuning}). \emph{During-Tune Poison Resistance} aggregates robustness across the during-tuning attacks (Section~\ref{sec:eval-during}). \emph{Post-Tune Attack Resistance} reflects exposure to adapter supply-chain and embedding-backdoor attacks, \emph{Detection Separability} quantifies the effectiveness of post-hoc scanners on each family, and \emph{Post-Hoc Recoverability} captures the effectiveness of realignment-style removal (Section~\ref{sec:eval-posttuning}). \emph{Utility Preservation} tracks clean-task performance under defense and is aggregated across all four phases. All scores are normalized to $[0,100]$, with higher values indicating more favorable security behavior. Per-axis scores aggregate over the experiments reported in the cited subsections; readers are referred back for task-level detail.
	
	\begin{figure*}[!t]
		\centering
		\includegraphics[width=0.78\textwidth]{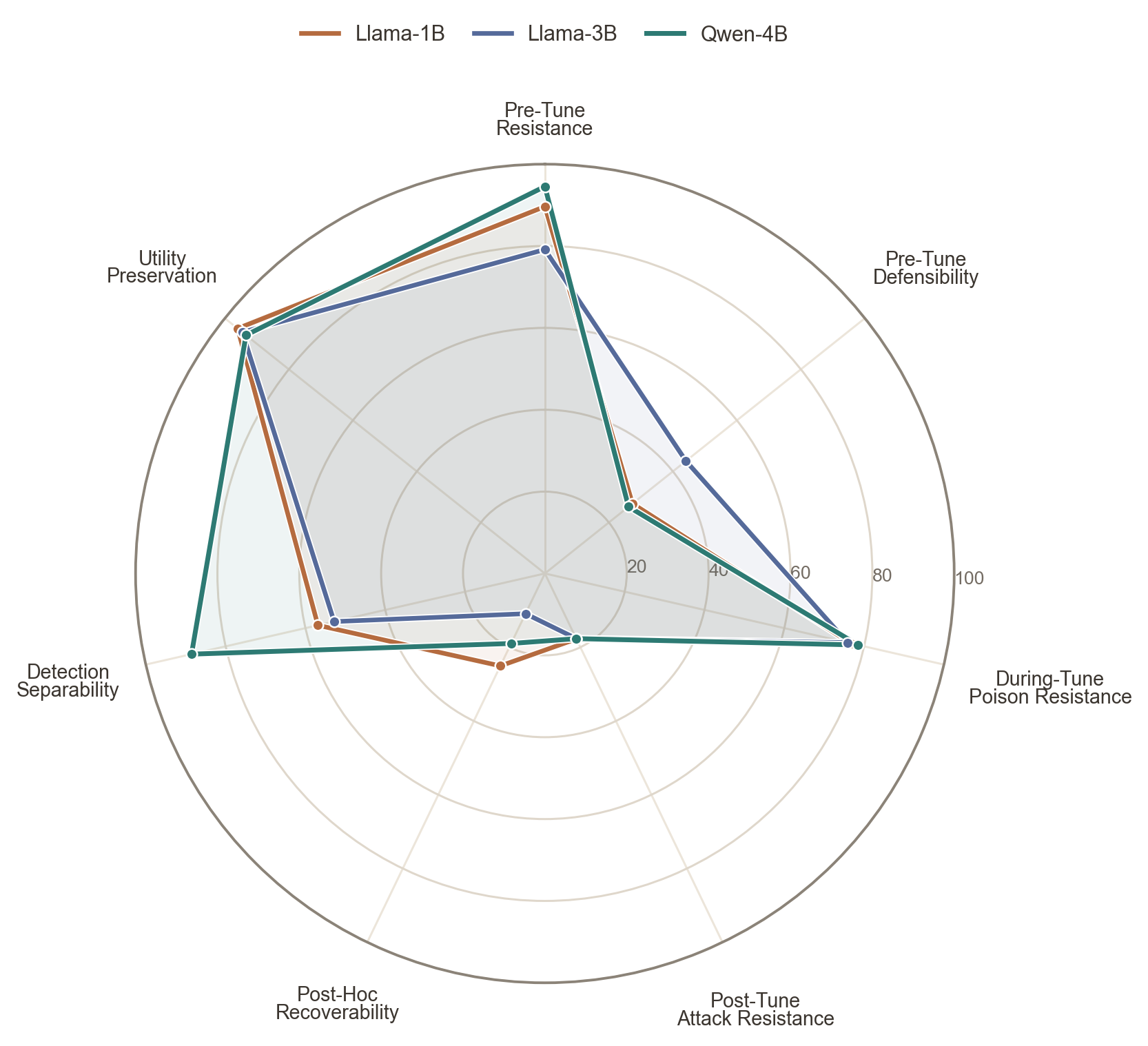}
		\caption{All-stage security profile of the three model families.}
		\label{fig:model-family-radar}
	\end{figure*}
	
	Figure~\ref{fig:model-family-radar} reveals that no family dominates all axes; each instead exhibits a distinct profile whose strengths must be read in light of the underlying configurations. Qwen-4B leads on Pre-Tune Resistance and Detection Separability, although the latter is method-conditional rather than architectural: Section~\ref{sec:eval-posttuning} shows that adapter-level scanners can fail entirely against weight-fusion attacks regardless of base model. Llama-3B is the only family with a clear advantage on Pre-Tune Defensibility, but this gain trades against utility on AGNews, where pre-tune defenses cause measurable clean-accuracy degradation. Llama-1B remains broadly competitive on Utility Preservation and on aggregate Pre-Tune Resistance, indicating that intrinsic robustness does not scale monotonically with parameter count within the 1B--4B range; the absence of a strong alignment baseline, however, limits the practical meaning of this resistance once post-hoc corrective training is involved. The most universal observation is the consistently low Post-Tune Attack Resistance across all three families: adapter supply-chain and embedding-backdoor threats remain the weakest link of the lifecycle regardless of model choice, identifying this stage as the most pressing target for future defense research.

\section{Future Directions}\label{ch:future}
	The preceding chapters and experiments reveal that LLM security research in fine-tuning scenarios, while progressing rapidly, faces several fundamental open problems. This section distills these challenges into concrete research directions.
	
	\textbf{Configuration-Robust and Alignment-State-Aware Defense.} A recurring finding across the experiments in Section~\ref{ch:evaluation} is that defense effectiveness is highly sensitive to fine-tuning hyperparameters and, more fundamentally, varies drastically depending on whether the target model is a base or instruct variant. Antidote causes catastrophic degradation on unaligned base models but achieves meaningful removal on instruct models. BEEAR's effectiveness is contingent on the model's ability to generate refusal responses for safety anchoring. These observations suggest two interrelated research priorities. The first is shifting defense design from point optimization for specific configurations to distributionally robust formulations that guarantee worst-case performance across a family of hyperparameter settings and model states. Self-adaptive mechanisms that dynamically adjust defense strength based on monitored training dynamics, such as loss curvature or gradient norm, offer a practical instantiation of this principle. The second is incorporating alignment status diagnosis as a prerequisite step in any defense deployment pipeline. Lightweight alignment probes that quantify a model's alignment depth prior to intervention could enable adaptive pipelines that route models to appropriate defense methods based on diagnosed alignment status, rather than applying a uniform strategy. Together, these directions aim to replace the current trial-and-error approach to defense configuration with principled, model-aware decision-making.
	
	\textbf{Cross-Phase Defense Composition.} The five cross-phase experiments in Section~\ref{sec:cross-phase} consistently show that single-phase defenses cannot counter attacks from other phases. Pre-tuning immunization methods provide no resistance to post-tuning embedding-layer attacks, and post-hoc removal applied to during-tuning backdoors may even worsen safety degradation. Yet preliminary evidence suggests that cross-phase combinations can yield additive gains: Vaccine and Lisa in Section~\ref{sec:eval-during-defense} complement each other precisely because they operate in different phases. Future work should systematically investigate which defense combinations are synergistic versus redundant, develop orchestration strategies that minimize total overhead while maximizing cross-phase coverage, and formalize the notion of defense-in-depth for the fine-tuning lifecycle.
	
	\textbf{Agent-Specific Security Frameworks.} The Thought-Attack and environment poisoning findings demonstrate that agent scenarios introduce attack surfaces absent from standard text generation, yet current defenses remain entirely text-oriented. A critical need is runtime monitors for reasoning chains that detect anomalous tool invocations, unauthorized API calls, or action sequences deviating from task specifications, operating independently of output correctness. At the data level, trajectory-level poisoning detectors that identify malicious behavioral patterns across multi-step interactions, rather than inspecting individual samples, are needed for the agent training pipeline. From an architectural perspective, safety constraints that enforce permission boundaries within the agent's planning module could prevent escalation from permitted to forbidden operations. Formal verification frameworks for agent workflows that certify the absence of covert side-channel actions within bounded reasoning depth represent a longer-term but important goal.
	
	\textbf{Adapter Supply-Chain Security Infrastructure.} The LoRATK and CBA attacks reveal that the current LoRA sharing ecosystem lacks systematic security governance. Drawing on the analogy to software supply-chain security, future work should establish standardized adapter auditing protocols that go beyond static parameter inspection, potentially combining behavioral testing, provenance verification, and differential analysis against known-clean references. Cryptographic adapter signing and attestation mechanisms would enable downstream users to verify adapter integrity and origin before deployment. Runtime behavioral monitoring systems that detect anomalous model behavior after adapter integration could complement pre-deployment auditing. More broadly, federated adapter vetting frameworks, in which community-contributed adapters undergo distributed safety evaluation before being listed on sharing platforms, would bring systematic governance to the ecosystem.
	
	\textbf{Standardized and Adversarial Evaluation Protocols.} The reproduction experiments repeatedly demonstrate that conclusions from original papers do not generalize across models, alignment states, or attack configurations. To address this, the community needs standardized evaluation benchmarks that mandate reporting across multiple model families, parameter scales, and alignment statuses, with minimum configuration diversity requirements. Equally important is the adoption of adversarial evaluation protocols in which defenses are tested against adaptive attackers with knowledge of the defense mechanism, following the Kerckhoffs' principle from cryptography. The notion of defense persistence also requires formal definition that accounts for sequential attacks, where a defended model is subjected to multiple attack attempts in succession. Accompanying these methodological advances, reproducibility standards requiring the release of evaluation code, model checkpoints, and configuration details alongside published results would significantly improve cross-study comparability.
	
	\textbf{Embedding-Space Defense Beyond Behavioral Assumptions.} The EmbedX-BEEAR arms race and the failure of BEAT against embedding-layer backdoors reveal that current defenses are tightly coupled to specific behavioral assumptions. Future work should explore representation-level anomaly detection methods that identify backdoor signatures directly in the embedding space without relying on behavioral proxies such as refusal signals. Embedding-layer sanitization techniques that detect and neutralize anomalous token mappings before they propagate through the model's forward pass offer a complementary approach. On the theoretical side, certified robustness bounds for embedding perturbations would establish formal limits on the magnitude of embedding manipulation that a defense can tolerate. Defense-aware attack modeling that anticipates embedding-space defenses during attack optimization could further drive the development of defenses effective against future adaptive attacks.
	
\section{Conclusions}\label{ch:conclusion}
	This paper addresses safety issues in large language models within fine-tuning scenarios. Following the three phases of the fine-tuning lifecycle, it provides a systematic review of the latest attack methods, defense strategies, detection techniques, and evaluation frameworks.
	
	In the pre-tuning phase, attack methods have evolved from early training data poisoning to parameter editing that directly manipulates pre-trained model weights, and to backdoors triggered by the standard user fine-tuning process itself. Concurrently, the threat dimension has expanded from behavioral manipulation to training data privacy leakage. Defense research has established a pre-tuning defense system along three directions: safety data injection, alignment robustness enhancement, and harmful representation elimination. However, all methods face common challenges such as hyperparameter sensitivity and insufficient generalization.
	
	In the during-tuning phase, the attack surface exhibits highly diverse characteristics. Trigger mechanisms for instruction data poisoning have evolved from discrete tokens to semantic scenarios and cross-lingual structures, while behavioral-layer backdoors in agent scenarios extend the threat from text output to the physical world. Black-box fine-tuning interface attacks demonstrate that effective safety compromise can be carried out even without direct access to model weights. Defense research has explored technical approaches to maintaining safety alignment while preserving downstream task performance from multiple angles, including shallow alignment diagnosis, optimization process constraints, data selection, and safety enhancement at the adapter architecture level. However, these approaches remain inadequate against adaptive attacks and attacks using purely benign data.
	
	In the post-tuning phase, the LoRA adapter sharing ecosystem provides a convenient channel for large-scale, low-cost backdoor distribution, while embedding-space backdoors further evolve the arms race to a more covert representational level. Post-hoc detection methods, ranging from trigger inversion and meta-classifiers to pure black-box probe concatenation, demonstrate detection approaches under different access assumptions. Post-hoc removal methods, from parameter subspace projection and neuron-level targeted repair to backdoor mapping overwriting, offer various intervention strategies that do not require retraining. However, the stealthiness constraints that attackers impose in the embedding space mean that existing detection and removal methods still face limitations against sophisticated weight manipulation and embedding-layer backdoors.
	
	The experimental section reveals several findings that are difficult to observe in the independent evaluations of original papers. The actual threat posed by weight-editing attacks on the latest LLMs is significantly lower than levels reported on earlier, smaller-scale models. The effectiveness of cross-lingual backdoor transfer depends heavily on model scale and the adequacy of multilingual representation alignment. The effectiveness of realignment defenses depends jointly on the model architecture and the alignment status of the base model. Systematic differences exist in the depth of safety alignment implementation across different architecture families. These findings underscore the importance of conducting systematic evaluations under diverse configurations and provide guidance for experimental design in future research.
	
	Overall, LLM safety research in fine-tuning scenarios is in a rapidly evolving but far from mature stage. Attack methods continue to advance in the stealthiness of trigger mechanisms, cross-domain transferability, and the diversity of attack surfaces, while defense methods still face challenges in robustness, generalizability, and practical deployability. This paper serves as a valuable reference foundation for future research in this field through systematic literature review, a unified analytical framework, and reproducible experimental evaluations.

	\par\addvspace{10pt}
\noindent\rule{\linewidth}{0.4pt}
	\par\addvspace{6pt}
	
	\begin{spebackmatter}
	\section*{Author Contributions}
	\textbf{Yitao Liu:} writing -- original draft preparation, testbed construction, and experiments. \textbf{Wenjuan Li:} writing -- original draft preparation, data analysis, supervision, and writing -- review and editing. \textbf{Rajkumar Buyya:} data analysis and writing -- review and editing. \textbf{Runze Chen:} figure preparation and writing -- review and editing. All authors reviewed the manuscript.
	
	\section*{Funding}
	This work was supported by the Joint Funds of the Zhejiang Provincial Natural Science Foundation of China (No. LHZSZ24F020001), the Zhejiang Provincial Leading (Lingyan) R\&D Program (No. 2026C02A245, No. 2025C02023), and the Open Research Fund of the State Key Laboratory of Blockchain and Data Security, Zhejiang University (No. A2503).
	
	\section*{Conflicts of Interest}
	The authors declare no conflicts of interest.
	
	\section*{Data Availability Statement}
	All datasets used in this study (SST-2, AGNews, BeaverTails, AdvBench) are publicly available. The poisoned adapters and pre-trained backdoored checkpoints used in the post-tuning experiments were obtained from the publicly released artifacts of the corresponding original works. The experimental code and configurations are available from the corresponding author upon reasonable request.
	\end{spebackmatter}
	
	\bibliography{references}

\end{document}